\documentclass[11pt]{article}

\usepackage[preprint]{acl}
\usepackage{times}
\usepackage{latexsym}

\usepackage[utf8]{inputenc} 
\usepackage[T1]{fontenc}    
\usepackage{hyperref}       
\usepackage{url}            
\usepackage{booktabs}       
\usepackage{amsfonts}       
\usepackage{nicefrac}       
\usepackage{microtype}      

\usepackage{amsmath}
\usepackage{cuted}
\usepackage{multirow}
\usepackage{array}
\usepackage[table, dvipsnames, svgnames]{xcolor}  
\usepackage{enumitem}
\usepackage{colortbl}
\usepackage{adjustbox}
\usepackage{makecell}
\usepackage[noend]{algpseudocode}
\usepackage{algorithm}
\usepackage{bm}
\usepackage{tcolorbox} 
\usepackage{subcaption}
\usepackage{arydshln}
\usepackage{caption}
\usepackage{pifont}

\definecolor{lightgray}{RGB}{242,242,242}
\definecolor{ourscolor}{RGB}{217,234,255}
\definecolor{blockgray}{RGB}{220,220,220}
\definecolor{oursblue}{RGB}{218,232,252}
\definecolor{blockgray}{RGB}{220,220,220}
\definecolor{oursblue}{RGB}{218,232,252}

\tcbuselibrary{listings,breakable,skins}
\usepackage{listings}

\definecolor{correctgreen}{RGB}{46,139,87}
\definecolor{incorrectred}{RGB}{205,51,51}
\definecolor{casebg}{RGB}{245,245,252}
\definecolor{oursorange}{RGB}{230,126,34} 
\definecolor{skillpurple}{RGB}{128,0,128}   
\lstdefinestyle{modeloutput}{
  basicstyle=\ttfamily\scriptsize,
  breaklines=true,
  breakautoindent=false,
  breakatwhitespace=false,
  columns=fullflexible,
  keepspaces=true,
  escapeinside={(*@}{@*)},
  frame=none,
  aboveskip=2pt,
  belowskip=2pt,
  xleftmargin=0pt,
  postbreak=\mbox{\textcolor{gray}{$\hookrightarrow$}\space},
  literate={—}{{\textemdash}}1 {–}{{\textendash}}1 {…}{{\ldots}}1,
}

\title{\texttt{MetaForge}: A Self-Evolving Multimodal Agent that Retrieves, Adapts, and Forges Tools On Demand}

\author{
\textbf{Shouang Wei\textsuperscript{1,*}, Houcheng Min\textsuperscript{1,*}, Xinpeng Dong\textsuperscript{2}, Xin Lin\textsuperscript{1}, Sen Cui\textsuperscript{3},} \\
\textbf{Bo Jiang\textsuperscript{1}, Zhongxiang Dai\textsuperscript{4}, Kun Kuang\textsuperscript{2}, Guandong Xu\textsuperscript{5}, Fei Wu\textsuperscript{2}, Min Zhang\textsuperscript{1,\ensuremath{\dagger},\ensuremath{\ddagger}}} \\[1.0em]
\textsuperscript{1}East China Normal University \quad
\textsuperscript{2}Zhejiang University \quad
\textsuperscript{3}Tsinghua University \\
\textsuperscript{4}The Chinese University of Hong Kong, Shenzhen \quad
\textsuperscript{5}The Education University of Hong Kong \\
\texttt{\{52285901019@stu.ecnu.edu.cn, mzhang@cs.ecnu.edu.cn\}}
}

\begin{document}
\maketitle

\renewcommand{\thefootnote}{\fnsymbol{footnote}}\footnotetext[1]{Equal contribution. \, \ensuremath{\dagger} Corresponding author. \, \ensuremath{\ddagger} Project leader.}\renewcommand{\thefootnote}{\arabic{footnote}}\setcounter{footnote}{0}

\begin{abstract}
Multimodal agents have achieved notable progress on complex reasoning tasks through tool use, yet remain limited by two issues: \textit{statically predefined tool inventories} fail to generalize to unseen scenarios, and \textit{indiscriminate tool invocation} incurs redundant cost and noise-induced errors. We propose \texttt{MetaForge}, a multimodal agent framework that learns when to invoke tools and how to evolve its toolset on demand. MetaForge factorizes agentic behavior into four coupled stages: \textbf{Decide} (judging whether tool use is warranted), \textbf{Retrieve} (selecting suitable tools), \textbf{Adapt} (grounding tool parameters in task context), and \textbf{Forge} (synthesizing new skills online and recycling them into the tool library for reuse), forming a closed judge–retrieve–adapt–forge–recycle loop. A unified orchestration policy enables the agent to choose among answering directly, reusing existing tools, or forging new ones. We jointly optimize invocation necessity, retrieval accuracy, execution effectiveness, and forged-skill reusability via reinforcement learning, with an explicit invocation-cost penalty discouraging redundant calls. Across 12 benchmarks, MetaForge consistently surpasses 16 baselines in accuracy,  efficiency, and generalization, validating a paradigm shift from static tool inventories to on-demand self-evolution.
\end{abstract}

\section{Introduction}

Multimodal agents have made significant progress in recent years in visual understanding and complex reasoning tasks~\cite{jian2025teaching, ning2025gns, liao2025doclayllm}. However, many real-world scenarios still exceed their built-in capabilities, such as fine-grained text recognition and complex visual analysis. To bridge these capability gaps, existing studies have integrated external tools, such as OCR and image-processing modules, into multimodal agents to enhance their task-solving abilities.

\begin{figure}[htpb]
  \centering
  \includegraphics[width=0.47\textwidth]{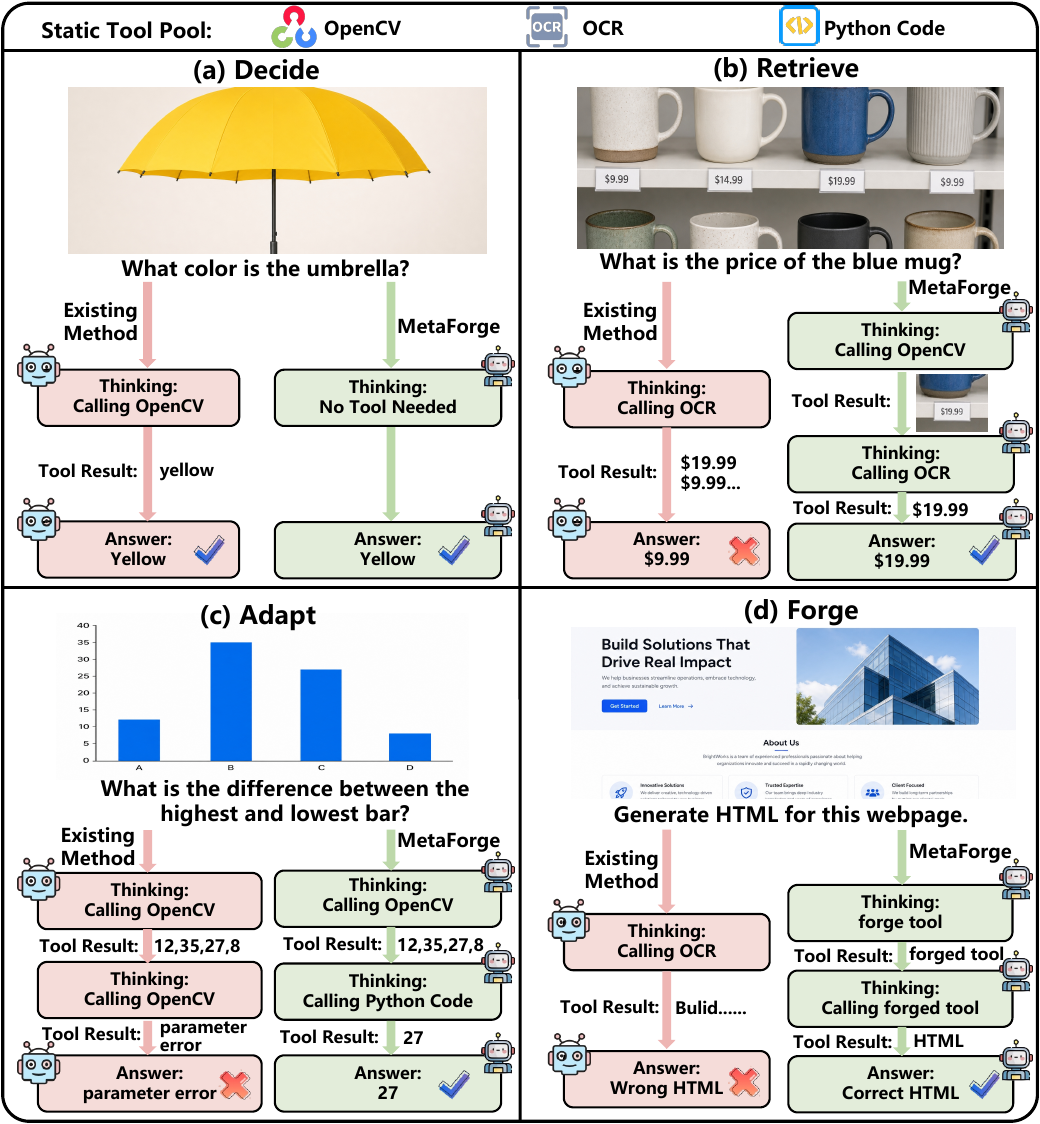}
\caption{Comparison examples between existing tool-augmented methods and our method.}
  \label{fig:motivation_decisions}
\end{figure}

Existing work on tool-augmented multimodal agents mainly falls into two categories. One type uses a single tool intensively to obtain fine-grained visual evidence and external knowledge through cropping, zooming, or multimodal retrieval~\cite{fangrit, su2025openthinkimg, zheng2025deepeyes, narayan2025deepmmsearch, wu2025mmsearch}. The other type builds multi-tool orchestration frameworks that integrate code interpreters, specialized vision models, and search engines into a reasoning loop~\cite{zhang2025thyme, zhaopyvision, lu2025octotools}.

Although these methods improve their performance on specific tasks, they still face two core limitations when they use tools for complex reasoning tasks. First, statically predefined tool inventories fail to generalize to unseen scenarios. Second, indiscriminate tool invocation incurs redundant cost and noise-induced errors. As Fig.~\ref{fig:motivation_decisions} shows, these limitations lead to four failure modes. In (a), the model cannot accurately judge whether it needs a tool. It therefore makes redundant calls in scenarios that allow direct answers. In (b), the model retrieves a tool, but the selected tool does not match the task requirement. This mismatch causes execution failure. In (c), the model identifies relevant tools, but it cannot adapt the call parameters to the current task context. This causes parameter errors. In (d), the task falls outside the coverage of the statically predefined tool inventory. The model lacks an on-demand tool forging mechanism, so it cannot close the capability gap and ultimately fails. 
Overall, these cases show that existing methods lack a unified mechanism for deciding whether to invoke tools, adapting tool calls, and expanding the toolset when capability gaps arise.

To this end, we propose \texttt{MetaForge}, a self-evolving multimodal agent that retrieves, adapts, and forges tools on demand. \texttt{MetaForge} models tool use as a closed-loop capability orchestration process consisting of \textbf{Decide–Retrieve–Adapt–Forge}: the agent first determines whether an external tool is needed; if so, it retrieves suitable tools from a dynamic tool library and adapts the invocation according to the task context. When existing tools cannot satisfy the requirement, the system forges new skills online and, after validation, registers them back into the tool library for future reuse. Through this mechanism, \texttt{MetaForge} can continuously expand its capabilities across different tasks and tools, rather than being limited to a predefined set of tools.

To train the model to learn such an adaptive orchestration strategy, we adopt multi-turn GRPO training and design a composite reward function that jointly optimizes answer correctness, tool retrieval, and parameter adaptation, invocation necessity, skill forging and reuse, and format parsability. In particular, we introduce cost-related rewards to reduce unnecessary tool calls, thereby improving inference efficiency while maintaining performance. In addition, to address the potential expansion of the tool library caused by online skill accumulation, we design a capacity-constrained dynamic tool-library management mechanism. The main contributions are as follows:
\begin{itemize}
    \item We propose \texttt{MetaForge}, a framework for static tool users into on-demand, self-evolving capability orchestrators. With an online skill-forging mechanism and dynamic tool management, the agent can identify capability, generate new tools, and reuse them efficiently.
    \item We propose a GRPO training paradigm for multi-turn tool interaction, using a composite reward to jointly optimize tool invocation, adaptation, forging, reuse, and efficiency.
    \item We conduct systematic experiments on multiple multimodal benchmarks under both IID and OOD tool settings, demonstrating the advantages of \texttt{MetaForge} in terms of accuracy, efficiency, and generalization.
\end{itemize}

\begin{figure*}[htpb]
  \centering
  \includegraphics[width=1.0\textwidth]{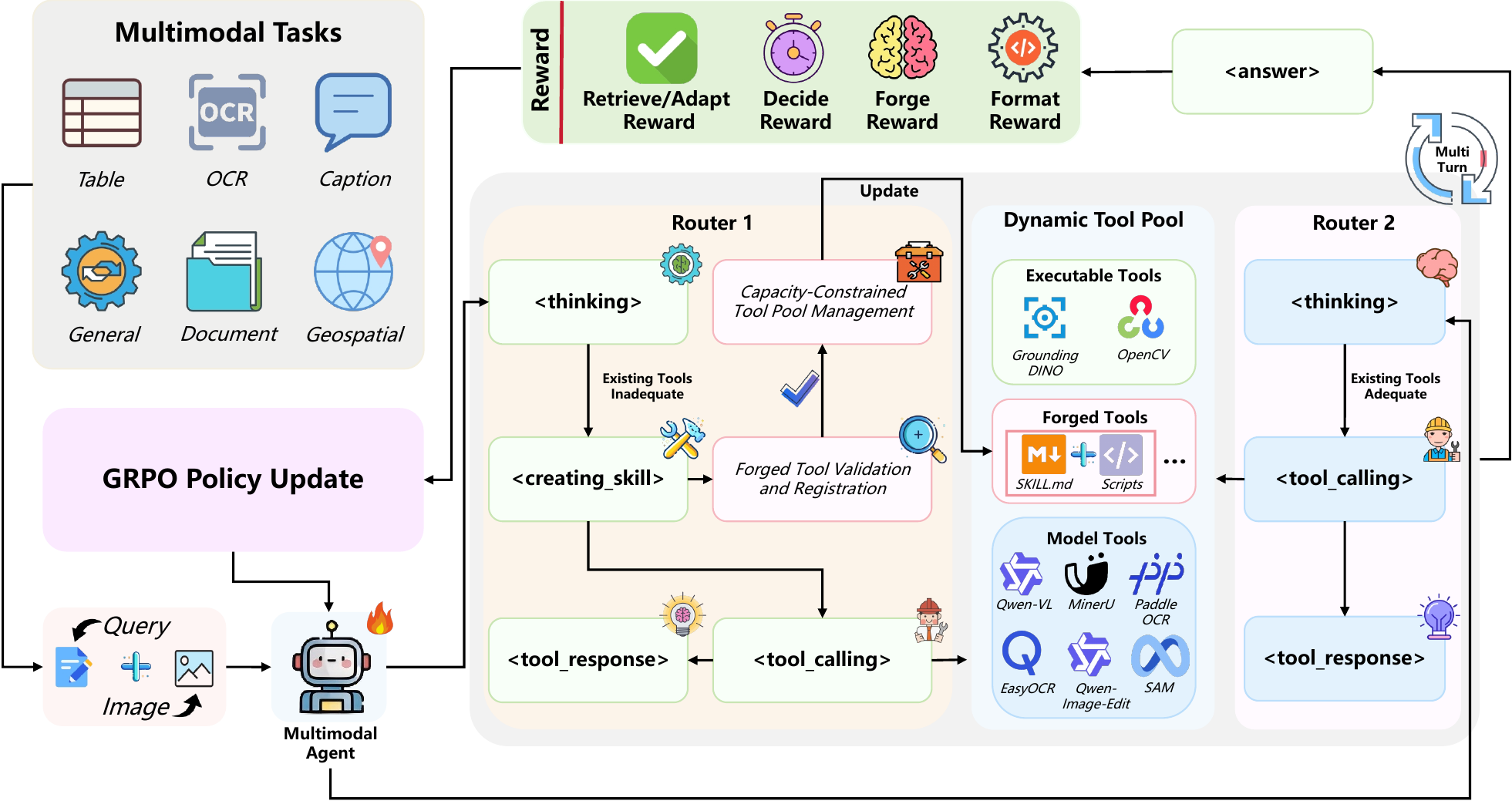}
  \caption{Overview of MetaForge, a self-evolving multimodal agent that retrieves, adapts, and forges tools on demand. At each reasoning turn, \textbf{Router~1} triggers \texttt{<creating\_skill>} when existing tools are inadequate, and the forged tool is validated and registered into the Dynamic Tool Pool; \textbf{Router~2} issues \texttt{<tool\_calling>} when existing tools are adequate. The Dynamic Tool Pool contains Executable Tools, Model Tools, and Forged Tools. After multi-turn interaction, MetaForge outputs the final \texttt{<answer>}, while Retrieve/Adapt, Decide, Forge, and Format rewards supervise GRPO policy updates.}
  \label{fig:frame}
\end{figure*}

\section{Related Work}

\subsection{Multimodal Tool-Augmented Reasoning}
Multimodal tool-augmented reasoning mainly follows two lines. The first line focuses on deep use of one or a few visual tools, using cropping, zooming, or retrieval to obtain fine-grained visual evidence and external knowledge~\cite{fangrit, su2025openthinkimg, zheng2025deepeyes, narayan2025deepmmsearch, wu2025mmsearch}. The second line builds tool-augmented or multi-tool orchestration frameworks that integrate code execution, search, vision models, and specialized perception tools into the reasoning loop. Some methods further train models to improve tool invocation~\cite{zhang2025thyme, zhaopyvision, zhao2026pyvision, lu2025octotools, song2026adareasoner, wu2025vtool, yang2025r1, hong2025deepeyesv2, su2025pixel, ding2026tikart, wei2026zooming}. These methods improve model performance on specific tasks, but they usually operate within statically predefined tool inventories and fixed invocation patterns in practice. As a result, they may struggle to decide whether tool use is warranted, retrieve a suitable tool, or adapt the invocation to the current task context. Our work instead learns adaptive orchestration, allowing the model to decide whether to answer directly, reuse existing tools, adapt tool calls, or forge new ones on demand when needed.

\subsection{Dynamic Skill Synthesis and Evolution}
To overcome the capability ceiling of statically predefined tool inventories, prior work explores reusable skill synthesis and evolution. Offline skill synthesis generates reusable skills from task descriptions before deployment~\cite{qian2023creator, yuancraft, cailarge, wolflein2025llm}, while experience-based skill evolution uses interaction histories to retain, revise, or remove skills~\cite{wangvoyager, qiu2025alita, huang2025toolace, sun2025seagent}. Recent work also extracts structured skills from historical multi-path trajectories for retrieval and adaptation during inference~\cite{jiang2026xskill}. These approaches provide useful reusable knowledge, but often depend on predefined task descriptions or accumulated historical trajectories, and do not address cases where a current multimodal task exposes a capability gap in the available tool pool. In contrast, our online tool forging mechanism identifies such gaps during interaction, generates executable skills through planning and implementation, validates them, and registers them back into the tool pool for future reuse.

\section{Methodology}

\label{sec:method}

\subsection{Problem Formulation}
\label{sec:formulation}

We propose MetaForge, a self-evolving multimodal agent for adaptive tool orchestration. Given $x = [I, T]$, where $I \in \mathcal{I}$ is visual input and $T$ the textual prompt, MetaForge outputs $\hat{y}$ through multi-turn interaction with a dynamic tool pool.

MetaForge divides the tool pool into three complementary types:
\begin{itemize}[nosep,leftmargin=1.2em]
\item \emph{Executable tools}: deterministic OCR, code interpretation, and symbolic computation;
\item \emph{Model tools}: specialized vision models for object detection and visual understanding;
\item \emph{Forged tools}: reusable programs that the agent creates on demand when existing tools cannot satisfy the current task. In implementation, each forged tool is a skill with a Markdown specification, a JSON schema, and executable scripts.
\end{itemize}
All tools share a unified interface $f = (\mathit{name}, \mathit{desc}, \mathit{schema}, \mathit{type})$, e.g.:
\begin{equation*}
\begin{split}
f_{\text{ex}} = \big(\;\underbrace{\texttt{Qwen3-VL-32B}}_{\mathit{name}},\;\underbrace{\text{``reasoning\,\ldots''}}_{\mathit{desc}}, \\[6pt]
\quad\underbrace{\{\texttt{prompt}\!:\!\text{str},\,\texttt{img\_idx}\!:\!\text{int}\}}_{\mathit{schema}},\,\underbrace{\text{model tool}}_{\mathit{type}}\;\big).
\end{split}
\end{equation*}

At turn $t$, the policy selects an action given the history $h_t = \{x, a_1, o_1, \dots, a_{t-1}, o_{t-1}\}$ and the tool pool $\mathcal{T}_t$:
\begin{equation}
a_t \sim \pi_\theta(a \mid h_t, \mathcal{T}_t),
\label{eq:tool_policy}
\end{equation}
where $a_t$ has three forms: invoking a tool from $\mathcal{T}_t$, forging a task-specific tool, or outputting the final answer. Each action yields an observation $o_t$ appended to history, and $\mathcal{T}_t$ expands when validated forged tools are added. The process ends when the policy answers or reaches $N$ turns.

\subsection{Online Tool Forging}
\label{sec:skill_creation}

Static tool sets cannot cover all requirements of diverse multimodal tasks. We introduce an online tool forging mechanism that enables the policy to expand the tool pool during interaction. When no existing tool can handle a task, the policy triggers a \texttt{create\_skill} action with a natural language requirement $d$. A tool-forging model $G_\phi$ then produces a forged tool through a two-stage process.

\subsubsection{Two-Stage Tool Forging}
\label{sec:two_stage}

\paragraph{Planning stage.} $G_\phi$ takes the requirement $d$ and produces a structured plan:
\begin{equation}
z = G_\phi^{\mathrm{plan}}(d),
\end{equation}
where $z$ specifies the forged tool name, description, parameter interface, and required scripts.

\paragraph{Implementation stage.} $G_\phi$ generates code for each planned script conditioned on $d$ and $z$:
\begin{equation}
c_i = G_\phi^{\mathrm{code}}(d, z, i), \quad i = 1, \ldots, N_z,
\end{equation}
where $N_z$ is the number of scripts.

Each forged tool is a triplet $s = (m_s, p_s, e_s)$: a Markdown description $m_s$ recording name, function, scenarios, and calling conventions; a JSON schema $p_s$ defining parameters, types, and constraints; and a script set $e_s$ containing Python scripts with parameterized inputs and outputs.

\subsubsection{Forged Tool Validation and Registration}
\label{sec:validation}

Each candidate forged tool must pass a two-part validation gate before registration:
\begin{equation}
v(\hat{s}; d) = v_{\mathrm{exec}}(\hat{s}) \cdot v_{\mathrm{sem}}(\hat{s}, d),
\end{equation}
where $v_{\mathrm{exec}} \in \{0,1\}$ checks whether all scripts compile and execute without errors, and $v_{\mathrm{sem}} \in \{0,1\}$ uses a judge model to verify that the forged tool's behavior matches $d$. Only candidates who pass both checks join the tool pool:
\begin{equation}
\mathcal{T}_{t+1} = \mathcal{T}_t \cup \mathcal{Q}^*_t,
\end{equation}
where $\mathcal{Q}^*_t$ denotes validated forged tools at $t$.

\subsection{Composite Reward Training via Multi-Turn GRPO}
\label{sec:optimization}

Online tool forging addresses missing tools, but the policy must also learn when to answer directly, when to retrieve an existing tool, how to adapt the tool call to the task, and when to forge or reuse a forged tool. We optimize the policy with Group Relative Policy Optimization (GRPO) and design a composite reward with four components: Retrieve/Adapt, Decide, Forge, and Format rewards.

\subsubsection{Reward Function Design}
\label{sec:reward}

For each input $x$, we sample a group of rollouts $\mathcal{G}(x) = \{\tau_1, \dots, \tau_M\}$ from the current policy, where each $\tau_i$ is a multi-turn trajectory. The reward for rollout $\tau$ decomposes as:
\begin{equation}
R(\tau) = r_{\mathrm{ra}}(\tau) + r_{\mathrm{dec}}(\tau) + r_{\mathrm{forge}}(\tau) + r_{\mathrm{fmt}}(\tau).
\end{equation}
Each component targets a distinct aspect of desirable agent behavior.

\noindent\textbf{retrieve/adapt reward.}
The retrieve/adapt reward is a binary signal for whether the answer matches ground truth:
\begin{equation}
r_{\mathrm{ra}}(\tau) =
\begin{cases}
1, & \text{if } \hat{y} = y, \\
0, & \text{otherwise}.
\end{cases}
\end{equation}
This reward optimizes the result of retrieving useful tools, adapting their calls to the task context, forging tools when needed, or answering directly when no tool is necessary.

\noindent\textbf{Decide reward.}
The decide reward provides an efficiency ranking within correct rollouts. It encourages the policy to decide when tool use is necessary and avoid redundant calls. Let $\mathcal{G}^+(x) = \{\tau \in \mathcal{G}(x) \mid \hat{y}_\tau = y\}$ denote the correct rollouts. For $\tau \in \mathcal{G}^+(x)$, the normalized latency is:
\begin{equation}
\hat{\ell}(\tau) =
\begin{cases}
\dfrac{\ell(\tau) - L_{\max}^+}{L_{\min}^+ - L_{\max}^+}, & L_{\max}^+ > L_{\min}^+, \\[6pt]
0, & \text{otherwise},
\end{cases}
\end{equation}
where $\ell(\tau)$ denotes the total tool-call latency of $\tau$, and $L_{\min}^+ = L_{\min}^+(x)$, $L_{\max}^+ = L_{\max}^+(x)$ are the minimum and maximum latencies in $\mathcal{G}^+(x)$. The Decide reward is:
\begin{equation}
r_{\mathrm{dec}}(\tau) =
\begin{cases}
\hat{\ell}(\tau), & \tau \in \mathcal{G}^+(x), \\
0, & \text{otherwise}.
\end{cases}
\end{equation}
By normalizing within the correct group, the Decide reward is mapped to the range $[0, 1]$, maintaining a consistent range with other reward components while assigning higher scores to faster, correctly completed trajectories. It is zero for incorrect rollouts, ensuring the policy is never incentivized to sacrifice correctness for speed.

\noindent\textbf{Forge reward.}
The forge reward encourages the policy to forge and reuse forged tools, but only within correct trajectories, preventing exploitation through trivial tool generation. Let $\mathcal{A}_\tau$ denote all tool actions in trajectory $\tau$:
\begin{equation}
\begin{split}
r_{\mathrm{forge}}(\tau)
&= \mathbf{1}[\mathrm{Correct}(\tau)]
\quad \cdot \big(\mathbf{1}[\mathrm{Forge}(\tau)] \\
&\qquad + \mathbf{1}[\mathrm{ReuseForged}(\tau)]\big)/2,
\end{split}
\end{equation}
where $\mathrm{Correct}(\tau)$ indicates that $\tau$ produces the correct final answer. The $\mathrm{Forge}$ term indicates successful online tool forging, requiring the policy to invoke \texttt{create\_skill} and produce a forged tool that passes validation (\S\ref{sec:validation}):
\begin{equation}
\mathrm{Forge}(\tau) \!=\! \mathbf{1}\big[\exists\, c \!\in\! \mathcal{A}_\tau : \mathrm{Create}(c) \!\land\! \mathrm{Valid}(c)\big].
\end{equation}
The $\mathrm{ReuseForged}$ term indicates successful invocation of previously registered forged tools:
\begin{equation}
\begin{split}
\mathrm{ReuseForged}(\tau)
&= \mathbf{1}\big[\exists\, c \!\in\! \mathcal{A}_\tau : \\
&\quad \mathrm{ForgedTool}(c) \!\land\! \mathrm{NoError}(c)\big],
\end{split}
\end{equation}
where $\mathrm{ForgedTool}(c)$ indicates that $c$ invokes a registered forged tool, and $\mathrm{NoError}(c)$ indicates error-free execution. Together, these terms incentivize a forge-and-reuse lifecycle: forging tools when existing tools are insufficient, and reusing them in later tasks to avoid redundant construction.

\noindent\textbf{Format reward.}
Each response must follow a structured format (e.g., tool-call syntax and answer delimiters). The format reward enforces this:
\begin{equation}
r_{\mathrm{fmt}}(\tau) =
\begin{cases}
0.2, & \text{format-compliant}, \\
-0.2, & \text{otherwise}.
\end{cases}
\end{equation}
Format-compliant trajectories receive a positive reward, while violations incur a mild penalty, encouraging the policy to maintain a consistent output structure throughout the interaction.

\subsubsection{Policy Optimization}
\label{sec:grpo}

We adopt GRPO to optimize the policy. For input $x$, the policy $\pi_{\theta_{\mathrm{old}}}$ samples $M$ rollouts and computes their rewards $R(\tau_i)$. Rewards are standardized within the group to obtain advantages:
\begin{equation}
A(\tau_i) = \frac{R(\tau_i) - \mathrm{mean}_{j=1}^{M} R(\tau_j)}{\mathrm{std}_{j=1}^{M} R(\tau_j) + \epsilon}.
\end{equation}
The policy is updated by maximizing a clipped surrogate with KL regularization. Let $\rho_i = \pi_\theta / \pi_{\theta_{\mathrm{old}}}$ denote the importance ratio and $\bar{\rho}_i = \mathrm{clip}(\rho_i, 1\!-\!\epsilon, 1\!+\!\epsilon)$ the clipped ratio:
\begin{equation}
\begin{split}
& \mathcal{J}_{\mathrm{GRPO}}(\theta) =
\mathbb{E}_{\substack{x \sim \mathcal{D} \\ \{\tau_i\} \sim \pi_{\theta_{\mathrm{old}}}}}
\bigg[
\frac{1}{M} \sum_{i=1}^{M}
\Big( \\
& \quad \min\!\big(
\rho_i A(\tau_i),\,
\bar{\rho}_i A(\tau_i)
\big)
- \beta\, \mathbb{D}_{\mathrm{KL}}(\pi_\theta \| \pi_{\mathrm{ref}})
\Big)
\bigg].
\end{split}
\end{equation}

\subsection{Capacity-Constrained Tool Pool Management}
\label{sec:pool_management}

As forged tools accumulate through online forging (\S\ref{sec:skill_creation}), the dynamic tool pool grows. Since all tool descriptions are injected into the model's context, an excessively large pool occupies the limited context window and dilutes attention to critical information. 
Meanwhile, overlapping forged tools waste the exploration budget during training. To address this, we manage the forged-tool portion of the pool through two mechanisms: semantic deduplication and usage-based retention, keeping it compact and non-redundant.

Let $\mathcal{P}_n$ denote the forged-tool pool before the $n$-th training step, and $K$ the maximum capacity. Each training step processes a batch of samples, each explored through multiple rollout trajectories with multi-turn interactions.

\begin{table*}[ht]
\centering
\resizebox{\textwidth}{!}{%
\setlength{\tabcolsep}{3.5pt}
\renewcommand{\arraystretch}{1.3}
\begin{tabular}{c|cccccc|c|cccccc|c|c}
\hline
\multirow{2}{*}{\textbf{Models\,($\downarrow$)\,/\,Datasets\,($\rightarrow$)}}
 & \multicolumn{7}{c|}{\textbf{IID Datasets}}
 & \multicolumn{7}{c|}{\textbf{OOD Datasets}}
 & \textbf{ALL} \\
\cline{2-16}
 & \textbf{DocVQA} & \textbf{TallyQA} & \textbf{OCRVQA} & \textbf{RoBUT} & \textbf{MapQA} & \textbf{Loc.Narr.} & \textbf{AVG.}
   & \textbf{ChartQA} & \textbf{AI2D} & \textbf{WebSight} & \textbf{MathVista} & \textbf{ScienceQA} & \textbf{CLEVR-M} & \textbf{AVG.} & \textbf{AVG.} \\
\hline
\multicolumn{16}{c}{\cellcolor{blockgray}\textbf{w/ IID Tools}} \\
\hline
 GPT-5.4 & 96.85&72.95&80.25&77.35&57.75&84.00&78.19 & 90.96&84.03&\underline{97.95}&72.00&87.45&36.40&78.13&78.16 \\
 Claude-Sonnet-4-6 & \textbf{98.18}&\underline{73.80}&82.20&\textbf{88.24}&81.40&83.71&\underline{84.59} & \textbf{92.12}&\textbf{92.22}&97.49&\underline{72.18}&\underline{87.90}&\textbf{93.15}&\textbf{89.18}&\underline{86.88} \\
 Gemini-3.1-Pro-Preview & 96.95&72.65&81.48&81.98&68.60&88.50&81.69 & \underline{91.68}&88.77&97.92&69.53&87.03&37.15&78.68&80.19 \\
 Qwen2.5-VL-7B & 93.43&68.22&71.83&70.24&45.31&99.10&74.69 & 74.95&48.94&71.47&57.04&73.73&64.40&65.09&69.89 \\
 Qwen3-VL-8B & 89.80&70.20&77.40&61.70&70.65&95.78&77.59 & 82.16&84.58&51.15&60.70&78.03&63.50&70.02&73.80 \\
 Qwen3-VL-32B & 94.70&72.75&\underline{82.70}&85.15&69.10&\underline{99.14}&83.92 & 86.76&58.32&83.15&71.90&87.50&69.35&76.16&80.04 \\
 InternVL3.5-14B & 87.60&71.18&66.46&69.98&59.20&72.27&71.12 & 80.83&67.08&44.00&35.07&69.06&66.88&60.49&65.80 \\
 InternVL3.5-38B & 91.80&73.50&72.90&70.20&\textbf{86.75}&78.35&78.92 & 85.76&74.95&49.15&28.70&85.56&76.75&66.81&72.86 \\
 VTool-R1-7B & 93.10&73.45&79.30&66.75&77.85&94.57&80.84 & 86.80&52.23&62.35&62.30&79.12&69.00&68.63&74.74 \\
 R1-Onevision-7B & 83.20&69.20&76.10&61.60&54.10&87.00&71.87 & 76.88&48.93&49.70&52.40&81.65&46.75&59.39&65.63 \\
 PyVision-Image-7B-RL & 69.95&41.50&37.25&47.45&37.45&48.00&46.93 & 47.28&38.89&25.15&33.20&31.73&50.75&37.83&42.38 \\
 DeepEyesV2-7B & 89.90&70.25&58.70&72.75&\underline{82.60}&50.71&70.82 & 85.16&27.04&28.30&66.80&62.86&71.10&56.88&63.85 \\
 PixelReasoner-RL-v1-7B & 92.30&71.90&77.25&67.15&74.65&93.35&79.43 & 81.44&83.45&70.30&63.10&79.12&63.70&73.52&76.48 \\
 AdaReasoner-7B & 89.40&66.25&73.20&68.70&67.75&91.71&76.17 & 82.16&85.39&66.55&55.50&74.71&46.35&68.44&72.31 \\
 TikArt-8B & 96.70&73.05&82.40&83.05&63.55&98.78&82.92 & 86.44&\underline{91.09}&64.95&70.30&84.08&66.95&77.30&80.11 \\
 ZwZ-8B & 96.80&73.65&81.50&80.35&67.60&98.92&83.14 & 85.52&72.63&56.25&70.50&86.01&62.55&72.24&77.69 \\
 \cellcolor{oursblue}\textbf{MetaForge} & \cellcolor{oursblue}\underline{97.90}&\cellcolor{oursblue}\textbf{80.52}&\cellcolor{oursblue}\textbf{86.51}&\cellcolor{oursblue}\underline{87.36}&\cellcolor{oursblue}77.98&\cellcolor{oursblue}\textbf{99.62}&\cellcolor{oursblue}\textbf{88.32} & \cellcolor{oursblue}90.49&\cellcolor{oursblue}89.60&\cellcolor{oursblue}\textbf{98.34}&\cellcolor{oursblue}\textbf{78.01}&\cellcolor{oursblue}\textbf{89.68}&\cellcolor{oursblue}\underline{81.70}&\cellcolor{oursblue}\underline{87.97}&\cellcolor{oursblue}\textbf{88.14} \\
\hline
\multicolumn{16}{c}{\cellcolor{blockgray}\textbf{w/ OOD Tools}} \\
\hline
 GPT-5.4 & 96.55&72.80&79.45&77.70&59.10&83.85&78.24 & 90.44&83.67&\underline{97.82}&73.00&87.20&35.95&78.01&78.13 \\
 Claude-Sonnet-4-6 & \textbf{98.08}&72.05&\underline{82.55}&\textbf{87.99}&81.30&83.50&84.25 & \textbf{92.12}&\textbf{92.81}&97.49&\underline{74.50}&86.16&\textbf{93.85}&\textbf{89.49}&\underline{86.87} \\
 Gemini-3.1-Pro-Preview & 96.13&72.70&81.25&81.80&70.25&87.57&81.62 & \underline{91.60}&88.65&97.70&69.65&86.82&39.25&78.95&80.28 \\
 Qwen2.5-VL-7B & 93.63&67.46&78.18&70.66&50.01&98.66&76.43 & 75.41&51.95&73.15&56.88&74.42&66.60&66.40&71.42 \\
 Qwen3-VL-8B & 90.20&71.15&76.95&65.35&72.85&95.71&78.70 & 82.04&84.68&51.70&58.80&77.93&63.20&69.73&74.21 \\
 Qwen3-VL-32B & 95.15&\underline{74.40}&78.65&83.40&77.35&98.78&\underline{84.62} & 86.16&59.74&83.05&70.00&\underline{88.20}&79.25&77.73&81.18 \\
 InternVL3.5-14B & 88.20&63.85&59.10&59.26&53.90&70.64&65.83 & 74.44&63.47&39.50&29.90&65.34&68.95&56.93&61.38 \\
 InternVL3.5-38B & 92.10&73.40&71.75&70.90&\underline{86.65}&76.21&78.50 & 85.80&75.65&36.65&29.70&84.08&\underline{80.55}&65.41&71.95 \\
 VTool-R1-7B & 93.25&73.10&77.90&69.40&78.25&92.14&80.67 & 86.40&53.88&62.10&63.50&81.01&70.85&69.62&75.15 \\
 R1-Onevision-7B & 83.90&69.10&75.75&61.30&53.35&87.21&71.77 & 77.08&50.06&49.35&52.70&80.61&48.65&59.74&65.76 \\
 PyVision-Image-7B-RL & 77.00&53.30&52.75&55.85&49.80&68.00&59.45 & 69.24&59.94&27.30&45.50&41.59&48.25&48.64&54.04 \\
 DeepEyesV2-7B & 93.05&73.10&56.75&74.60&84.05&70.21&75.29 & 86.60&29.56&47.00&64.20&63.55&74.55&60.91&68.10 \\
 PixelReasoner-RL-v1-7B & 91.85&73.55&78.45&66.10&71.80&93.78&79.26 & 80.24&82.90&75.75&61.30&80.56&65.65&74.40&76.83 \\
 AdaReasoner-7B & 92.30&66.80&71.25&71.10&69.60&81.85&75.48 & 81.04&86.56&71.60&58.10&75.45&49.10&70.31&72.90 \\
 TikArt-8B & 94.70&74.30&79.30&81.15&70.60&98.42&83.08 & 81.36&\underline{91.25}&69.35&70.30&83.83&68.70&77.47&80.27 \\
 ZwZ-8B & 95.60&73.90&81.70&80.85&75.50&\underline{98.92}&84.41 & 85.20&75.64&77.35&69.30&85.87&76.35&78.29&81.35 \\
 \cellcolor{oursblue}\textbf{MetaForge} & \cellcolor{oursblue}\underline{97.22}&\cellcolor{oursblue}\textbf{81.17}&\cellcolor{oursblue}\textbf{86.00}&\cellcolor{oursblue}\underline{87.67}&\cellcolor{oursblue}\textbf{89.14}&\cellcolor{oursblue}\textbf{99.58}&\cellcolor{oursblue}\textbf{90.13} & \cellcolor{oursblue}90.18&\cellcolor{oursblue}88.15&\cellcolor{oursblue}\textbf{98.72}&\cellcolor{oursblue}\textbf{79.78}&\cellcolor{oursblue}\textbf{89.73}&\cellcolor{oursblue}80.40&\cellcolor{oursblue}\underline{87.83}&\cellcolor{oursblue}\textbf{88.98} \\
\hline
\end{tabular}%
}
\caption{Accuracy (\%) across 12 multimodal benchmarks under IID and OOD tool settings. Results are grouped into IID and OOD datasets with per-group and overall averages. \textbf{Bold} and \underline{underline} denote the best and second-best per tool setting. All reported results are averaged over three independent runs.} 
\label{tab:main}
\end{table*}

\subsubsection{Forged Tool Merging}
\label{sec:merging}

During training step $n$, trajectories across samples may independently forge tools for similar needs, producing a validated forged-tool set $\mathcal{Q}^*_n$ with potential functional overlaps. 

We perform semantic deduplication to eliminate this redundancy: a judge model extracts each forged tool's semantic summary, groups forged tools with overlapping functionality (the number of forged tools in each group is the \emph{group size}), evaluates members within each group for completeness and generality, and retains the highest-quality one:
\begin{equation}
\tilde{\mathcal{Q}}_n = \mathrm{Merge}(\mathcal{Q}^*_n),
\end{equation}
where $\tilde{\mathcal{Q}}_n$ contains the highest-quality forged tool from each functional group.

\subsubsection{Usage-Based Retention}
\label{sec:retention}

For each forged tool $s$ in the pool, we track how many times it is invoked during each interaction turn and maintain a cumulative usage count $u_n(s)$ across training steps.

The deduplicated forged tools join the current pool to form $\mathcal{U}_n = \mathcal{P}_n \cup \tilde{\mathcal{Q}}_n$. If $|\mathcal{U}_n| \le K$, we set $\mathcal{P}_{n+1} = \mathcal{U}_n$. When the capacity is exceeded, we apply a usage-based retention policy that prioritizes two types of forged tools: well-established ones that the policy has repeatedly relied on, and newly forged ones addressing frequent needs. This allows the pool to preserve tools with either strong historical utility or high current demand. Specifically, we first retain existing forged tools whose cumulative usage count exceeds an upper-quartile threshold $\delta_n = Q_3 + 1.5\,\mathrm{IQR}$, computed over $\{u_n(s) \mid s \in \mathcal{P}_n\}$, forming the high-priority set:
\begin{equation}
\mathcal{H}_n = \{s \in \mathcal{P}_n \mid u_n(s) \ge \delta_n\}.
\end{equation}
The remaining capacity is filled with newly forged tools from $\tilde{\mathcal{Q}}_n$, ranked in descending order by group size from the merging stage. A larger group size indicates more training samples independently requested similar functionality, reflecting higher demand and thus higher retention priority.

\section{Experiments}

\subsection{Experimental Setup}


\textbf{Dataset and tool set splits.}
To evaluate task and tool generalization, we split datasets and tools into IID and OOD settings. The detailed dataset, tool, and split compositions appear in Appendices~\ref{app:splits}.

\paragraph{Metrics.}
We use accuracy (ACC) as the primary metric. Multiple-choice answers are evaluated by exact option match, and open-ended answers are judged by GPT-5~\cite{singh2025openai}. We also report tool usage behavior and inference efficiency. Tool Usage Entropy (TUE) measures the concentration of tool selection:
\begin{equation}
\label{eq:tue}
\mathrm{TUE} = -\sum_{i:\,w_i>0} \frac{w_i}{\sum_j w_j} \log_2 \frac{w_i}{\sum_j w_j},
\end{equation}
where $w_i$ is the number of calls to tool $i$ during evaluation, including forged tools, and lower TUE indicates more concentrated selection. Tool SR is the ratio of successful tool calls to all tool calls. Forged Tool Success Rate (FSR) is the ratio of forged tools that pass validation and enter the tool pool to all forging attempts. AET, RLA, and ITC measure interaction turns, end-to-end latency, and generated tokens.

\paragraph{Implementation details.}
We implement MetaForge on VerlTool~\cite{jiang2025verltool} and train Qwen3-VL-8B-Instruct~\cite{bai2025qwen3} with GRPO on 8 H200 GPUs. Each input uses $M{=}8$ rollouts and at most $T{=}10$ turns. Appendix~\ref{app:hyperparams} provides sampling, optimization, and reward configuration.

\paragraph{Baselines.}
We compare 16 baselines in three groups. The first includes closed-source models: GPT-5.4~\cite{openai2026gpt54}, Claude Sonnet 4.6~\cite{anthropic2026claude}, and Gemini 3.1 Pro~\cite{google2026gemini31pro}. The second includes open-source multimodal models from Qwen2.5-VL~\cite{Qwen2.5-VL}, Qwen3-VL~\cite{bai2025qwen3}, and InternVL3.5~\cite{wang2025internvl3} families across scales. The third contains tool-augmented multimodal models closest to our setting, including VTool-R1-7B~\cite{wu2025vtool}, R1-Onevision-7B~\cite{yang2025r1}, PyVision-Image-7B-RL~\cite{zhao2026pyvision}, DeepEyesV2-7B~\cite{hong2025deepeyesv2}, PixelReasoner-RL-vl-7B~\cite{su2025pixel}, AdaReasoner-7B~\cite{song2026adareasoner}, TikArt-8B~\cite{ding2026tikart}, and ZwZ-8B~\cite{wei2026zooming}. Appendix~\ref{app:baselines} gives descriptions.

\begin{table}[!t]
\centering
\scriptsize
\resizebox{\columnwidth}{!}{%
\setlength{\tabcolsep}{1.8pt}
\renewcommand{\arraystretch}{1.2}
\begin{tabular}{c|cccc|c|c|c|c|c}
\hline
\multirow{2}{*}{\textbf{Model}}
 & \multicolumn{5}{c|}{\textbf{Tool Forging}}
 & \multicolumn{4}{c}{\textbf{Inference Efficiency}} \\
\cline{2-10}
 & \textbf{Att.} & \textbf{Reg.} & \textbf{Err.} & \textbf{Dup.} & \textbf{FSR}
 & \textbf{ACC} & \textbf{AET} & \textbf{RLA} & \textbf{ITC} \\
\hline
\multicolumn{10}{c}{\cellcolor{blockgray}\textbf{w/ IID Tools}} \\
\hline
Qwen2.5-VL-7B             & 145  & 66  & 68  & 11 & 45.5 & 69.89 & 1.07 & 2.31  & 176.41 \\
Qwen3-VL-8B              & 34   & 5   & 26  & 3  & 14.7 & 73.80 & 1.46 & 3.75  & 239.06 \\
VTool-R1-7B               & 110  & 27  & 69  & 14 & 24.5 & 74.74 & 1.39 & 6.44  & 266.05 \\
R1-Onevision-7B            & 9    & 5   & 2   & 2  & \underline{55.6} & 65.63 & 1.01 & 0.14  & 302.71 \\
PyVision-Image-7B-RL & 65   & 19  & 45  & 1  & 29.2 & 42.38 & 5.02 & 75.50 & 1148.74 \\
DeepEyesV2-7B      & 1286 & 269 & 935 & 82 & 20.9 & 63.85 & 3.01 & 10.11 & 537.88 \\
PixelReasoner-RL-v1-7B    & 49   & 25  & 19  & 5  & 51.0 & 76.48 & 1.55 & 10.22 & 312.36 \\
AdaReasoner-7B    & 166  & 75  & 64  & 27 & 45.2 & 72.31 & 2.84 & 18.09 & 329.56 \\
TikArt-8B              & 291  & 130 & 133 & 28 & 44.7 & \underline{80.11} & 2.22 & 11.77 & 298.99 \\
ZwZ-8B                 & 609  & 138 & 379 & 92 & 22.7 & 77.69 & 1.65 & 12.79 & 223.31 \\
\cellcolor{oursblue}\textbf{MetaForge}
                 & \cellcolor{oursblue}190 & \cellcolor{oursblue}123 & \cellcolor{oursblue}51 & \cellcolor{oursblue}16 & \cellcolor{oursblue}\textbf{64.7}
                 & \cellcolor{oursblue}\textbf{88.14} & \cellcolor{oursblue}2.28 & \cellcolor{oursblue}10.99 & \cellcolor{oursblue}280.87 \\
\hline
\multicolumn{10}{c}{\cellcolor{blockgray}\textbf{w/ OOD Tools}} \\
\hline
Qwen2.5-VL-7B             & 230  & 131 & 83  & 16 & 57.0 & 71.42 & 1.07 & 0.97 & 174.28 \\
Qwen3-VL-8B              & 28   & 5   & 22  & 1  & 17.9 & 74.21 & 1.40 & 2.17 & 272.68 \\
VTool-R1-7B               & 86   & 24  & 57  & 5  & 27.9 & 75.15 & 1.41 & 2.27 & 262.26 \\
R1-Onevision-7B            & 6    & 3   & 3   & 0  & 50.0 & 65.76 & 1.01 & 0.01 & 304.99 \\
PyVision-Image-7B-RL & 32   & 15  & 17  & 0  & 46.9 & 54.04 & 5.83 & 2.50 & 1087.73 \\
DeepEyesV2-7B      & 404  & 114 & 260 & 30 & 28.2 & 68.10 & 2.60 & 2.44 & 642.37 \\
PixelReasoner-RL-v1-7B    & 67   & 33  & 22  & 12 & 49.3 & 76.83 & 1.58 & 1.82 & 306.81 \\
AdaReasoner-7B    & 200  & 50  & 132 & 18 & 25.0 & 72.90 & 2.52 & 4.39 & 312.68 \\
TikArt-8B              & 292  & 161 & 100 & 31 & 55.1 & 80.27 & 2.56 & 7.70 & 383.35 \\
ZwZ-8B                 & 44   & 28  & 14  & 2  & \underline{63.6} & \underline{81.35} & 1.45 & 0.90 & 236.67 \\
\cellcolor{oursblue}\textbf{MetaForge}
                 & \cellcolor{oursblue}12 & \cellcolor{oursblue}11 & \cellcolor{oursblue}0 & \cellcolor{oursblue}1 & \cellcolor{oursblue}\textbf{91.7}
                 & \cellcolor{oursblue}\textbf{88.98} & \cellcolor{oursblue}2.30 & \cellcolor{oursblue}4.97 & \cellcolor{oursblue}336.65 \\
\hline
\end{tabular}%
}
\caption{Tool forging and inference efficiency. Att., Reg., Err., and Dup.\ denote attempts, registrations, generation errors, and removed duplicates.}
\label{tab:forging_efficiency_singlecol}
\end{table}

\subsection{Performance Comparison}

\noindent\textbf{Adaptive tool orchestration improves overall accuracy.}
Table~\ref{tab:main} shows that MetaForge achieves the best accuracy across 12 benchmarks under both IID and OOD tool settings.
The gains hold across IID and OOD datasets, suggesting that the model does not merely learn dataset-specific tool routines.
We attribute this robustness to online tool forging, multi-turn GRPO with composite rewards, and tool pool management, which supply missing tools, improve call validity, and reduce calls.

\noindent\textbf{Fixed tool-use policies lack adaptive invocation.}
Several tool-augmented baselines underperform their base models.
R1-Onevision-7B, PyVision-Image-7B-RL, and DeepEyesV2-7B are all built on Qwen2.5-VL-7B, yet achieve lower overall accuracy.
This suggests that policies trained with a fixed tool set can produce erroneous calls when the available tools change, offsetting the benefits of tool augmentation.
In contrast, MetaForge's unified tool interface and online tool forging allow the model to invoke available tools adaptively and forge tools when existing tools are insufficient.

\noindent\textbf{Tool complementarity matters.}
Closed-source models remain stable under IID/OOD tools, likely because their intrinsic strengths reduce reliance on external tools.
By contrast, open-source multimodal models are more sensitive to tool changes: QwenVL benefits more from OOD tools, whereas InternVL benefits more from IID tools.
Because both settings use the same online tool forging mechanism, the main differences lie in executable tools and model tools.
For executable tools, IID tools emphasize perception functions (OCR, grounding, segmentation), whereas OOD tools emphasize reasoning functions (symbolic computation, code execution, chart parsing), suggesting that tools help more when they match model gaps.
For model tools, QwenVL can call InternVL/Step3 under OOD tools, while InternVL can call QwenVL under IID tools, indicating that cross-family multimodal models provide complementary gains.
These results show that tool augmentation depends not only on tool functionality, but also on complementarity between tools and the calling model.

\subsection{Tool Usage Behavior Analysis}
\begin{figure}[htpb]
  \centering
  \includegraphics[width=0.47\textwidth]{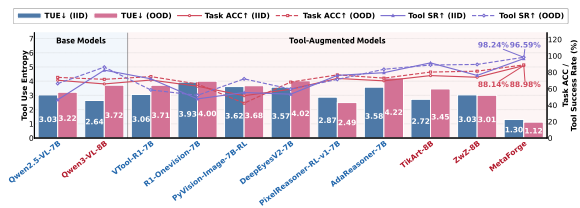}
  \caption{Comparison of TUE, Tool SR, and Task ACC across methods.}
  \label{fig:ablation_1}
\end{figure}

We analyze tool usage behavior with TUE, Tool SR, and task accuracy (Task ACC). Figure~\ref{fig:ablation_1} shows that methods with more dispersed selection and lower call success usually obtain lower task accuracy. Base models and existing tool-augmented methods often show this pattern because fixed tool sets generalize poorly to unseen tools. When tools change, they tend to make mistakes or issue invalid calls. Without tool-forging training, their forged tools are also more likely to be non-executable or misaligned with task needs. These errors reduce Tool SR and disrupt subsequent reasoning, thereby lowering task accuracy.

In contrast, MetaForge maintains a more concentrated selection, higher call success, and better task performance under both IID and OOD tool settings. This advantage stems from improved call quality and enhanced generalization. The Retrieve/Adapt, Format, and Decide rewards encourage useful tool use, valid formatting, and concise interaction. The unified tool interface helps the model understand and invoke unseen tools, while the Forge reward encourages forging and reusing forged tools when existing tools are insufficient. Together, these designs improve selection precision and the reliability of forged-tool construction.

\subsection{Tool Forging Behavior Analysis}
All methods use the same inference prompt and therefore have the same opportunity to forge tools. Table~\ref{tab:forging_efficiency_singlecol} shows that base models and existing tool-augmented methods obtain low FSR, indicating that prompting alone is insufficient for reliable tool forging. Without dedicated training, these methods struggle to decide when a new forged tool is needed and may create redundant ones. Even when they trigger forging, low generation quality can cause validation failure.

In contrast, MetaForge achieves the highest FSR under both tool settings. This improvement comes from the combined effect of the Forge reward, two-stage tool forging, and decide reward. The Forge reward favors forged tools that help solve the task, two-stage forging improves alignment with task needs, and the Decide reward suppresses unnecessary creation. Together, these components improve forged-tool quality and registration success.

\subsection{Inference Efficiency Analysis}
Table~\ref{tab:forging_efficiency_singlecol} shows three efficiency-accuracy patterns among existing methods. Some methods, such as R1-Onevision-7B, have AET close to 1 and low latency, but bypass tools and gain little from augmentation. Others, such as PyVision-Image-7B-RL, incur many turns and high token costs without accuracy gains, indicating redundant interaction. A third group, such as TikArt-8B, has efficiency comparable to MetaForge but shows an accuracy gap. Under similar overhead, precise tool selection and invocation determine final performance.

In contrast, MetaForge achieves the highest accuracy with efficient inference under IID/OOD settings.
AET stays around 2.3 turns, showing stability.
Its RLA (IID: 10.99s, OOD: 4.97s) and ITC (IID: 280.87, OOD: 336.65) remain relatively low among tool-augmented models.
This efficiency comes from the Decide reward, which penalizes redundancy.
Among correct trajectories, those with fewer calls and lower latency receive higher rewards, guiding efficiency during training.


\section{Conclusion}
We propose MetaForge, a self-evolving multimodal agent for adaptive tool orchestration. MetaForge addresses the two limitations of tool-augmented multimodal agents, namely statically predefined tool inventories and indiscriminate tool invocation, by learning a closed Decide-Retrieve-Adapt-Forge loop. Combined with multi-turn GRPO, a composite reward that balances answer correctness, tool retrieval, parameter adaptation, invocation necessity, and forged-skill reusability, and capacity-constrained tool pool management, MetaForge learns when to answer, when to reuse existing tools, when to adapt tool calls, and when to forge new skills on demand.
Experiments on 12 benchmarks against 16 baselines show consistent gains in accuracy, efficiency, and generalization under both IID and OOD dataset/tool settings. Further analysis confirms more reliable tool selection, higher tool-call and skill-registration success, and effective reuse of forged skills. Overall, MetaForge shifts multimodal agents from static tool inventories toward on-demand self-evolution.

\section*{Limitations}

Although MetaForge achieves strong results in adaptive tool orchestration and on-demand toolset evolution, several limitations remain.

\begin{itemize}
    \item In our implementation, each forged tool is implemented as an executable skill, consisting of a Markdown specification, an interface schema, and Python scripts. This form is mainly suited for programmatic task-level operations, such as code processing, chart analysis, and numerical computation. Therefore, Forge does not imply that the model can solve all multimodal tasks. Tasks requiring additional models or external resources, such as video synthesis, may remain beyond the current skill design. Future work can explore stronger tool-forging models, richer skill forms, and more reliable validation mechanisms to broaden the supported task range.

    \item Our evaluation covers 12 benchmarks but remains centered on image-text tasks. We have not evaluated settings involving audio, video, temporal reasoning, or interactive environments. While the results show strong generalization across the evaluated tasks, whether the same gains transfer to richer multimodal settings remains unclear. Future work can extend the decide-retrieve-adapt-forge loop to richer inputs and study whether online skill forging remains effective for temporal and audio-visual reasoning.

    \item MetaForge uses the decide reward and capacity-constrained tool pool management to reduce redundant calls and tool library growth, but the efficiency objective focuses on latency. In deployment, tool calls also introduce monetary cost and context usage. Future work can incorporate these factors into tool invocation, balancing monetary cost, context usage, and response time while preserving accuracy.
\end{itemize}

\bibliography{custom}

@inproceedings{fangrit,
  title={GRIT: Teaching MLLMs to Think with Images},
  author={Fan, Yue and He, Xuehai and Yang, Diji and Zheng, Kaizhi and Kuo, Ching-Chen and Zheng, Yuting and Guan, Xinze and Wang, Xin Eric},
  booktitle={The Thirty-ninth Annual Conference on Neural Information Processing Systems}
}

@article{su2025openthinkimg,
  title={Openthinkimg: Learning to think with images via visual tool reinforcement learning},
  author={Su, Zhaochen and Li, Linjie and Song, Mingyang and Hao, Yunzhuo and Yang, Zhengyuan and Zhang, Jun and Chen, Guanjie and Gu, Jiawei and Li, Juntao and Qu, Xiaoye and others},
  journal={arXiv preprint arXiv:2505.08617},
  year={2025}
}

@article{zheng2025deepeyes,
  title={Deepeyes: Incentivizing" thinking with images" via reinforcement learning},
  author={Zheng, Ziwei and Yang, Michael and Hong, Jack and Zhao, Chenxiao and Xu, Guohai and Yang, Le and Shen, Chao and Yu, Xing},
  journal={arXiv preprint arXiv:2505.14362},
  year={2025}
}

@article{narayan2025deepmmsearch,
  title={Deepmmsearch-r1: Empowering multimodal llms in multimodal web search},
  author={Narayan, Kartik and Xu, Yang and Cao, Tian and Nerella, Kavya and Patel, Vishal M and Shiee, Navid and Grasch, Peter and Jia, Chao and Yang, Yinfei and Gan, Zhe},
  journal={arXiv preprint arXiv:2510.12801},
  year={2025}
}

@article{wu2025mmsearch,
  title={Mmsearch-r1: Incentivizing lmms to search},
  author={Wu, Jinming and Deng, Zihao and Li, Wei and Liu, Yiding and You, Bo and Li, Bo and Ma, Zejun and Liu, Ziwei},
  journal={arXiv preprint arXiv:2506.20670},
  year={2025}
}

@inproceedings{zhang2025thyme,
  title={Thyme: Think Beyond Images},
  author={Zhang, YiFan and Lu, Xingyu and Yin, Shukang and Fu, Chaoyou and Chen, Wei and Hu, Xiao and Wen, Bin and Jiang, Kaiyu and Liu, Changyi and Zhang, Tianke and others},
  booktitle={NeurIPS 2025 Fourth Workshop on Deep Learning for Code}
}

@inproceedings{zhaopyvision,
  title={PyVision: Agentic Vision with Dynamic Tooling},
  author={Zhao, Shitian and Zhang, Haoquan and Lin, Shaoheng and Li, Ming and Wu, Qilong and Zhang, Kaipeng and Wei, Chen},
  booktitle={First Workshop on Multi-Turn Interactions in Large Language Models}
}

@inproceedings{lu2025octotools,
  title={OctoTools: An Agentic Framework with Extensible Tools for Complex Reasoning},
  author={Lu, Pan and Chen, Bowen and Liu, Sheng and Thapa, Rahul and Boen, Joseph and Zou, James},
  booktitle={ICLR 2025 Workshop on Foundation Models in the Wild}
}

@inproceedings{qian2023creator,
  title={Creator: Tool creation for disentangling abstract and concrete reasoning of large language models},
  author={Qian, Cheng and Han, Chi and Fung, Yi and Qin, Yujia and Liu, Zhiyuan and Ji, Heng},
  booktitle={Findings of the Association for Computational Linguistics: EMNLP 2023},
  pages={6922--6939},
  year={2023}
}

@inproceedings{yuancraft,
  title={CRAFT: Customizing LLMs by Creating and Retrieving from Specialized Toolsets},
  author={Yuan, Lifan and Chen, Yangyi and Wang, Xingyao and Fung, Yi and Peng, Hao and Ji, Heng},
  booktitle={The Twelfth International Conference on Learning Representations}
}

@inproceedings{cailarge,
  title={Large Language Models as Tool Makers},
  author={Cai, Tianle and Wang, Xuezhi and Ma, Tengyu and Chen, Xinyun and Zhou, Denny},
  booktitle={The Twelfth International Conference on Learning Representations}
}

@inproceedings{wolflein2025llm,
  title={Llm agents making agent tools},
  author={W{\"o}lflein, Georg and Ferber, Dyke and Truhn, Daniel and Arandjelovic, Ognjen and Kather, Jakob Nikolas},
  booktitle={Proceedings of the 63rd Annual Meeting of the Association for Computational Linguistics (Volume 1: Long Papers)},
  pages={26092--26130},
  year={2025}
}

@article{wangvoyager,
  title={Voyager: An Open-Ended Embodied Agent with Large Language Models},
  author={Wang, Guanzhi and Xie, Yuqi and Jiang, Yunfan and Mandlekar, Ajay and Xiao, Chaowei and Zhu, Yuke and Fan, Linxi and Anandkumar, Anima},
  journal={Transactions on Machine Learning Research}
}

@article{qiu2025alita,
  title={Alita: Generalist agent enabling scalable agentic reasoning with minimal predefinition and maximal self-evolution},
  author={Qiu, Jiahao and Qi, Xuan and Zhang, Tongcheng and Juan, Xinzhe and Guo, Jiacheng and Lu, Yifu and Wang, Yimin and Yao, Zixin and Ren, Qihan and Jiang, Xun and others},
  journal={arXiv preprint arXiv:2505.20286},
  year={2025}
}

@article{huang2025toolace,
  title={Toolace-dev: Self-improving tool learning via decomposition and evolution},
  author={Huang, Xu and Liu, Weiwen and Zeng, Xingshan and Huang, Yuefeng and Hao, Xinlong and Wang, Yuxian and Zeng, Yirong and Wu, Chuhan and Wang, Yasheng and Tang, Ruiming and others},
  journal={arXiv preprint arXiv:2505.07512},
  year={2025}
}

@article{sun2025seagent,
  title={Seagent: Self-evolving computer use agent with autonomous learning from experience},
  author={Sun, Zeyi and Liu, Ziyu and Zang, Yuhang and Cao, Yuhang and Dong, Xiaoyi and Wu, Tong and Lin, Dahua and Wang, Jiaqi},
  journal={arXiv preprint arXiv:2508.04700},
  year={2025}
}

@inproceedings{jian2025teaching,
  title={Teaching vision-language models to ask: Resolving ambiguity in visual questions},
  author={Jian, Pu and Yu, Donglei and Yang, Wen and Ren, Shuo and Zhang, Jiajun},
  booktitle={Proceedings of the 63rd Annual Meeting of the Association for Computational Linguistics (Volume 1: Long Papers)},
  pages={3619--3638},
  year={2025}
}

@inproceedings{ning2025gns,
  title={Gns: Solving plane geometry problems by neural-symbolic reasoning with multi-modal llms},
  author={Ning, Maizhen and Zhou, Zihao and Wang, Qiufeng and Huang, Xiaowei and Huang, Kaizhu},
  booktitle={Proceedings of the AAAI Conference on Artificial Intelligence},
  volume={39},
  number={23},
  pages={24957--24965},
  year={2025}
}

@inproceedings{liao2025doclayllm,
  title={Doclayllm: An efficient multi-modal extension of large language models for text-rich document understanding},
  author={Liao, Wenhui and Wang, Jiapeng and Li, Hongliang and Wang, Chengyu and Huang, Jun and Jin, Lianwen},
  booktitle={Proceedings of the IEEE/CVF Conference on Computer Vision and Pattern Recognition},
  pages={4038--4049},
  year={2025}
}

@inproceedings{zhao2023robut,
  title={RobuT: A systematic study of table QA robustness against human-annotated adversarial perturbations},
  author={Zhao, Yilun and Zhao, Chen and Nan, Linyong and Qi, Zhenting and Zhang, Wenlin and Tang, Xiangru and Mi, Boyu and Radev, Dragomir},
  booktitle={Proceedings of the 61st Annual Meeting of the Association for Computational Linguistics (Volume 1: Long Papers)},
  pages={6064--6081},
  year={2023}
}

@inproceedings{mathew2021docvqa,
  title={Docvqa: A dataset for vqa on document images},
  author={Mathew, Minesh and Karatzas, Dimosthenis and Jawahar, CV},
  booktitle={Proceedings of the IEEE/CVF winter conference on applications of computer vision},
  pages={2200--2209},
  year={2021}
}

@inproceedings{lindstrom2022clevr,
  title={CLEVR-Math: A Dataset for Compositional Language, Visual and Mathematical Reasoning},
  author={Lindstr{\"o}m, AD and Abraham, SS},
  booktitle={Proceedings of the 16th International Workshop on Neural-Symbolic Learning and Reasoning},
  volume={3212},
  pages={155--170},
  year={2022}
}

@article{lu2023mathvista,
  title={Mathvista: Evaluating mathematical reasoning of foundation models in visual contexts},
  author={Lu, Pan and Bansal, Hritik and Xia, Tony and Liu, Jiacheng and Li, Chunyuan and Hajishirzi, Hannaneh and Cheng, Hao and Chang, Kai-Wei and Galley, Michel and Gao, Jianfeng},
  journal={arXiv preprint arXiv:2310.02255},
  year={2023}
}

@article{saikh2022scienceqa,
  title={Scienceqa: A novel resource for question answering on scholarly articles},
  author={Saikh, Tanik and Ghosal, Tirthankar and Mittal, Amish and Ekbal, Asif and Bhattacharyya, Pushpak},
  journal={International Journal on Digital Libraries},
  volume={23},
  number={3},
  pages={289--301},
  year={2022},
  publisher={Springer}
}

@article{bhathal2025websight,
  title={Websight: A vision-first architecture for robust web agents},
  author={Bhathal, Tanvir and Gupta, Asanshay},
  journal={arXiv preprint arXiv:2508.16987},
  year={2025}
}

@inproceedings{kembhavi2016diagram,
  title={A diagram is worth a dozen images},
  author={Kembhavi, Aniruddha and Salvato, Mike and Kolve, Eric and Seo, Minjoon and Hajishirzi, Hannaneh and Farhadi, Ali},
  booktitle={European conference on computer vision},
  pages={235--251},
  year={2016},
  organization={Springer}
}

@inproceedings{masry2022chartqa,
  title={Chartqa: A benchmark for question answering about charts with visual and logical reasoning},
  author={Masry, Ahmed and Do, Xuan Long and Tan, Jia Qing and Joty, Shafiq and Hoque, Enamul},
  booktitle={Findings of the association for computational linguistics: ACL 2022},
  pages={2263--2279},
  year={2022}
}

@article{chang2022mapqa,
  title={Mapqa: A dataset for question answering on choropleth maps},
  author={Chang, Shuaichen and Palzer, David and Li, Jialin and Fosler-Lussier, Eric and Xiao, Ningchuan},
  journal={arXiv preprint arXiv:2211.08545},
  year={2022}
}

@inproceedings{pont2020connecting,
  title={Connecting vision and language with localized narratives},
  author={Pont-Tuset, Jordi and Uijlings, Jasper and Changpinyo, Soravit and Soricut, Radu and Ferrari, Vittorio},
  booktitle={European conference on computer vision},
  pages={647--664},
  year={2020},
  organization={Springer}
}

@inproceedings{mishra2019ocr,
  title={Ocr-vqa: Visual question answering by reading text in images},
  author={Mishra, Anand and Shekhar, Shashank and Singh, Ajeet Kumar and Chakraborty, Anirban},
  booktitle={2019 international conference on document analysis and recognition (ICDAR)},
  pages={947--952},
  year={2019},
  organization={IEEE}
}

@inproceedings{acharya2019tallyqa,
  title={Tallyqa: Answering complex counting questions},
  author={Acharya, Manoj and Kafle, Kushal and Kanan, Christopher},
  booktitle={Proceedings of the AAAI conference on artificial intelligence},
  volume={33},
  number={01},
  pages={8076--8084},
  year={2019}
}

@inproceedings{goyal2017making,
  title={Making the v in vqa matter: Elevating the role of image understanding in visual question answering},
  author={Goyal, Yash and Khot, Tejas and Summers-Stay, Douglas and Batra, Dhruv and Parikh, Devi},
  booktitle={Proceedings of the IEEE conference on computer vision and pattern recognition},
  pages={6904--6913},
  year={2017}
}

@inproceedings{singh2019towards,
  title={Towards vqa models that can read},
  author={Singh, Amanpreet and Natarajan, Vivek and Shah, Meet and Jiang, Yu and Chen, Xinlei and Batra, Dhruv and Parikh, Devi and Rohrbach, Marcus},
  booktitle={Proceedings of the IEEE/CVF conference on computer vision and pattern recognition},
  pages={8317--8326},
  year={2019}
}

@inproceedings{biten2019scene,
  title={Scene text visual question answering},
  author={Biten, Ali Furkan and Tito, Ruben and Mafla, Andres and Gomez, Lluis and Rusinol, Mar{\c{c}}al and Valveny, Ernest and Jawahar, CV and Karatzas, Dimosthenis},
  booktitle={Proceedings of the IEEE/CVF international conference on computer vision},
  pages={4291--4301},
  year={2019}
}

@article{liu2023visual,
  title={Visual spatial reasoning},
  author={Liu, Fangyu and Emerson, Guy and Collier, Nigel},
  journal={Transactions of the Association for Computational Linguistics},
  volume={11},
  pages={635--651},
  year={2023},
  publisher={MIT Press One Broadway, 12th Floor, Cambridge, Massachusetts 02142, USA~…}
}

@inproceedings{tanaka2021visualmrc,
  title={Visualmrc: Machine reading comprehension on document images},
  author={Tanaka, Ryota and Nishida, Kyosuke and Yoshida, Sen},
  booktitle={Proceedings of the AAAI Conference on Artificial Intelligence},
  volume={35},
  number={15},
  pages={13878--13888},
  year={2021}
}

@inproceedings{zhu2021tat,
  title={TAT-QA: A question answering benchmark on a hybrid of tabular and textual content in finance},
  author={Zhu, Fengbin and Lei, Wenqiang and Huang, Youcheng and Wang, Chao and Zhang, Shuo and Lv, Jiancheng and Feng, Fuli and Chua, Tat-Seng},
  booktitle={Proceedings of the 59th annual meeting of the Association for Computational Linguistics and the 11th international joint conference on natural language processing (volume 1: long papers)},
  pages={3277--3287},
  year={2021}
}

@inproceedings{wang2021screen2words,
  title={Screen2words: Automatic mobile ui summarization with multimodal learning},
  author={Wang, Bryan and Li, Gang and Zhou, Xin and Chen, Zhourong and Grossman, Tovi and Li, Yang},
  booktitle={The 34th Annual ACM Symposium on User Interface Software and Technology},
  pages={498--510},
  year={2021}
}

@inproceedings{cheng2022hitab,
  title={Hitab: A hierarchical table dataset for question answering and natural language generation},
  author={Cheng, Zhoujun and Dong, Haoyu and Wang, Zhiruo and Jia, Ran and Guo, Jiaqi and Gao, Yan and Han, Shi and Lou, Jian-Guang and Zhang, Dongmei},
  booktitle={Proceedings of the 60th Annual Meeting of the Association for Computational Linguistics (Volume 1: Long Papers)},
  pages={1094--1110},
  year={2022}
}

@inproceedings{chen2021finqa,
  title={Finqa: A dataset of numerical reasoning over financial data},
  author={Chen, Zhiyu and Chen, Wenhu and Smiley, Charese and Shah, Sameena and Borova, Iana and Langdon, Dylan and Moussa, Reema and Beane, Matt and Huang, Ting-Hao and Routledge, Bryan R and others},
  booktitle={Proceedings of the 2021 Conference on Empirical Methods in Natural Language Processing},
  pages={3697--3711},
  year={2021}
}

@inproceedings{schwenk2022okvqa,
  title={A-okvqa: A benchmark for visual question answering using world knowledge},
  author={Schwenk, Dustin and Khandelwal, Apoorv and Clark, Christopher and Marino, Kenneth and Mottaghi, Roozbeh},
  booktitle={European conference on computer vision},
  pages={146--162},
  year={2022},
  organization={Springer}
}

@inproceedings{du2019cocoqa,
  title={Cocoqa: Question answering for coding conventions over knowledge graphs},
  author={Du, Tianjiao and Cao, Junming and Wu, Qinyue and Li, Wei and Shen, Beijun and Chen, Yuting},
  booktitle={2019 34th IEEE/ACM International Conference on Automated Software Engineering (ASE)},
  pages={1086--1089},
  year={2019},
  organization={IEEE}
}

@inproceedings{kembhavi2017you,
  title={Are you smarter than a sixth grader? textbook question answering for multimodal machine comprehension},
  author={Kembhavi, Aniruddha and Seo, Minjoon and Schwenk, Dustin and Choi, Jonghyun and Farhadi, Ali and Hajishirzi, Hannaneh},
  booktitle={Proceedings of the IEEE Conference on Computer Vision and Pattern recognition},
  pages={4999--5007},
  year={2017}
}

@inproceedings{mathew2022infographicvqa,
  title={Infographicvqa},
  author={Mathew, Minesh and Bagal, Viraj and Tito, Rub{\`e}n and Karatzas, Dimosthenis and Valveny, Ernest and Jawahar, CV},
  booktitle={Proceedings of the IEEE/CVF Winter Conference on Applications of Computer Vision},
  pages={1697--1706},
  year={2022}
}

@article{marti2002iam,
  title={The IAM-database: an English sentence database for offline handwriting recognition},
  author={Marti, U-V and Bunke, Horst},
  journal={International journal on document analysis and recognition},
  volume={5},
  number={1},
  pages={39--46},
  year={2002},
  publisher={Springer}
}

@article{lau2018dataset,
  title={A dataset of clinically generated visual questions and answers about radiology images},
  author={Lau, Jason J and Gayen, Soumya and Ben Abacha, Asma and Demner-Fushman, Dina},
  journal={Scientific data},
  volume={5},
  number={1},
  pages={180251},
  year={2018},
  publisher={Nature Publishing Group}
}

@inproceedings{zhu2016visual7w,
  title={Visual7w: Grounded question answering in images},
  author={Zhu, Yuke and Groth, Oliver and Bernstein, Michael and Fei-Fei, Li},
  booktitle={Proceedings of the IEEE conference on computer vision and pattern recognition},
  pages={4995--5004},
  year={2016}
}

@article{wang2025internvl3,
  title={Internvl3. 5: Advancing open-source multimodal models in versatility, reasoning, and efficiency},
  author={Wang, Weiyun and Gao, Zhangwei and Gu, Lixin and Pu, Hengjun and Cui, Long and Wei, Xingguang and Liu, Zhaoyang and Jing, Linglin and Ye, Shenglong and Shao, Jie and others},
  journal={arXiv preprint arXiv:2508.18265},
  year={2025}
}

@article{bai2025qwen3,
  title={Qwen3-vl technical report},
  author={Bai, Shuai and Cai, Yuxuan and Chen, Ruizhe and Chen, Keqin and Chen, Xionghui and Cheng, Zesen and Deng, Lianghao and Ding, Wei and Gao, Chang and Ge, Chunjiang and others},
  journal={arXiv preprint arXiv:2511.21631},
  year={2025}
}

@article{wu2025vtool,
  title={Vtool-r1: Vlms learn to think with images via reinforcement learning on multimodal tool use},
  author={Wu, Mingyuan and Yang, Jingcheng and Jiang, Jize and Li, Meitang and Yan, Kaizhuo and Yu, Hanchao and Zhang, Minjia and Zhai, Chengxiang and Nahrstedt, Klara},
  journal={arXiv preprint arXiv:2505.19255},
  year={2025}
}

@article{wei2026zooming,
  title={Zooming without Zooming: Region-to-Image Distillation for Fine-Grained Multimodal Perception},
  author={Wei, Lai and He, Liangbo and Lan, Jun and Dong, Lingzhong and Cai, Yutong and Li, Siyuan and Zhu, Huijia and Wang, Weiqiang and Kong, Linghe and Wang, Yue and others},
  journal={arXiv preprint arXiv:2602.11858},
  year={2026}
}

@article{zhao2026pyvision,
  title={PyVision-RL: Forging Open Agentic Vision Models via RL},
  author={Zhao, Shitian and Lin, Shaoheng and Li, Ming and Zhang, Haoquan and Peng, Wenshuo and Zhang, Kaipeng and Wei, Chen},
  journal={arXiv preprint arXiv:2602.20739},
  year={2026}
}

@article{song2026adareasoner,
  title={AdaReasoner: Dynamic Tool Orchestration for Iterative Visual Reasoning},
  author={Song, Mingyang and Sun, Haoyu and Gu, Jiawei and Li, Linjie and Xu, Luxin and Krishna, Ranjay and Cheng, Yu},
  journal={arXiv preprint arXiv:2601.18631},
  year={2026}
}

@inproceedings{su2025pixel,
  title={Pixel Reasoner: Incentivizing Pixel Space Reasoning via Curiosity-Driven Reinforcement Learning},
  author={Su, Alex and Wang, Haozhe and Ren, Weiming and Lin, Fangzhen and Chen, Wenhu},
  booktitle={The Thirty-ninth Annual Conference on Neural Information Processing Systems},
  year={2025}
}

@article{ding2026tikart,
  title={TikArt: Aperture-Guided Observation for Fine-Grained Visual Reasoning via Reinforcement Learning},
  author={Ding, Hao and Yang, Zhichuan and Ge, Weijie and Gao, Ziqin and Lu, Chaoyi and Zhao, Lei},
  journal={arXiv preprint arXiv:2602.14482},
  year={2026}
}

@article{hong2025deepeyesv2,
  title={Deepeyesv2: Toward agentic multimodal model},
  author={Hong, Jack and Zhao, Chenxiao and Zhu, ChengLin and Lu, Weiheng and Xu, Guohai and Yu, Xing},
  journal={arXiv preprint arXiv:2511.05271},
  year={2025}
}

@inproceedings{yang2025r1,
  title={R1-onevision: Advancing generalized multimodal reasoning through cross-modal formalization},
  author={Yang, Yi and He, Xiaoxuan and Pan, Hongkun and Jiang, Xiyan and Deng, Yan and Yang, Xingtao and Lu, Haoyu and Yin, Dacheng and Rao, Fengyun and Zhu, Minfeng and others},
  booktitle={Proceedings of the IEEE/CVF International Conference on Computer Vision},
  pages={2376--2385},
  year={2025}
}

@misc{anthropic2026claude,
  title={Introducing Claude Sonnet 4.6},
  author={Anthropic},
  year={2026},
  howpublished={\url{https://www.anthropic.com/news/claude-sonnet-4-6}},
  note={Accessed: 2026-02-17}
}

@misc{google2026gemini31pro,
  title={Gemini 3.1 Pro},
  author={Google DeepMind},
  year={2026},
  howpublished={\url{https://deepmind.google/models/model-cards/gemini-3-1-pro/}},
  note={Accessed: 2026-02-19}
}

@article{Qwen2.5-VL,
  title={Qwen2.5-VL Technical Report},
  author={Bai, Shuai and Chen, Keqin and Liu, Xuejing and Wang, Jialin and Ge, Wenbin and Song, Sibo and Dang, Kai and Wang, Peng and Wang, Shijie and Tang, Jun and Zhong, Humen and Zhu, Yuanzhi and Yang, Mingkun and Li, Zhaohai and Wan, Jianqiang and Wang, Pengfei and Ding, Wei and Fu, Zheren and Xu, Yiheng and Ye, Jiabo and Zhang, Xi and Xie, Tianbao and Cheng, Zesen and Zhang, Hang and Yang, Zhibo and Xu, Haiyang and Lin, Junyang},
  journal={arXiv preprint arXiv:2502.13923},
  year={2025}
}

@misc{openai2026gpt54,
  title   = {Introducing {GPT}-5.4},
  author  = {{OpenAI}},
  year    = {2026},
  howpublished = {\url{https://openai.com/index/introducing-gpt-5-4/}},
  note    = {Accessed: 2026-03-05}
}

@inproceedings{liu2024grounding,
  title={Grounding dino: Marrying dino with grounded pre-training for open-set object detection},
  author={Liu, Shilong and Zeng, Zhaoyang and Ren, Tianhe and Li, Feng and Zhang, Hao and Yang, Jie and Jiang, Qing and Li, Chunyuan and Yang, Jianwei and Su, Hang and others},
  booktitle={European conference on computer vision},
  pages={38--55},
  year={2024},
  organization={Springer}
}

@article{carion2025sam,
  title={Sam 3: Segment anything with concepts},
  author={Carion, Nicolas and Gustafson, Laura and Hu, Yuan-Ting and Debnath, Shoubhik and Hu, Ronghang and Suris, Didac and Ryali, Chaitanya and Alwala, Kalyan Vasudev and Khedr, Haitham and Huang, Andrew and others},
  journal={arXiv preprint arXiv:2511.16719},
  year={2025}
}

@article{cui2025paddleocr,
  title={Paddleocr-vl: Boosting multilingual document parsing via a 0.9 b ultra-compact vision-language model},
  author={Cui, Cheng and Sun, Ting and Liang, Suyin and Gao, Tingquan and Zhang, Zelun and Liu, Jiaxuan and Wang, Xueqing and Zhou, Changda and Liu, Hongen and Lin, Manhui and others},
  journal={arXiv preprint arXiv:2510.14528},
  year={2025}
}

@article{niu2025mineru2,
  title={Mineru2. 5: A decoupled vision-language model for efficient high-resolution document parsing},
  author={Niu, Junbo and Liu, Zheng and Gu, Zhuangcheng and Wang, Bin and Ouyang, Linke and Zhao, Zhiyuan and Chu, Tao and He, Tianyao and Wu, Fan and Zhang, Qintong and others},
  journal={arXiv preprint arXiv:2509.22186},
  year={2025}
}

@article{wu2025qwen,
  title={Qwen-image technical report},
  author={Wu, Chenfei and Li, Jiahao and Zhou, Jingren and Lin, Junyang and Gao, Kaiyuan and Yan, Kun and Yin, Sheng-ming and Bai, Shuai and Xu, Xiao and Chen, Yilei and others},
  journal={arXiv preprint arXiv:2508.02324},
  year={2025}
}

@article{wang2025step,
  title={Step-3 is large yet affordable: Model-system co-design for cost-effective decoding},
  author={Wang, Bin and Wang, Bojun and Wan, Changyi and Huang, Guanzhe and Hu, Hanpeng and Jia, Haonan and Nie, Hao and Li, Mingliang and Chen, Nuo and Chen, Siyu and others},
  journal={arXiv preprint arXiv:2507.19427},
  year={2025}
}

@inproceedings{masry2023unichart,
  title={Unichart: A universal vision-language pretrained model for chart comprehension and reasoning},
  author={Masry, Ahmed and Kavehzadeh, Parsa and Do, Xuan Long and Hoque, Enamul and Joty, Shafiq},
  booktitle={Proceedings of the 2023 conference on empirical methods in natural language processing},
  pages={14662--14684},
  year={2023}
}

@inproceedings{liu2023deplot,
  title={DePlot: One-shot visual language reasoning by plot-to-table translation},
  author={Liu, Fangyu and Eisenschlos, Julian and Piccinno, Francesco and Krichene, Syrine and Pang, Chenxi and Lee, Kenton and Joshi, Mandar and Chen, Wenhu and Collier, Nigel and Altun, Yasemin},
  booktitle={Findings of the Association for Computational Linguistics: ACL 2023},
  pages={10381--10399},
  year={2023}
}

@inproceedings{xuchartmoe,
  title={ChartMoE: Mixture of Diversely Aligned Expert Connector for Chart Understanding},
  author={Xu, Zhengzhuo and Qu, Bowen and Qi, Yiyan and Du, SiNan and Xu, Chengjin and Yuan, Chun and Guo, Jian},
  booktitle={The Thirteenth International Conference on Learning Representations}
}

@article{jiang2025verltool,
  title={Verltool: Towards holistic agentic reinforcement learning with tool use},
  author={Jiang, Dongfu and Lu, Yi and Li, Zhuofeng and Lyu, Zhiheng and Nie, Ping and Wang, Haozhe and Su, Alex and Chen, Hui and Zou, Kai and Du, Chao and others},
  journal={arXiv preprint arXiv:2509.01055},
  year={2025}
}

@article{singh2025openai,
  title={Openai gpt-5 system card},
  author={Singh, Aaditya and Fry, Adam and Perelman, Adam and Tart, Adam and Ganesh, Adi and El-Kishky, Ahmed and McLaughlin, Aidan and Low, Aiden and Ostrow, AJ and Ananthram, Akhila and others},
  journal={arXiv preprint arXiv:2601.03267},
  year={2025}
}

@article{jiang2026xskill,
  title={Xskill: Continual learning from experience and skills in multimodal agents},
  author={Jiang, Guanyu and Su, Zhaochen and Qu, Xiaoye and Fung, Yi R},
  journal={arXiv preprint arXiv:2603.12056},
  year={2026}
}

\newpage
\appendix
\clearpage
\section{Experimental Details}
\label{app:experimental_details}

\subsection{Dataset and Tool Set Splits}
\label{app:splits}
To evaluate task and tool generalization, we split both datasets and tools into In-Domain (IID) and Out-of-Domain (OOD) settings.

\begin{itemize}[leftmargin=1.2em,itemsep=0.35em]
\item \textbf{Dataset Split.} 
The training set contains 9,500 samples from 19 datasets across six categories: document understanding, general VQA, OCR, table understanding, image captioning, and geospatial reasoning. Table~\ref{tab:training_data} summarizes datasets, categories, and sample counts. For evaluation, IID datasets share these categories but come from different sources. They include DocVQA~\cite{mathew2021docvqa}, TallyQA~\cite{acharya2019tallyqa}, OCRVQA~\cite{mishra2019ocr}, RoBUT-WikiSQL~\cite{zhao2023robut}, Localized Narratives~\cite{pont2020connecting}, and MapQA~\cite{chang2022mapqa}. OOD datasets cover six task categories unseen during training: chart, diagram, code, math, science, and spatial math. They include ChartQA~\cite{masry2022chartqa}, AI2D~\cite{kembhavi2016diagram}, WebSight~\cite{bhathal2025websight}, MathVista~\cite{lu2023mathvista}, ScienceQA~\cite{saikh2022scienceqa}, and CLEVR-Math~\cite{lindstrom2022clevr}. We evaluate MetaForge on 12 benchmarks, with 24{,}005 test samples. Table~\ref{tab:test_data} reports benchmarks, splits, sample counts, and data sources. We construct test sets in three ways. For ChartQA, AI2D, MathVista, and ScienceQA, we use full official test sets. For DocVQA, TallyQA, OCRVQA, RoBUT-WikiSQL, MapQA, Localized Narratives, and CLEVR-Math, we sample subsets to reduce evaluation cost. WebSight does not provide an official test split, so we sample from its training split. Because our training set excludes WebSight and code-generation tasks, this evaluation does not overlap with our training data.

\item \textbf{Tool Set Split.}
The IID tool setting uses the 10 tools available during training, while the OOD setting replaces them with 8 unseen tools. Both settings keep the shared meta-tools \texttt{create\_skill} and \texttt{run\_skill}, so all agents can forge and execute reusable tools when existing tools are insufficient. Table~\ref{tab:tool_details_v2} lists all tools used in the IID and OOD tool settings.

\end{itemize}


\begin{table}[h]
\centering
\setlength{\tabcolsep}{6pt}
\renewcommand{\arraystretch}{1.15}
\resizebox{\linewidth}{!}{%
\begin{tabular}{l|l|r}
\hline
\textbf{Dataset} & \textbf{Category} & \textbf{Samples} \\
\hline
InfographicVQA~\cite{mathew2022infographicvqa} & Document & 500 \\
AOKVQA~\cite{schwenk2022okvqa} & General VQA & 500 \\
COCOQA~\cite{du2019cocoqa} & General VQA & 500 \\
VQAv2~\cite{goyal2017making} & General VQA & 500 \\
Visual7W~\cite{zhu2016visual7w} & General VQA & 500 \\
TQA~\cite{kembhavi2017you} & General VQA & 500 \\
VQARAD~\cite{lau2018dataset} & General VQA & 500 \\
VSR~\cite{liu2023visual} & Image Captioning & 500 \\
MapQA~\cite{chang2022mapqa} & Geospatial & 500 \\
STVQA~\cite{biten2019scene} & OCR & 500 \\
TextVQA~\cite{singh2019towards} & OCR & 500 \\
VisualMRC~\cite{tanaka2021visualmrc} & OCR & 500 \\
IAM~\cite{marti2002iam} & OCR & 500 \\
Screen2Words~\cite{wang2021screen2words} & OCR & 500 \\
TATQA~\cite{zhu2021tat} & Table & 500 \\
FinQA~\cite{chen2021finqa} & Table & 500 \\
HiTab~\cite{cheng2022hitab} & Table & 500 \\
RoBUT-WTQ~\cite{zhao2023robut} & Table & 500 \\
RoBUT-SQA~\cite{zhao2023robut} & Table & 500 \\
\hline
\textbf{Total} & \textbf{6 categories} & \textbf{9{,}500} \\
\hline
\end{tabular}
}
\caption{Composition of the training set across datasets and task categories.}
\label{tab:training_data}
\end{table}



\begin{table}[!t]
\centering
\setlength{\tabcolsep}{5pt}
\renewcommand{\arraystretch}{1.15}
\resizebox{\linewidth}{!}{%
\begin{tabular}{l|c|r|c}
\hline
\textbf{Dataset} & \textbf{Split} & \textbf{Samples} & \textbf{Source} \\
\hline
DocVQA~\cite{mathew2021docvqa} & IID & 2{,}000 & Subset \\
TallyQA~\cite{acharya2019tallyqa} & IID & 2{,}000 & Subset \\
OCRVQA~\cite{mishra2019ocr} & IID & 2{,}000 & Subset \\
RoBUT-WikiSQL~\cite{zhao2023robut} & IID & 2{,}000 & Subset \\
MapQA~\cite{chang2022mapqa} & IID & 2{,}000 & Subset \\
Localized Narratives~\cite{pont2020connecting} & IID & 1{,}400 & Subset \\
\hline
ChartQA~\cite{masry2022chartqa} & OOD & 2{,}500 & Full \\
AI2D~\cite{kembhavi2016diagram} & OOD & 3{,}088 & Full \\
WebSight~\cite{bhathal2025websight} & OOD & 2{,}000 & Train-split \\
MathVista~\cite{lu2023mathvista} & OOD & 1{,}000 & Full \\
ScienceQA~\cite{saikh2022scienceqa} & OOD & 2{,}017 & Full \\
CLEVR-Math~\cite{lindstrom2022clevr} & OOD & 2{,}000 & Subset \\
\hline
\textbf{Total} & & \textbf{24{,}005} & \\
\hline
\end{tabular}%
}
\caption{Test set composition. ``Full'' denotes the complete official test set; ``Subset'' denotes a random subset of the official test set; ``Train-split'' denotes a sample drawn from the training set due to the absence of an official test split.}
\label{tab:test_data}
\end{table}



\begin{table*}[!t]
\centering
\resizebox{\textwidth}{!}{%
\setlength{\tabcolsep}{5pt}
\renewcommand{\arraystretch}{1.18}
\begin{tabular}{l|l|l|p{8.2cm}}
\hline
\textbf{Setting} & \textbf{Tool} & \textbf{Type} & \textbf{Function} \\
\hline
\multicolumn{4}{c}{\cellcolor{blockgray}\textbf{IID Tools}} \\
\hline
IID & Qwen3-VL-8B~\cite{bai2025qwen3} & Model Tool & General visual understanding and VQA. \\
IID & Qwen3-VL-32B~\cite{bai2025qwen3} & Model Tool & Stronger multimodal reasoning for complex visual analysis. \\
IID & Qwen3-VL-235B-A22B~\cite{bai2025qwen3} & Model Tool & Large-scale visual reasoning and answer verification. \\
IID & GroundingDINO~\cite{liu2024grounding} & Model Tool & Open-vocabulary object localization. \\
IID & SAM3~\cite{carion2025sam} & Model Tool & Object and region segmentation. \\
IID & PaddleOCR~\cite{cui2025paddleocr} & Model Tool & Text detection and recognition in images. \\
IID & EasyOCR\footnotemark[1] & Model Tool & Lightweight OCR for scene text. \\
IID & MinerU2.5~\cite{niu2025mineru2} & Model Tool & Layout analysis and structured document extraction. \\
IID & Qwen-Image-Edit~\cite{wu2025qwen} & Model Tool & Instruction-based image editing. \\
IID & OpenCV\footnotemark[2] & Executable Tool & Classical image operations such as crop, resize, thresholding, and edge detection. \\
\hline
\multicolumn{4}{c}{\cellcolor{blockgray}\textbf{OOD Tools}} \\
\hline
OOD & InternVL3.5-38B~\cite{wang2025internvl3} & Model Tool & Large multimodal model from a different architecture family. \\
OOD & InternVL3.5-14B~\cite{wang2025internvl3} & Model Tool & Efficient multimodal model from a different architecture family. \\
OOD & Step3~\cite{wang2025step} & Model Tool & Multimodal reasoning over image-text inputs. \\
OOD & WolframAlpha\footnotemark[3] & Executable Tool & Math, science, unit conversion, and symbolic calculation. \\
OOD & UniChart~\cite{masry2023unichart} & Model Tool & Chart-specific visual question answering. \\
OOD & Deplot~\cite{liu2023deplot} & Model Tool & Chart-to-table conversion. \\
OOD & ChartMoE~\cite{xuchartmoe} & Model Tool & Chart understanding and reconstruction. \\
OOD & Python Code~\cite{jiang2025verltool} & Executable Tool & Sandboxed numerical computation and data processing. \\
\hline
\end{tabular}%
}
\caption{IID and OOD tools used in our experiments. IID tools are available during training, while OOD tools are unseen during training.}
\label{tab:tool_details_v2}
\end{table*}
\footnotetext[1]{\url{https://github.com/JaidedAI/EasyOCR}}
\footnotetext[2]{\url{https://github.com/opencv/opencv}}
\footnotetext[3]{\url{https://www.wolframalpha.com/}}

\subsection{Baseline Details}
\label{app:baselines}

Table~\ref{tab:baseline_details_v2} summarizes the 16 baselines used in our experiments.

\begin{table*}[!t]
\centering
\resizebox{\textwidth}{!}{%
\setlength{\tabcolsep}{4pt}
\renewcommand{\arraystretch}{1.18}
\begin{tabular}{l|l|p{10.8cm}}
\hline
\textbf{Group} & \textbf{Model} & \textbf{Description} \\
\hline
Closed-source & GPT-5.4~\cite{openai2026gpt54} & Proprietary multimodal model from OpenAI. \\
Closed-source & Claude Sonnet 4.6~\cite{anthropic2026claude} & Proprietary multimodal model from Anthropic. \\
Closed-source & Gemini 3.1 Pro~\cite{google2026gemini31pro} & Proprietary multimodal model from Google. \\
\hline
Open-source multimodal model & Qwen2.5-VL-7B-Instruct~\cite{Qwen2.5-VL} & Open-source QwenVL model used as a 7B-scale baseline. \\
Open-source multimodal model & Qwen3-VL-8B-Instruct~\cite{bai2025qwen3} & Base policy family and main 8B-scale multimodal baseline. \\
Open-source multimodal model & Qwen3-VL-32B-Instruct~\cite{bai2025qwen3} & Larger Qwen3-VL baseline for scale comparison. \\
Open-source multimodal model & Qwen3-VL-235B-A22B-Instruct~\cite{bai2025qwen3} & Large-scale Qwen3-VL baseline. \\
Open-source multimodal model & InternVL3.5-14B-Instruct~\cite{wang2025internvl3} & InternVL-family baseline with a different architecture. \\
Open-source multimodal model & InternVL3.5-38B-Instruct~\cite{wang2025internvl3} & Larger InternVL-family baseline. \\
\hline
Tool-augmented multimodal model & VTool-R1-7B~\cite{wu2025vtool} & Reinforcement learning method for multimodal chain-of-thought and tool use. \\
Tool-augmented multimodal model & R1-Onevision-7B~\cite{yang2025r1} & Converts visual inputs into structured text for cross-modal reasoning. \\
Tool-augmented multimodal model & PyVision-Image-7B-RL~\cite{zhao2026pyvision} & Learns multi-turn tool invocation with cumulative tool rewards. \\
Tool-augmented multimodal model & DeepEyesV2-7B~\cite{hong2025deepeyesv2} & Uses cold-start training and reinforcement learning for tool usage behavior. \\
Tool-augmented multimodal model & PixelReasoner-RL-vl-7B~\cite{su2025pixel} & Learns pixel-level zooming and frame selection through reinforcement learning. \\
Tool-augmented multimodal model & AdaReasoner-7B~\cite{song2026adareasoner} & Trains tool usage as general reasoning behavior with Tool-GRPO. \\
Tool-augmented multimodal model & TikArt-8B~\cite{ding2026tikart} & Interleaves language reasoning with zooming and segmentation operations. \\
Tool-augmented multimodal model & ZwZ-8B~\cite{wei2026zooming} & Distills inference-time zooming behaviors into single-forward-pass primitives. \\
\hline
\end{tabular}%
}
\caption{Baseline models used for comparison.}
\label{tab:baseline_details_v2}
\end{table*}

\subsection{Hyperparameters Configuration}
\label{app:hyperparams}

Table~\ref{tab:hyperparams_v2} summarizes the training configuration.

\begin{table}[!t]
\centering
\setlength{\tabcolsep}{5pt}
\renewcommand{\arraystretch}{1.12}
\resizebox{\linewidth}{!}{%
\begin{tabular}{l|l}
\hline
\textbf{Item} & \textbf{Value} \\
\hline
Framework & VerlTool~\cite{jiang2025verltool} \\
Base policy & Qwen3-VL-8B-Instruct~\cite{bai2025qwen3} \\
Tool-forging model & Qwen3-VL-235B-A22B-Instruct~\cite{bai2025qwen3} \\
RL algorithm & GRPO \\
Python version & 3.10 \\
PyTorch version & 2.6.0 \\
Transformers version & 4.51.3 \\
vLLM version & 0.8.4 \\
Hardware & 8$\times$ NVIDIA H200 GPUs, 141 GB each \\
Precision & BF16 mixed precision \\
Rollouts per input $M$ & 8 \\
Sampling temperature / top-$p$ & 1.0 / 1.0 \\
Clipping ratio $\epsilon$ & 0.2 \\
KL coefficient $\beta$ & 0.002 \\
Learning rate & $3\text{e-}6$ \\
Batch size & 128 \\
Epochs / update steps & 3 / 223 \\
Training time & approximately 80 hours \\
Maximum prompt length & 24{,}576 tokens \\
Maximum response length & 24{,}576 tokens \\
Maximum action length & 8{,}192 tokens \\
Maximum observation length & 16{,}384 tokens \\
Maximum interaction turns $T$ & 10 \\
Forged-tool pool capacity $K$ & 50 \\
LLM judge & GPT-5~\cite{singh2025openai} \\
\hline
\end{tabular}
}
\caption{Training hyperparameters configuration.}
\label{tab:hyperparams_v2}
\end{table}

\section{Extended Experimental Analysis}
\label{app:extended_analysis}

\subsection{Forged Tool Pool Evolution Analysis}
\label{app:forged_tool_pool}
\begin{figure*}[htpb]
  \centering
  \includegraphics[width=1.0\textwidth]{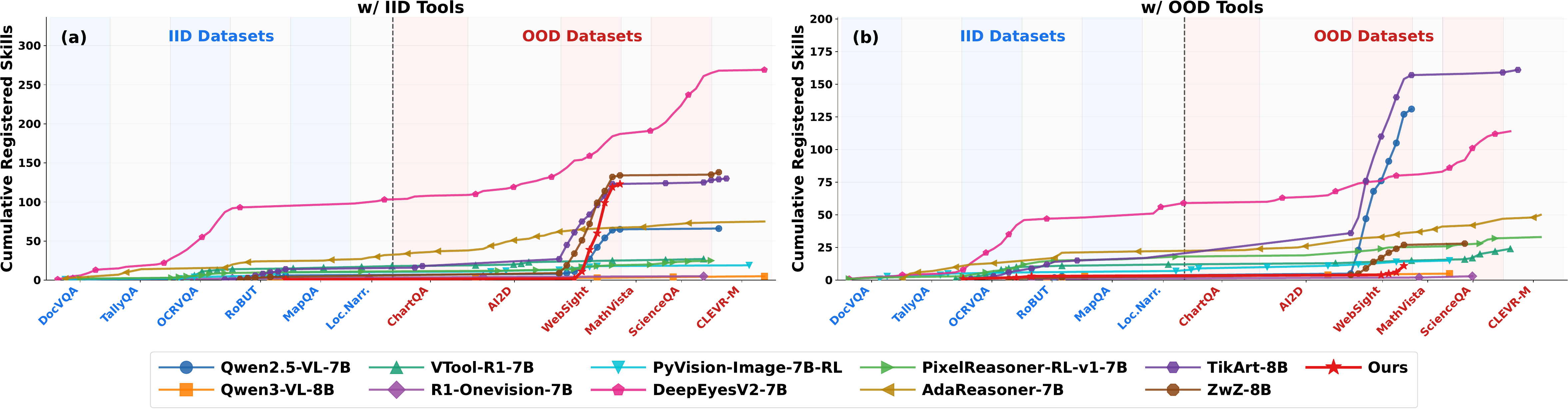}
  \caption{Cumulative registered forged tools over the evaluation process under IID and OOD tool settings.}
  \label{fig:ablation_2}
\end{figure*}

To analyze tool forging during evaluation, we track how the forged-tool pool grows as each method is evaluated on samples from the 12 datasets, ordered by dataset. As Figure~\ref{fig:ablation_2} shows, existing baselines exhibit two distinct forging patterns. One group continuously forges tools in large numbers, regardless of whether matching tools already exist for the current task. The other group barely forges any tools, even in OOD dataset regions where clear tool gaps exist. Both patterns reveal the same deficiency: these models fail to identify tool gaps between task requirements and the available tools, and thus cannot adapt their forging behavior.

In contrast, MetaForge adapts tool forging to the gap between task requirements and existing tools.
Under IID tools, it forges almost no tools on the first six datasets and then increases sharply on OOD datasets such as WebSight and MathVista before plateauing.
Under OOD tools, broader tool coverage leaves fewer gaps, so the model forges almost no tools.
This behavior comes from the reward design: when existing tools suffice, tool forging adds latency without improving accuracy.
When tools fall short, tool forging becomes necessary for the correct answer.
Through competition among sampled trajectories, the model learns to forge tools when needed.

\subsection{Reward Function Analysis}
\label{app:Reward} 
The Retrieve/Adapt reward is the core objective and activates the auxiliary rewards, so we ablate only the three auxiliary rewards.
As shown in Table~\ref{tab:ablation}, removing any auxiliary reward reduces accuracy, confirming that each component is useful.
Removing the format reward causes the largest drop across nearly all datasets.
This reward enforces valid tool-call syntax; without it, the model more often produces malformed calls that cannot be parsed or executed, blocking tool feedback for later decisions.
Removing the Forge reward mainly hurts OOD datasets, with a large drop on tasks such as WebSight.
These cases often contain clear tool gaps, showing that the Forge reward is important for encouraging on-demand tool forging.
Removing the Decide reward causes the smallest drop, but the decline appears under both IID and OOD tool settings.
This suggests that Decide supervision encourages concise calling paths and reduces redundant interaction rounds, which lowers the chance of intermediate errors.

\newcommand{\cmark}{\ding{51}}
\newcommand{\xmark}{\ding{55}}
\begin{table*}[!t]
\centering
\resizebox{\textwidth}{!}{%
\setlength{\tabcolsep}{3.5pt}
\renewcommand{\arraystretch}{1.25}
\begin{tabular}{cccc|cccccc|c|cccccc|c|c}
\hline
\multicolumn{4}{c|}{\textbf{Reward Components}}
 & \multicolumn{7}{c|}{\textbf{IID Datasets}}
 & \multicolumn{7}{c|}{\textbf{OOD Datasets}}
 & \textbf{ALL} \\
\cline{1-19}
 $r_\text{ra}$ & $r_\text{forge}$ & $r_\text{fmt}$ & $r_\text{dec}$
 & \textbf{DocVQA} & \textbf{TallyQA} & \textbf{OCRVQA} & \textbf{RoBUT} & \textbf{MapQA} & \textbf{Loc.Narr.} & \textbf{AVG.}
 & \textbf{ChartQA} & \textbf{AI2D} & \textbf{WebSight} & \textbf{MathVista} & \textbf{ScienceQA} & \textbf{CLEVR-M} & \textbf{AVG.} & \textbf{AVG.} \\
\hline
\multicolumn{19}{c}{\cellcolor{blockgray}\textbf{w/ IID Tools}} \\
\hline
 \multicolumn{4}{c|}{Qwen3-VL-8B~\cite{bai2025qwen3}} & 89.80&70.20&77.40&61.70&70.65&95.78&77.59 & 82.16&84.58&51.15&60.70&78.03&63.50&70.02&73.80 \\
\hline
 \cellcolor{oursblue}\cmark & \cellcolor{oursblue}\cmark & \cellcolor{oursblue}\cmark & \cellcolor{oursblue}\cmark & \cellcolor{oursblue}\textbf{97.90}&\cellcolor{oursblue}\textbf{80.52}&\cellcolor{oursblue}\textbf{86.51}&\cellcolor{oursblue}\textbf{87.36}&\cellcolor{oursblue}\textbf{77.98}&\cellcolor{oursblue}\textbf{99.62}&\cellcolor{oursblue}\textbf{88.32} & \cellcolor{oursblue}\textbf{90.49}&\cellcolor{oursblue}\textbf{89.60}&\cellcolor{oursblue}\textbf{98.34}&\cellcolor{oursblue}\textbf{78.01}&\cellcolor{oursblue}\textbf{89.68}&\cellcolor{oursblue}\underline{81.70}&\cellcolor{oursblue}\textbf{87.97}&\cellcolor{oursblue}\textbf{88.14} \\
 \cmark & \xmark & \cmark & \cmark & \underline{96.25}&\underline{74.25}&79.40&80.30&68.88&\underline{98.71}&82.97 & 85.88&87.82&73.45&70.20&85.91&\textbf{82.00}&80.88&81.92 \\
 \cmark & \cmark & \xmark & \cmark & 92.30&68.55&75.70&78.75&64.85&96.28&79.41 & 85.92&85.62&89.95&68.20&83.53&79.55&82.13&80.77 \\
 \cmark & \cmark & \cmark & \xmark & 96.00&73.65&\underline{83.05}&\underline{83.80}&\underline{76.35}&98.57&\underline{85.24} & \underline{87.64}&\underline{89.28}&\underline{90.80}&\underline{72.50}&\underline{86.56}&78.05&\underline{84.14}&\underline{84.69} \\
\hline
\multicolumn{19}{c}{\cellcolor{blockgray}\textbf{w/ OOD Tools}} \\
\hline
 \multicolumn{4}{c|}{Qwen3-VL-8B~\cite{bai2025qwen3}} & 90.20&71.15&76.95&65.35&72.85&95.71&78.70 & 82.04&84.68&51.70&58.80&77.93&63.20&69.73&74.21 \\
\hline
 \cellcolor{oursblue}\cmark & \cellcolor{oursblue}\cmark & \cellcolor{oursblue}\cmark & \cellcolor{oursblue}\cmark & \cellcolor{oursblue}\textbf{97.22}&\cellcolor{oursblue}\textbf{81.17}&\cellcolor{oursblue}\textbf{86.00}&\cellcolor{oursblue}\textbf{87.67}&\cellcolor{oursblue}\textbf{89.14}&\cellcolor{oursblue}\textbf{99.58}&\cellcolor{oursblue}\textbf{90.13} & \cellcolor{oursblue}\textbf{90.18}&\cellcolor{oursblue}\underline{88.15}&\cellcolor{oursblue}\textbf{98.72}&\cellcolor{oursblue}\textbf{79.78}&\cellcolor{oursblue}\textbf{89.73}&\cellcolor{oursblue}\underline{80.40}&\cellcolor{oursblue}\textbf{87.83}&\cellcolor{oursblue}\textbf{88.98} \\
 \cmark & \xmark & \cmark & \cmark & \underline{95.85}&\underline{74.80}&78.70&78.40&75.00&\underline{99.21}&83.66 & \underline{85.72}&87.50&75.00&71.00&85.37&\textbf{81.00}&80.93&82.30 \\
 \cmark & \cmark & \xmark & \cmark & 92.20&65.80&71.85&76.95&65.30&97.50&78.27 & 84.64&87.07&\underline{91.25}&67.80&85.27&78.65&82.45&80.36 \\
 \cmark & \cmark & \cmark & \xmark & 94.40&73.95&\underline{83.60}&\underline{83.75}&\underline{78.25}&98.21&\underline{85.36} & 85.48&\textbf{89.50}&90.65&\underline{71.70}&\underline{86.12}&78.55&\underline{83.67}&\underline{84.51} \\
\hline
\end{tabular}%
}
\caption{Ablation study of reward components under IID and OOD tool settings. \cmark\ and \xmark\ indicate the inclusion and removal of a reward component, respectively. We compare the full model with all four rewards against three variants, each removing one of $r_\text{forge}$, $r_\text{fmt}$, and $r_\text{dec}$.}
\label{tab:ablation}
\end{table*}

\subsection{Backbone Generalization Analysis}
\label{app:Backbone} 
We train MetaForge on three backbones: Qwen2.5-VL-7B, Qwen3-VL-8B, and InternVL3.5-14B, covering architectures (QwenVL and InternVL) and scales (7B/8B/14B).
As shown in Table~\ref{tab:backbone_generalization}, MetaForge brings gains across all backbones, indicating that it does not depend on specific architecture, scale, or pretraining strategy.
Weaker backbones benefit more.
InternVL3.5-14B has the lowest baseline ALL AVG (65.80/61.38), but improves to 85.54/84.69 after training, yielding the largest gain.
The stronger Qwen3-VL-8B reaches the best overall performance of 88.14/88.98. 
These results demonstrate that adaptive tool call and online tool forging are backbone-agnostic. They compensate for the inherent limitations of different backbone models through tool interaction and forging.

\begin{table*}[!t]
\centering
\resizebox{\textwidth}{!}{%
\setlength{\tabcolsep}{3.5pt}
\renewcommand{\arraystretch}{1.25}
\begin{tabular}{c|cccccc|c|cccccc|c|c}
\hline
\multirow{2}{*}{\textbf{Models\,($\downarrow$)\,/\,Datasets\,($\rightarrow$)}}
 & \multicolumn{7}{c|}{\textbf{IID Datasets}}
 & \multicolumn{7}{c|}{\textbf{OOD Datasets}}
 & \textbf{ALL} \\
\cline{2-16}
 & \textbf{DocVQA} & \textbf{TallyQA} & \textbf{OCRVQA} & \textbf{RoBUT} & \textbf{MapQA} & \textbf{Loc.Narr.} & \textbf{AVG.}
   & \textbf{ChartQA} & \textbf{AI2D} & \textbf{WebSight} & \textbf{MathVista} & \textbf{ScienceQA} & \textbf{CLEVR-M} & \textbf{AVG.} & \textbf{AVG.} \\
\hline
\multicolumn{16}{c}{\cellcolor{blockgray}\textbf{w/ IID Tools}} \\
\hline
 Qwen2.5-VL-7B~\cite{Qwen2.5-VL} & 93.43 & 68.22 & 71.83 & 70.24 & 45.31 & 99.10 & 74.69 & 74.95 & 48.94 & 71.47 & 57.04 & 73.73 & 64.40 & 65.09 & 69.89 \\
 \cellcolor{oursblue}\textbf{MetaForge} & \cellcolor{oursblue}\textbf{95.80} & \cellcolor{oursblue}\textbf{75.85} & \cellcolor{oursblue}\textbf{82.75} & \cellcolor{oursblue}\textbf{82.35} & \cellcolor{oursblue}\textbf{87.15} & \cellcolor{oursblue}\textbf{99.21} & \cellcolor{oursblue}\textbf{87.19} & \cellcolor{oursblue}\textbf{84.36} & \cellcolor{oursblue}\textbf{87.24} & \cellcolor{oursblue}\textbf{96.20} & \cellcolor{oursblue}\textbf{64.30} & \cellcolor{oursblue}\textbf{83.83} & \cellcolor{oursblue}\textbf{78.15} & \cellcolor{oursblue}\textbf{82.35} & \cellcolor{oursblue}\textbf{84.77} \\
\hline
 Qwen3-VL-8B~\cite{bai2025qwen3} & 89.80 & 70.20 & 77.40 & 61.70 & 70.65 & 95.78 & 77.59 & 82.16 & 84.58 & 51.15 & 60.70 & 78.03 & 63.50 & 70.02 & 73.80 \\
 \cellcolor{oursblue}\textbf{MetaForge} & \cellcolor{oursblue}\textbf{97.90} & \cellcolor{oursblue}\textbf{80.52} & \cellcolor{oursblue}\textbf{86.51} & \cellcolor{oursblue}\textbf{87.36} & \cellcolor{oursblue}\textbf{77.98} & \cellcolor{oursblue}\textbf{99.62} & \cellcolor{oursblue}\textbf{88.32} & \cellcolor{oursblue}\textbf{90.49} & \cellcolor{oursblue}\textbf{89.60} & \cellcolor{oursblue}\textbf{98.34} & \cellcolor{oursblue}\textbf{78.01} & \cellcolor{oursblue}\textbf{89.68} & \cellcolor{oursblue}\textbf{81.70} & \cellcolor{oursblue}\textbf{87.97} & \cellcolor{oursblue}\textbf{88.14} \\
\hline
 InternVL3.5-14B~\cite{wang2025internvl3} & 87.60 & 71.18 & 66.46 & 69.98 & 59.20 & 72.27 & 71.12 & 80.83 & 67.08 & 44.00 & 35.07 & 69.06 & 66.88 & 60.49 & 65.80 \\
 \cellcolor{oursblue}\textbf{MetaForge} & \cellcolor{oursblue}\textbf{95.90} & \cellcolor{oursblue}\textbf{72.90} & \cellcolor{oursblue}\textbf{83.25} & \cellcolor{oursblue}\textbf{83.70} & \cellcolor{oursblue}\textbf{78.60} & \cellcolor{oursblue}\textbf{99.00} & \cellcolor{oursblue}\textbf{85.56} & \cellcolor{oursblue}\textbf{87.32} & \cellcolor{oursblue}\textbf{87.75} & \cellcolor{oursblue}\textbf{96.60} & \cellcolor{oursblue}\textbf{78.20} & \cellcolor{oursblue}\textbf{85.07} & \cellcolor{oursblue}\textbf{78.20} & \cellcolor{oursblue}\textbf{85.52} & \cellcolor{oursblue}\textbf{85.54} \\
\hline
\multicolumn{16}{c}{\cellcolor{blockgray}\textbf{w/ OOD Tools}} \\
\hline
 Qwen2.5-VL-7B~\cite{Qwen2.5-VL} & 93.63 & 67.46 & 78.18 & 70.66 & 50.01 & 98.66 & 76.43 & 75.41 & 51.95 & 73.15 & 56.88 & 74.42 & 66.60 & 66.40 & 71.42 \\
 \cellcolor{oursblue}\textbf{MetaForge} & \cellcolor{oursblue}\textbf{95.45} & \cellcolor{oursblue}\textbf{76.75} & \cellcolor{oursblue}\textbf{83.10} & \cellcolor{oursblue}\textbf{82.90} & \cellcolor{oursblue}\textbf{87.00} & \cellcolor{oursblue}\textbf{99.35} & \cellcolor{oursblue}\textbf{87.43} & \cellcolor{oursblue}\textbf{84.56} & \cellcolor{oursblue}\textbf{89.24} & \cellcolor{oursblue}\textbf{96.00} & \cellcolor{oursblue}\textbf{64.10} & \cellcolor{oursblue}\textbf{84.38} & \cellcolor{oursblue}\textbf{79.40} & \cellcolor{oursblue}\textbf{82.95} & \cellcolor{oursblue}\textbf{85.19} \\
\hline
 Qwen3-VL-8B~\cite{bai2025qwen3} & 90.20 & 71.15 & 76.95 & 65.35 & 72.85 & 95.71 & 78.70 & 82.04 & 84.68 & 51.70 & 58.80 & 77.93 & 63.20 & 69.73 & 74.21 \\
 \cellcolor{oursblue}\textbf{MetaForge} & \cellcolor{oursblue}\textbf{97.22} & \cellcolor{oursblue}\textbf{81.17} & \cellcolor{oursblue}\textbf{86.00} & \cellcolor{oursblue}\textbf{87.67} & \cellcolor{oursblue}\textbf{89.14} & \cellcolor{oursblue}\textbf{99.58} & \cellcolor{oursblue}\textbf{90.13} & \cellcolor{oursblue}\textbf{90.18} & \cellcolor{oursblue}\textbf{88.15} & \cellcolor{oursblue}\textbf{98.72} & \cellcolor{oursblue}\textbf{79.78} & \cellcolor{oursblue}\textbf{89.73} & \cellcolor{oursblue}\textbf{80.40} & \cellcolor{oursblue}\textbf{87.83} & \cellcolor{oursblue}\textbf{88.98} \\
\hline
 InternVL3.5-14B~\cite{wang2025internvl3} & 88.20 & 63.85 & 59.10 & 59.26 & 53.90 & 70.64 & 65.83 & 74.44 & 63.47 & 39.50 & 29.90 & 65.34 & 68.95 & 56.93 & 61.38 \\
 \cellcolor{oursblue}\textbf{MetaForge} & \cellcolor{oursblue}\textbf{94.35} & \cellcolor{oursblue}\textbf{73.75} & \cellcolor{oursblue}\textbf{81.65} & \cellcolor{oursblue}\textbf{80.80} & \cellcolor{oursblue}\textbf{78.50} & \cellcolor{oursblue}\textbf{98.57} & \cellcolor{oursblue}\textbf{84.60} & \cellcolor{oursblue}\textbf{85.52} & \cellcolor{oursblue}\textbf{86.32} & \cellcolor{oursblue}\textbf{94.85} & \cellcolor{oursblue}\textbf{79.50} & \cellcolor{oursblue}\textbf{83.93} & \cellcolor{oursblue}\textbf{78.55} & \cellcolor{oursblue}\textbf{84.78} & \cellcolor{oursblue}\textbf{84.69} \\
\hline
\end{tabular}%
}
\caption{Generalization across backbone multimodal models. MetaForge is applied to three backbones spanning two architecture families (QwenVL, InternVL) and different parameter scales (7B/8B/14B).}
\label{tab:backbone_generalization}
\end{table*}

\subsection{Tool Usage Entropy Analysis}
\label{app:tool_usage_distribution}
\begin{figure*}[htpb]
  \centering
  \includegraphics[width=1.0\textwidth]{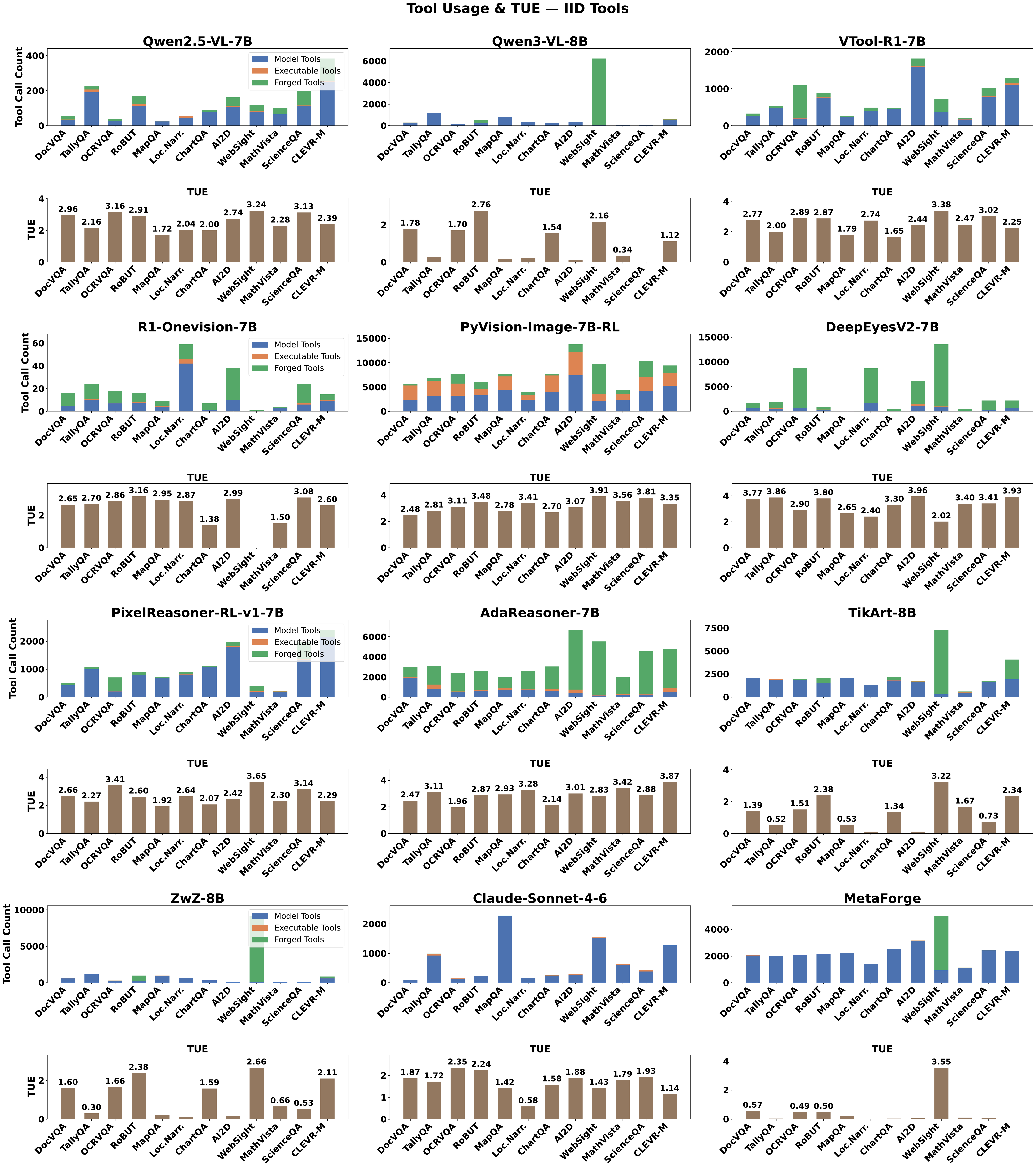}
  \caption{Tool call distribution and per-dataset Tool Usage Entropy (TUE) for each model under the IID tool setting.}
  \label{fig:ablation_5}
\end{figure*}

\begin{figure*}[htpb]
  \centering
  \includegraphics[width=1.0\textwidth]{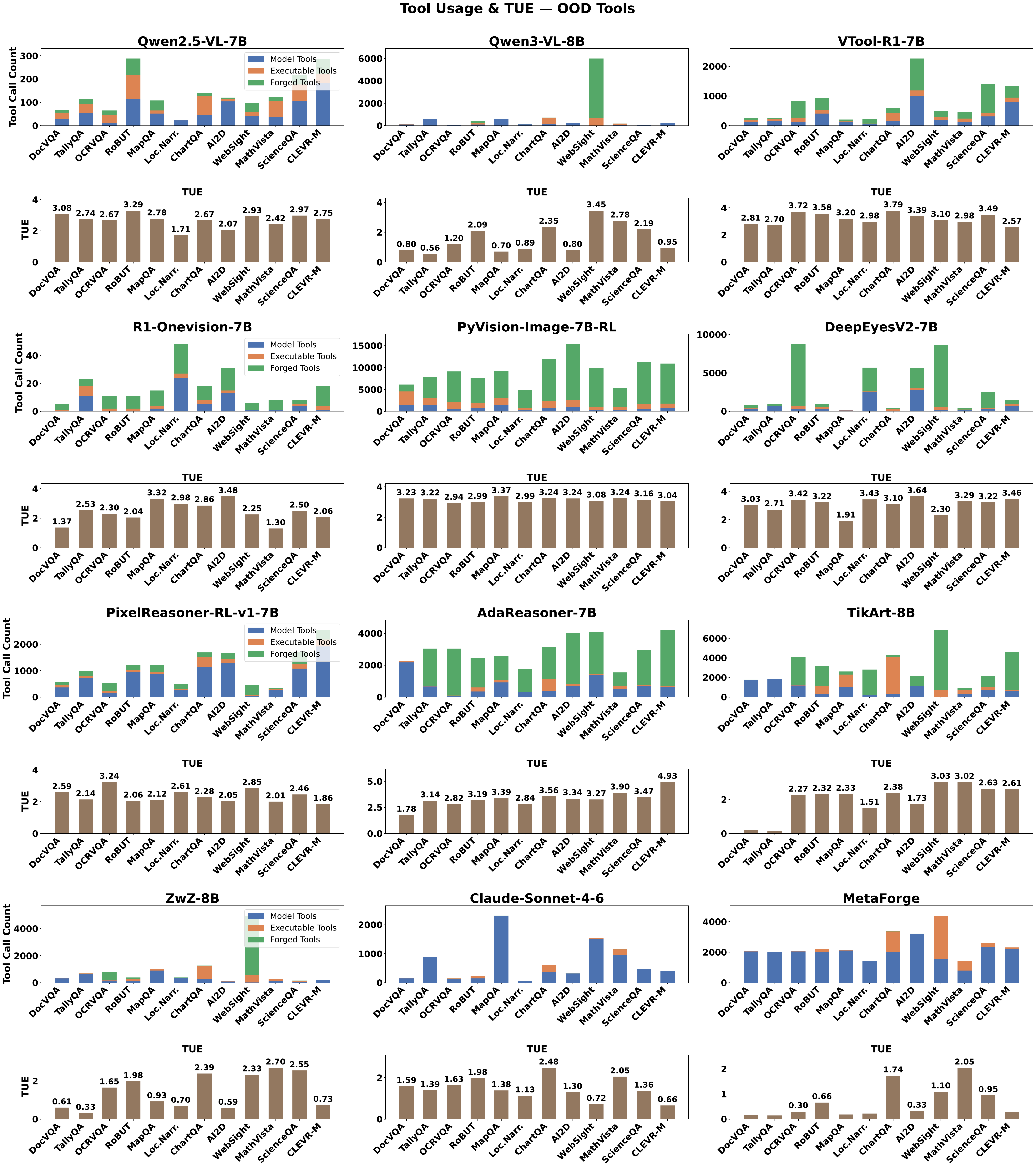}
 \caption{Tool call distribution and per-dataset Tool Usage Entropy (TUE) for each model under the OOD tool setting.}
  \label{fig:ablation_6}
\end{figure*}

To compare tool selection, we report call distributions and per-dataset TUE over 12 datasets (Figures~\ref{fig:ablation_5} and \ref{fig:ablation_6}). Distributions are grouped into Model tools, Executable tools, and Forged tools.

Existing methods show three patterns. First, some methods over-invoke tools. PyVision-Image-7B-RL and DeepEyesV2-7B make thousands to tens of thousands of calls per dataset, with TUE around 3.5--4.0. This suggests that they compensate for weak selection by trying many tools rather than choosing a concise combination. Second, some methods underuse external tools. R1-Onevision-7B issues only single-digit calls on most datasets, indicating limited ability or tendency to use the available tool pool. Third, some methods are biased toward tool categories. TikArt-8B and ZwZ-8B reach low TUE on some IID datasets (0.1--0.5), but their TUE rises to 2.6--3.2 on WebSight and MathVista, with overall TUE of 2.7--3.5. This indicates partial task differentiation, but not a stable orchestration strategy across tasks.

In contrast, MetaForge shows task-adaptive tool orchestration under both IID and OOD tool settings. Under IID tools (Figure~\ref{fig:ablation_5}), calls on 11 datasets are dominated by Model tools, suggesting that the policy recognizes their broad coverage in this setting. WebSight is the exception: forged-tool calls dominate because the existing tools lack the required website-generation function, so the model forges task-specific tools. Under OOD tools (Figure~\ref{fig:ablation_6}), the pattern changes. Executable tools increase on ChartQA, WebSight, and MathVista, showing that the policy can use newly available executable tools when they better match the task. Forged-tool calls nearly disappear, indicating that the expanded OOD tool set already covers most needs. Claude-Sonnet-4-6 also obtains relatively low TUE (IID 1.81, OOD 1.63), but it remains Model-dominant in both settings, suggesting focused selection without the same adaptive shift across tool settings.

These results align with MetaForge's design. The unified tool interface places Model tools, Executable tools, and Forged tools in one action space. Multi-turn GRPO optimizes the policy to retrieve, adapt, forge, and reuse tools according to task needs. Online tool forging expands the tool pool when existing tools are insufficient. Capacity-constrained tool pool management deduplicates overlapping forged tools and retains high-value ones. Together, these mechanisms lead to a compact and task-sensitive tool selection strategy.

\subsection{Analysis of Tool Invocation Success Rates} 
\label{app:tool_success_rates}
To thoroughly assess invocation robustness across different task scenarios and tool configurations, we report per-dataset tool call success rates for each model and conduct a comparative analysis under both the IID and OOD tool settings.
Under the IID tool setting, per-dataset tool call success rates reveal significant stratification among models. Existing baselines span a wide range of overall success rates, from below 50\% (e.g., R1-Onevision-7B at 47.6\%) to around 90\%. Most baselines exhibit large cross-dataset fluctuations, with some dropping to 0\% on specific datasets. 

Under the OOD tool setting, most baselines exhibit performance shifts between the two tool sets: some degrade substantially (e.g., VTool-R1-7B drops from 72.1\% to 58.5\%), while others improve (e.g., Qwen2.5-VL-7B rises from 46.5\% to 66.2\%). This sensitivity to tool set composition exposes the lack of robust tool generalization in existing methods. MetaForge never encounters the OOD tools during training, yet its success rate improves from 96.6\% to 98.2\%. All 12 datasets exceed 84.6\%, and 9 of them surpass 98\%. This result indicates that MetaForge learns general-purpose tool orchestration rather than memorizing specific tool configurations. The unified tool interface enables flexible dispatch of unfamiliar tools based on functional descriptions. Furthermore, online tool forging enables the model to forge task-specific tools on demand, effectively compensating for the limitations of the predefined tool set. Together, these two components ensure consistently high tool call success rates under arbitrary tool configurations.

\begin{figure*}[htpb]
  \centering
  \includegraphics[width=1.0\textwidth]{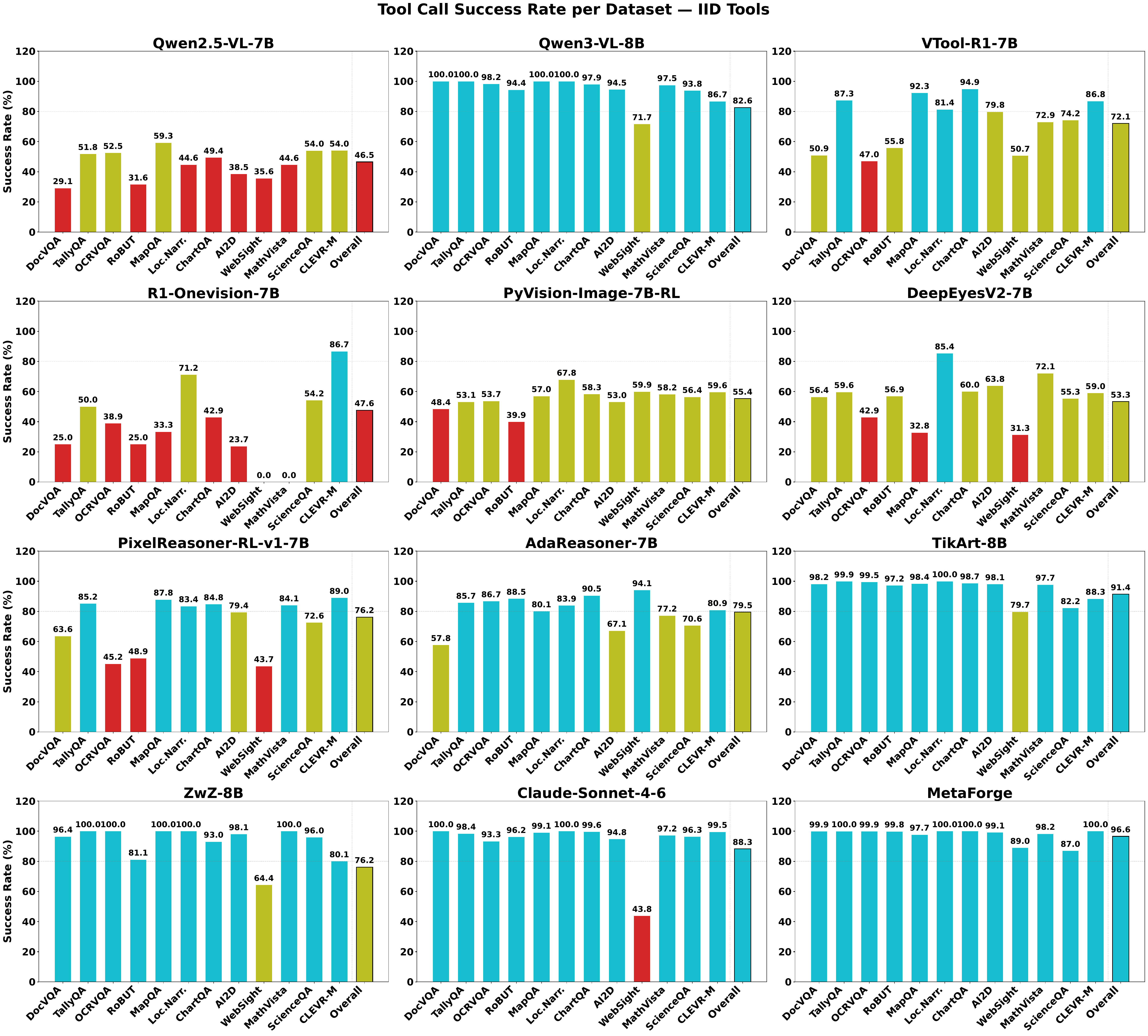}
  \caption{Per-dataset tool call success rates for each model under the IID tool setting. Existing baselines exhibit large cross-dataset variance, with some dropping to near 0\% on specific datasets. MetaForge achieves at least 87\% on all 12 datasets and exhibits the smallest cross-dataset variance.}
  \label{fig:ablation_7}
\end{figure*}

\begin{figure*}[htpb]
  \centering
  \includegraphics[width=1.0\textwidth]{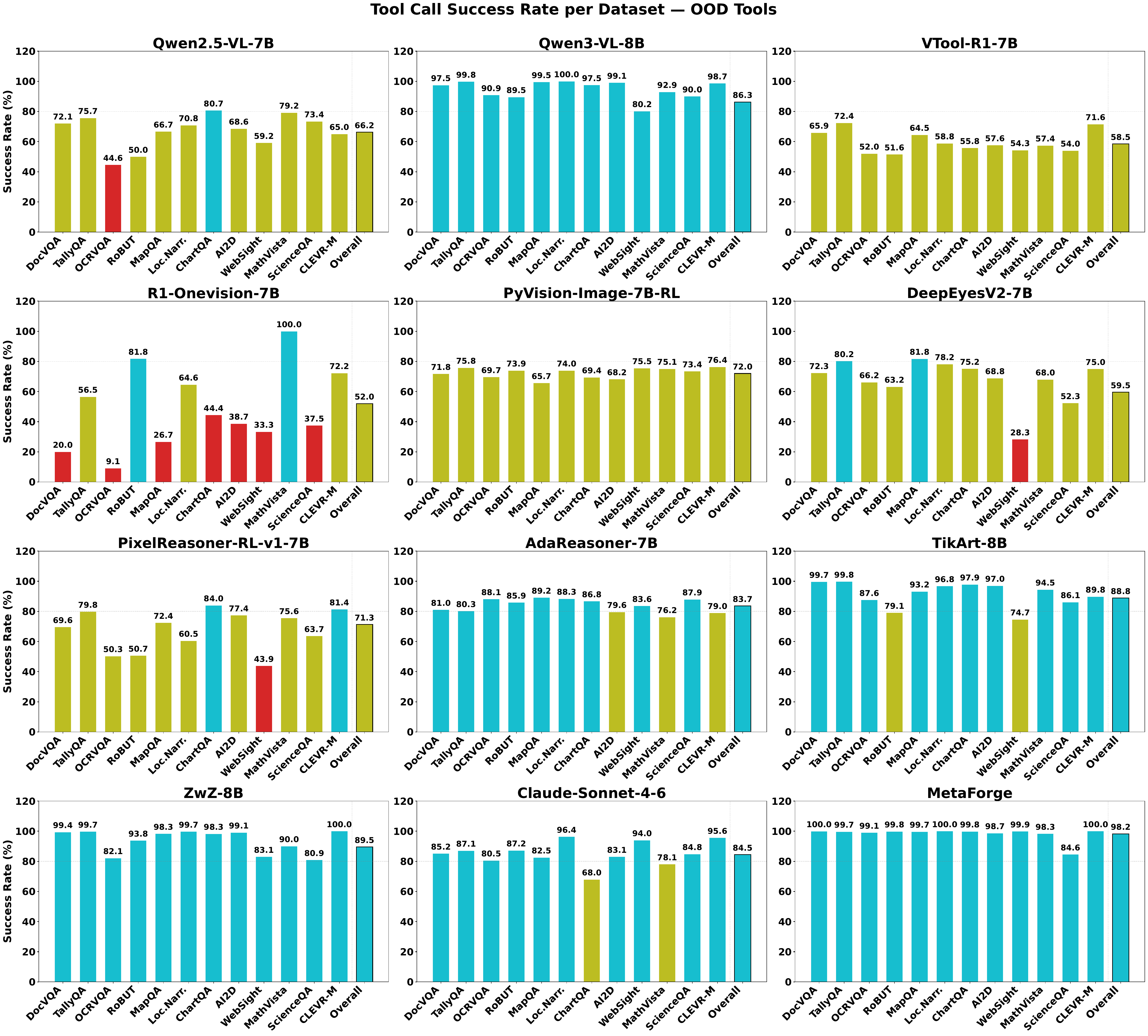}
  \caption{Per-dataset tool call success rates for each model under the OOD tool setting. Most baselines show notable performance shifts between the two tool sets, whereas MetaForge improves from an overall 96.6\% (IID) to 98.2\% (OOD), with all datasets exceeding 84.6\%.}
  \label{fig:ablation_8}
\end{figure*}



\subsection{Error Pattern Analysis}
\label{app:error_pattern}
To investigate the root causes of invocation failures across methods, we broadly categorize the 11 error types into three groups:
(1)~\textbf{Hallucination errors}, including \texttt{unknown\_skill} (invoking a non-existent forged tool), \texttt{missing\_entrypoint} (invoking a non-existent script within a forged tool, where the model fabricates script names without consulting the tool description), and \texttt{invalid\_entrypoint} (specifying a non-compliant script entry point, as only \texttt{.py} and \texttt{.sh} formats are supported), reflecting generally poor awareness of the available tool pool;
(2)~\textbf{Parameter errors}, including \texttt{invalid\_parameters} / \texttt{invalid\_arguments} (passing invalid values), \texttt{missing\_parameters} (omitting required parameters), and \texttt{missing\_image\_index} / \texttt{invalid\_image\_index} (missing or out-of-range image indices), reflecting a limited understanding of tool interface specifications;
(3)~\textbf{Execution errors}, including \texttt{runtime\_error} (internal script execution failure, typically due to missing internal arguments), \texttt{request\_failed} (external service exceptions such as network instability), and \texttt{timeout} (subprocess execution timeout), which typically arise mainly from environmental factors or forged-tool code quality rather than model behavior defects.

As shown in the figure, the majority of existing methods concentrate their failures on hallucination and parameter errors. PyVision-Image-7B-RL produces over 41K failed calls under the IID setting, with \texttt{invalid\_parameters} alone accounting for 19,785 instances, indicating a severe lack of understanding of tool parameter formats. DeepEyesV2-7B accumulates over 21K failures, with \texttt{missing\_entrypoint} reaching 7,610, revealing frequent hallucinated invocations of non-existent scripts within forged tools. R1-Onevision-7B exhibits the most severe dysfunction, issuing only 231 calls in total with a success rate of merely 47.62\%. Of its failed calls, \texttt{unknown\_skill} alone accounts for 59 instances, suggesting that the model lacks the basic ability to recognize and select from the available tool pool. Under the OOD setting, \texttt{unknown\_skill} errors increase substantially for most methods (VTool rises from 472 to 1,096; PixelReasoner rises from 607 to 1,424), indicating that methods trained on fixed tool sets lack generalization to unseen tools.

MetaForge achieves the highest overall success rates of 96.59\% (IID) and 98.24\% (OOD) among all methods. The errors consist primarily of \texttt{runtime\_error} (515 instances, concentrated on WebSight forged-tool code execution anomalies) and \texttt{invalid\_image\_index} (357/411 instances, concentrated on ScienceQA image index handling), both of which are execution-class errors attributable to environmental factors rather than deficiencies in the model's tool understanding. More importantly, hallucination errors (\texttt{missing\_entrypoint} + \texttt{unknown\_skill}) total only 14 and 9 under IID and OOD, respectively, validating the generalized invocation behavior endowed by the unified tool interface and Forge reward. Notably, Claude-Sonnet-4-6, serving as the closed-source upper bound, concentrates its errors almost exclusively on \texttt{request\_failed} (915 under IID, 1,293 under OOD), which stem from external API service instability rather than any deficiency in tool understanding. This further corroborates that once a model acquires robust tool understanding, the dominant failure mode shifts from model-intrinsic deficits (e.g., hallucinated invocations and parameter misspecification) to external execution conditions beyond the model's control.
\begin{figure*}[htpb]
  \centering
  \includegraphics[width=1.0\textwidth]{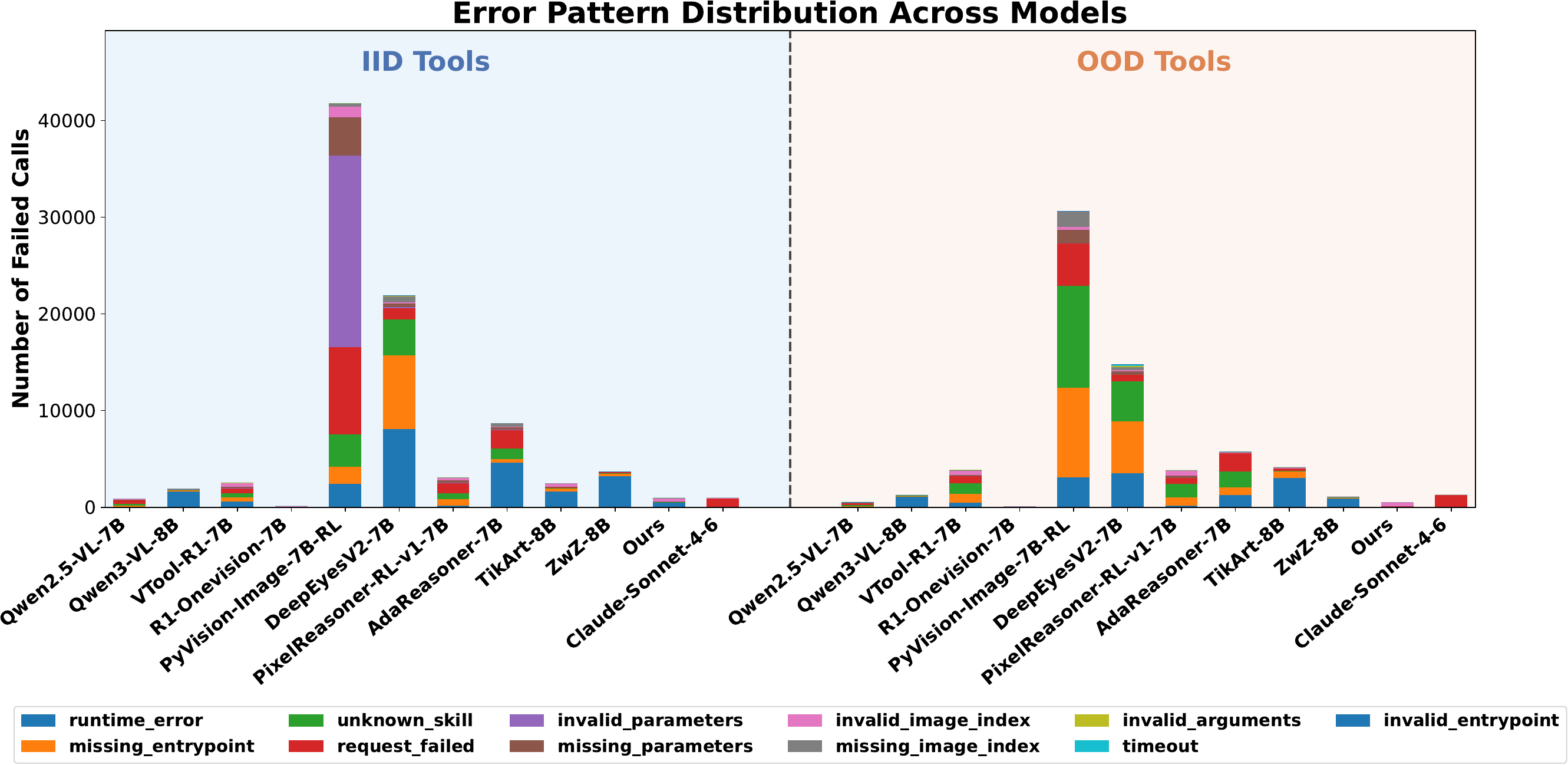}
  \caption{Distribution of error types across models under IID (left) and OOD (right) tool settings. Errors are categorized into Hallucination (e.g., invoking non-existent tools), Parameter (e.g., missing or invalid arguments), and Execution (e.g., runtime exceptions) errors. MetaForge produces the fewest total errors, with residual failures concentrated in execution-class errors attributable to environmental factors rather than tool-understanding deficiencies.}
  \label{fig:ablation_9}
\end{figure*}

\subsection{Joint Sensitivity Analysis of Learning Rate and Format Reward}
\label{app:joint_sensitivity}

The actor learning rate and the format reward magnitude are two key hyperparameters that jointly govern training dynamics. The learning rate controls the step size of policy gradient updates, while the format reward magnitude determines how strongly the model is penalized or rewarded for syntactic compliance in tool calls. We conduct a grid search over learning rates $\mathrm{lr} \in \{1\text{e-}6,\, 2\text{e-}6,\, 3\text{e-}6,\, 4\text{e-}6,\, 5\text{e-}6\}$ and format reward magnitudes $\alpha \in \{0.1,\, 0.2,\, 0.3,\, 0.4,\, 0.5\}$, yielding 25 configurations evaluated under both IID and OOD tool settings.

As shown in Table~\ref{tab:joint_sensitivity}, the combination of $\mathrm{lr}{=}3\text{e-}6$ and $\alpha{=}0.2$ achieves the highest overall accuracy under both tool settings (IID: 88.14\%, OOD: 88.98\%). Several observations emerge from the grid. First, for any fixed format reward magnitude, $\mathrm{lr}{=}3\text{e-}6$ consistently yields the best or near-best accuracy, confirming it as the optimal learning rate across a range of format reward strengths. Second, for any fixed learning rate, $\alpha{=}0.2$ consistently outperforms other magnitudes, indicating that a moderate format penalty provides the best balance between syntactic compliance and behavioral flexibility. Third, the two hyperparameters exhibit a mild interaction effect: at lower learning rates (e.g., $1\text{e-}6$), performance is relatively insensitive to $\alpha$ because the small policy updates limit the influence of any individual reward signal; at higher learning rates (e.g., $5\text{e-}6$), the choice of $\alpha$ becomes more consequential, as large policy updates amplify the impact of the format reward, and an overly strong penalty ($\alpha \geq 0.4$) combined with aggressive optimization leads to compounded performance degradation.

\begin{table*}[!t]
\centering
\resizebox{\textwidth}{!}{%
\setlength{\tabcolsep}{3.5pt}
\renewcommand{\arraystretch}{1.25}
\begin{tabular}{c|c|cccccc|c|cccccc|c|c}
\hline
\multirow{2}{*}{lr}
 & \multirow{2}{*}{\textbf{$\alpha$}}
 & \multicolumn{7}{c|}{\textbf{IID Datasets}}
 & \multicolumn{7}{c|}{\textbf{OOD Datasets}}
 & \textbf{ALL} \\
\cline{3-17}
 & & \textbf{DocVQA} & \textbf{TallyQA} & \textbf{OCRVQA} & \textbf{RoBUT} & \textbf{MapQA} & \textbf{Loc.Narr.} & \textbf{AVG.}
   & \textbf{ChartQA} & \textbf{AI2D} & \textbf{WebSight} & \textbf{MathVista} & \textbf{ScienceQA} & \textbf{CLEVR-M} & \textbf{AVG.} & \textbf{AVG.} \\
\hline
\multicolumn{17}{c}{\cellcolor{blockgray}\textbf{w/ IID Tools}} \\
\hline
 \multirow{5}{*}{$1\text{e-}6$} & $0.1$ & 94.81&75.59&81.86&82.13&71.61&97.69&83.95&85.81&85.35&89.58&70.92&84.90&75.03&81.93&82.94 \\
  & $0.2$ & 95.43&76.84&82.73&83.14&73.42&98.17&84.96&86.72&86.37&91.24&72.43&86.14&76.85&83.29&84.12 \\
  & $0.3$ & 95.23&76.19&82.14&82.62&72.66&97.91&84.46&86.10&85.92&90.24&71.53&85.47&75.75&82.50&83.48 \\
  & $0.4$ & 94.42&75.24&81.58&81.46&71.79&97.66&83.69&85.35&85.07&89.31&70.24&84.94&74.48&81.57&82.63 \\
  & $0.5$ & 94.25&73.75&80.25&80.65&70.08&97.40&82.73&83.86&83.60&86.72&67.28&83.57&73.85&79.81&81.27 \\
\hline
 \multirow{5}{*}{$2\text{e-}6$} & $0.1$ & 96.44&77.75&84.54&85.43&76.90&98.89&86.66&88.04&87.72&95.74&75.19&87.43&80.69&85.80&86.23 \\
  & $0.2$ & 97.12&79.34&85.63&86.28&\textbf{78.42}&99.28&87.68&89.57&88.93&97.48&76.84&88.92&\textbf{82.34}&87.35&87.51 \\
  & $0.3$ & 96.99&78.96&85.43&86.05&\underline{78.13}&99.21&87.46&89.35&88.77&96.95&76.39&88.67&\underline{81.97}&87.02&87.24 \\
  & $0.4$ & 96.62&78.27&84.69&85.16&77.44&98.91&86.85&88.54&87.85&95.89&74.94&87.89&80.84&85.99&86.42 \\
  & $0.5$ & 95.60&77.50&83.56&84.14&74.26&98.55&85.60&87.26&86.81&93.31&73.12&87.89&79.67&84.68&85.14 \\
\hline
 \multirow{5}{*}{$3\text{e-}6$} & $0.1$ & 97.23&78.84&85.44&85.43&75.80&99.29&87.01&89.06&88.50&95.70&76.31&88.66&80.27&86.42&86.71 \\
 \cellcolor{oursblue} & \cellcolor{oursblue}$\mathbf{0.2}$ & \cellcolor{oursblue}\textbf{97.90}&\cellcolor{oursblue}\textbf{80.52}&\cellcolor{oursblue}\textbf{86.51}&\cellcolor{oursblue}\textbf{87.36}&\cellcolor{oursblue}77.98&\cellcolor{oursblue}\textbf{99.62}&\cellcolor{oursblue}\textbf{88.32}&\cellcolor{oursblue}\textbf{90.49}&\cellcolor{oursblue}\textbf{89.60}&\cellcolor{oursblue}\textbf{98.34}&\cellcolor{oursblue}\textbf{78.01}&\cellcolor{oursblue}\textbf{89.68}&\cellcolor{oursblue}81.70&\cellcolor{oursblue}\textbf{87.97}&\cellcolor{oursblue}\textbf{88.14} \\
  & $0.3$ & \underline{97.62}&\underline{80.09}&\underline{86.08}&\underline{86.88}&77.54&\underline{99.45}&\underline{87.94}&\underline{90.11}&\underline{89.28}&\underline{97.55}&\underline{77.37}&\underline{89.42}&81.13&\underline{87.48}&\underline{87.71} \\
  & $0.4$ & 97.35&79.20&85.56&85.99&76.72&99.24&87.34&89.29&88.63&95.66&76.51&88.82&80.18&86.52&86.93 \\
  & $0.5$ & 96.74&78.25&84.48&85.69&74.70&99.16&86.50&88.78&88.15&95.02&74.68&87.91&78.69&85.54&86.02 \\
\hline
 \multirow{5}{*}{$4\text{e-}6$} & $0.1$ & 96.31&77.92&84.32&84.63&74.87&98.88&86.15&88.10&87.14&95.70&74.32&87.34&78.01&85.10&85.63 \\
  & $0.2$ & 97.04&79.17&85.27&86.14&76.73&99.14&87.25&89.47&88.34&97.06&76.42&88.13&79.83&86.54&86.90 \\
  & $0.3$ & 96.85&78.77&84.96&85.88&76.38&99.05&86.98&89.17&88.04&96.45&75.68&87.71&79.53&86.10&86.54 \\
  & $0.4$ & 96.49&78.01&84.59&85.10&74.93&98.90&86.34&88.47&87.19&95.55&74.53&87.42&78.18&85.22&85.78 \\
  & $0.5$ & 95.95&77.66&82.82&84.45&73.46&98.40&85.46&86.58&86.55&92.49&73.15&86.06&76.78&83.60&84.53 \\
\hline
 \multirow{5}{*}{$5\text{e-}6$} & $0.1$ & 96.03&77.59&83.87&84.98&73.71&98.37&85.76&87.77&86.49&94.78&73.02&86.86&77.50&84.40&85.08 \\
  & $0.2$ & 96.84&78.93&84.87&85.72&75.63&98.84&86.81&88.94&87.83&96.43&75.24&87.63&78.93&85.83&86.32 \\
  & $0.3$ & 96.59&78.25&84.40&85.28&75.01&98.79&86.39&88.29&87.39&95.31&74.17&87.18&78.21&85.09&85.74 \\
  & $0.4$ & 96.31&77.79&83.66&84.51&73.09&98.62&85.66&87.84&86.89&94.16&73.36&86.76&77.26&84.38&85.02 \\
  & $0.5$ & 95.45&76.54&83.18&83.31&72.18&98.34&84.83&86.65&86.35&91.16&71.87&85.86&76.15&83.01&83.92 \\
\hline
\multicolumn{17}{c}{\cellcolor{blockgray}\textbf{w/ OOD Tools}} \\
\hline
 \multirow{5}{*}{$1\text{e-}6$} & $0.1$ & 93.75&75.64&81.24&82.32&82.75&97.29&85.50&84.55&84.36&90.76&71.81&84.46&74.50&81.74&83.62 \\
  & $0.2$ & 94.82&77.23&82.37&83.54&84.28&97.83&86.68&86.42&85.74&93.47&74.38&85.83&76.12&83.66&85.17 \\
  & $0.3$ & 94.59&76.70&82.01&83.03&83.75&97.77&86.31&85.96&85.31&92.90&73.55&85.53&75.67&83.15&84.73 \\
  & $0.4$ & 94.37&76.15&81.19&82.25&82.60&97.36&85.65&85.36&84.43&91.33&71.71&84.72&74.61&82.03&83.84 \\
  & $0.5$ & 93.28&75.15&80.28&80.85&81.06&97.07&84.62&84.11&83.80&90.24&71.23&83.10&73.19&80.95&82.78 \\
\hline
 \multirow{5}{*}{$2\text{e-}6$} & $0.1$ & 95.83&78.60&83.94&85.57&87.18&98.79&88.32&87.76&86.68&94.86&77.96&87.60&77.32&85.36&86.84 \\
  & $0.2$ & 96.53&80.28&85.24&86.83&\textbf{89.47}&99.21&89.59&89.43&87.62&97.84&\textbf{80.14}&88.94&79.52&\underline{87.25}&\underline{88.42} \\
  & $0.3$ & 96.32&79.86&84.95&86.47&88.93&99.10&89.27&89.11&87.24&97.30&79.52&88.56&78.99&86.79&88.03 \\
  & $0.4$ & 96.18&78.95&84.58&85.67&88.58&98.92&88.81&88.65&86.79&96.43&78.33&88.10&78.57&86.15&87.48 \\
  & $0.5$ & 95.26&77.64&83.22&84.46&86.97&98.59&87.69&87.44&85.98&94.03&76.42&87.08&76.95&84.65&86.17 \\
\hline
 \multirow{5}{*}{$3\text{e-}6$} & $0.1$ & 96.43&79.61&84.08&86.19&87.37&99.00&88.78&89.01&86.68&96.64&75.80&88.40&77.82&85.73&87.25 \\
 \cellcolor{oursblue} & \cellcolor{oursblue}$\mathbf{0.2}$ & \cellcolor{oursblue}\textbf{97.22}&\cellcolor{oursblue}\textbf{81.17}&\cellcolor{oursblue}\textbf{86.00}&\cellcolor{oursblue}\textbf{87.67}&\cellcolor{oursblue}\underline{89.14}&\cellcolor{oursblue}\textbf{99.58}&\cellcolor{oursblue}\textbf{90.13}&\cellcolor{oursblue}\textbf{90.18}&\cellcolor{oursblue}\textbf{88.15}&\cellcolor{oursblue}\textbf{98.72}&\cellcolor{oursblue}\underline{79.78}&\cellcolor{oursblue}\textbf{89.73}&\cellcolor{oursblue}\textbf{80.40}&\cellcolor{oursblue}\textbf{87.83}&\cellcolor{oursblue}\textbf{88.98} \\
  & $0.3$ & \underline{96.95}&\underline{80.59}&\underline{85.52}&\underline{87.06}&88.45&\underline{99.39}&\underline{89.66}&\underline{89.61}&\underline{87.72}&\underline{97.97}&78.69&\underline{89.40}&\underline{79.68}&87.18&88.42 \\
  & $0.4$ & 96.80&80.12&84.77&86.83&87.57&99.23&89.22&89.46&87.40&96.50&78.03&89.01&78.86&86.54&87.88 \\
  & $0.5$ & 96.41&78.92&84.13&85.40&86.50&99.07&88.40&88.15&86.38&95.72&76.49&88.13&77.26&85.36&86.88 \\
\hline
 \multirow{5}{*}{$4\text{e-}6$} & $0.1$ & 95.70&78.66&83.50&85.36&85.59&98.59&87.90&87.94&85.56&94.54&75.39&86.93&76.40&84.46&86.18 \\
  & $0.2$ & 96.38&79.84&84.72&86.34&87.82&99.04&89.02&89.17&87.04&97.42&77.83&88.42&78.84&86.45&87.74 \\
  & $0.3$ & 96.13&79.48&84.42&85.85&87.19&98.96&88.67&88.62&86.62&96.51&77.12&88.14&78.21&85.87&87.27 \\
  & $0.4$ & 95.61&78.46&83.74&85.24&85.86&98.56&87.91&88.36&86.03&95.73&75.78&87.65&77.23&85.13&86.52 \\
  & $0.5$ & 95.39&77.95&83.05&84.86&83.98&98.51&87.29&87.33&84.91&93.48&73.75&85.73&75.13&83.39&85.34 \\
\hline
 \multirow{5}{*}{$5\text{e-}6$} & $0.1$ & 94.87&77.87&82.36&84.11&84.38&98.17&86.96&86.79&85.57&94.15&75.36&86.49&76.23&84.10&85.53 \\
  & $0.2$ & 96.13&79.43&84.18&85.82&86.73&98.74&88.51&88.53&86.47&96.84&76.93&87.82&78.23&85.80&87.15 \\
  & $0.3$ & 95.87&79.08&83.62&85.42&85.97&98.58&88.09&88.13&85.87&96.11&75.98&87.39&77.67&85.19&86.64 \\
  & $0.4$ & 95.24&78.09&83.24&84.51&85.02&98.24&87.39&87.19&85.57&94.74&74.81&86.73&76.58&84.27&85.83 \\
  & $0.5$ & 95.25&76.39&82.13&83.70&83.16&98.48&86.52&86.49&84.73&91.95&72.87&85.93&75.68&82.94&84.73 \\
\hline
\end{tabular}%
}
\caption{Joint sensitivity analysis of learning rate and format reward magnitude. We vary the learning rate from $1\text{e-}6$ to $5\text{e-}6$ and the format reward magnitude $\alpha$ from $0.1$ to $0.5$, yielding 25 configurations. We report accuracy (\%) across all 12 benchmarks under IID and OOD tool settings.}
\label{tab:joint_sensitivity}
\end{table*}

\subsection{Case Study}
\label{app:case_study}
To more intuitively demonstrate the advantages of MetaForge, we select six representative cases from the validation set and compare it against three types of baselines: the base model (Qwen3-VL-8B), a tool-augmented model (TikArt-8B), and a closed-source model (Claude-Sonnet-4-6). These cases are organized into three categories: (1) \textbf{Type~A}: only MetaForge answers correctly while all three baselines fail, demonstrating superior accuracy through adaptive tool orchestration; (2) \textbf{Type~B}: all models answer correctly but MetaForge requires significantly fewer interaction turns, demonstrating its ability to achieve an optimal trade-off between accuracy and inference cost through the Decide reward; and (3) \textbf{Type~C}: MetaForge forges task-specific tools when existing tools are insufficient and succeeds, while baselines either fail to forge tools or create ineffective ones, demonstrating its ability to extend the tool pool on demand.

\paragraph{Type~A: MetaForge Achieves Higher Accuracy via Adaptive Tool Orchestration.}
Case A1 is a ChartQA sample evaluated under the OOD dataset and OOD tool setting, requiring extraction of a numerical difference from a bar chart (ground truth: 52). MetaForge completes the task in 4~turns by orchestrating three tools: it first calls ChartMoe to extract the PET bottle collection rates, recognizes that the returned country names do not match the chart, then switches to the suitable tool InternVL3.5-14B-Instruct for re-extraction, and finally invokes \texttt{python\_code} to compute the difference. This trajectory illustrates a key behavior acquired through multi-turn GRPO training: the ability to detect tool failure from observations and adaptively switch to suitable tools. In contrast, Qwen3-VL-8B calls UniChart, then \texttt{python\_code}, but computes from incorrect values. TikArt-8B attempts to solve the problem using only \texttt{python\_code} without chart extraction. Claude-Sonnet-4-6 submits an answer directly without invoking tools, producing an incorrect result.

Case A2 is a MathVista sample evaluated under the OOD dataset and OOD tool setting, requiring computation of the average between the longest light-blue bar and the shortest gray bar in a chart (ground truth: 273). MetaForge decomposes the task into two stages: first calling UniChart to extract bar values (415 and 131), then invoking \texttt{python\_code} to compute the mean, yielding the correct answer. In contrast, Qwen3-VL-8B attempts to answer without tool usage and misidentifies the bar values, producing an incorrect result; TikArt-8B uses \texttt{python\_code} for computation but operates on incorrect values, also getting the wrong answer; and Claude-Sonnet-4-6 answers directly without tool invocation, producing an incorrect result. These cases demonstrate that multi-turn GRPO training enables MetaForge to learn adaptive tool selection that recovers from tool failures, yielding higher task accuracy.

\paragraph{Type~B: Optimizing Accuracy-Efficiency Trade-off via Decide Reward.}
Case B1 is a MathVista sample evaluated under the OOD dataset and OOD tool setting, requiring identification of the accuracy difference between the most and least accurate algorithms in a bar chart (ground truth: 80). MetaForge solves it in just 2 turns by directly calling InternVL3.5-14B-Instruct, which accurately extracts all four bar values in a single invocation, enabling immediate answer computation. In comparison, Qwen3-VL-8B requires 4~turns, sequentially trying UniChart, Deplot, and ChartMoe before obtaining usable values; TikArt-8B requires 5~turns, calling UniChart twice, then Deplot and \texttt{python\_code}; Claude-Sonnet-4-6 requires 3~turns, first calling ChartMoe and then cross-validating with InternVL3.5-14B-Instruct before computing the answer.

Case B2 is a CLEVR-Math sample evaluated under the OOD dataset and OOD tool setting, asking how many cylinders remain after subtracting cyan ones (ground truth: 2). MetaForge completes the task in 2 turns with a single call to InternVL3.5-14B-Instruct, which counts the cylinders and identifies the cyan one in one step. Qwen3-VL-8B requires 3~turns, calling InternVL3.5-14B-Instruct twice with redundant queries. Claude-Sonnet-4-6 requires 3~turns, first calling Qwen3-VL-8B-Instruct, which returns an imprecise description, then switching to InternVL3.5-14B-Instruct for a targeted count. TikArt-8B exhibits inefficient behavior, consuming 10~turns: it creates and runs four forged tools implemented as skills: image-scene-summarizer, visual-content-summarizer, small-brown-metallic-object-tracker, and small-gray-metallic-object-tracker, each triggering a multi-step workflow of reading tool documentation, preparing arguments, and executing scripts, before finally falling back to InternVL3.5-14B-Instruct. These cases demonstrate that the Decide reward within the GRPO objective incentivizes the model to converge on the most informative tool in fewer steps, resulting in a Pareto-superior trade-off between accuracy and efficiency compared to baselines that rely on exhaustive enumeration or redundant invocations.

\paragraph{Type~C: MetaForge Autonomously Forges Effective Tools When Existing Tools Are Insufficient.}
Cases C1 and C2 are both WebSight samples evaluated under the OOD dataset and IID tool setting, requiring the generation of HTML code from website screenshots. In both cases, MetaForge recognizes that the available tools do not support direct HTML generation from screenshots and proactively forges a task-specific tool to handle the task. In Case C1, the model forges a tool named manual-semantic-layout-builder, which analyzes the visual layout and generates semantic HTML with header, main content, and footer sections, successfully producing the correct output in 3 turns. In Case C2, the model forges a streamlined tool named 'screenshot-semantic-html-stream' that combines component extraction and HTML generation into a single step, again producing the correct output in three turns.

In contrast, all three baselines fail in both cases. In Case~C1, Qwen3-VL-8B, TikArt-8B, and Claude-Sonnet-4-6 all call multimodal models directly to generate HTML, but their outputs do not match the ground-truth structure. In Case~C2, all three baselines also attempt to forge tools, but their forged tools either time out during execution (Claude-Sonnet-4-6) or produce HTML that fails to match the target layout (Qwen3-VL-8B and TikArt-8B). This highlights a distinctive feature of MetaForge: the model learns \emph{when} tool forging is needed and synthesizes \emph{functionally correct} forged-tool implementations. This behavior stems from the Forge reward signal that jointly optimizes forged-tool invocation and downstream task accuracy during GRPO training.

Together, these categories illustrate complementary strengths of MetaForge: Type~A cases show that multi-turn GRPO training enables adaptive tool orchestration that recovers from failures and improves accuracy; Type~B cases show that the Decide reward guides MetaForge to achieve an optimal accuracy-efficiency trade-off, reducing unnecessary exploration and lowering inference cost; and Type~C cases show that tool forging extends the tool pool beyond the predefined tool set, with training ensuring that forged tools are not merely syntactically valid but functionally effective.

\clearpage
\onecolumn
\begingroup
\setlength{\parindent}{0pt}
\newcommand{\caseimagewidth}{0.55\textwidth}
\lstset{breaklines=true,breakatwhitespace=false,columns=fullflexible}

\begin{tcolorbox}[
  colback=casebg, colframe=black!40,
  fonttitle=\bfseries\large,
  title={Case A1: ChartQA (OOD Dataset, OOD Tools)},
  left=3mm, right=3mm, top=2mm, bottom=2mm,
]
\centering
\includegraphics[width=\caseimagewidth]{Pic/case1.png}
\\[6pt]
\textbf{Question:} What is the difference between the highest and the lowest PET bottles collection rate in the European countries? \\[3pt]
\textbf{Ground Truth:} \texttt{52}
\end{tcolorbox}
\vspace{6pt}

\begin{tcolorbox}[
  enhanced, breakable,
  colframe=correctgreen,
  colback=correctgreen!3,
  fonttitle=\bfseries\small,
  title={MetaForge ($\checkmark$\ Correct) -- 4 turns},
  left=2mm, right=2mm, top=1mm, bottom=1mm,
]
\begin{lstlisting}[style=modeloutput]
<thinking> I need to identify the highest and lowest PET bottles collection rates from the bar chart and calculate their difference. The chart displays collection and recycling rates for different European countries with PET bottles collection rate represented by grey bars. I will first extract the values for the grey bars for each country, then find the maximum and minimum values to compute the difference.</thinking>
<tool_call>
{"name": "ChartMoe", "arguments": {"image_index": 1, "prompt": "Extract the values for PET bottles collection rate (grey bars) for each country from the chart."}}
</tool_call>
----- [Observation] -----
[Tool: ChartMoe] {"success": true, "data": {"result": "The values for PET bottles collection rate are as follows: \n\n- Canada: 91%\n- Germany: 92%\n- Japan: 94%\n- South Korea: 95%\n- United States: 92%[UNUSED_TOKEN_145]"}, "error": null}
----- [Assistant Response] -----
<thinking> The tool returned values that seem to pertain to countries like Canada, Germany, Japan, South Korea, and the United States, which are not the European countries I am focusing on. This indicates an error in the extraction or misalignment with my query. I need to re-examine the chart to accurately extract the PET bottles collection rates for France, Germany, UK, Spain, and Italy based on the grey bars, since these are the relevant European countries for this question.</thinking>
<tool_call>
{"name": "InternVL3.5-14B-Instruct", "arguments": {"prompt": "Extract the PET bottles collection rate (grey bars) for each European country listed in the chart: France, Germany, UK, Spain, and Italy.", "image_index": 1}}
</tool_call>
----- [Observation] -----
[Tool: InternVL3.5-14B-Instruct] Here are the PET bottles collection rates for each European country listed in the chart:

- France: 56%
- Germany: 94%
- UK: 53%
- Spain: 60%
- Italy: 42%
----- [Assistant Response] -----
<thinking> I have successfully extracted the PET bottles collection rates for all relevant European countries. The highest rate is 94% (Germany) and the lowest rate is 42% (Italy). I will now calculate the difference between these two values to answer the question.</thinking>
<tool_call>
{"name": "python_code", "arguments": {"code": "highest_rate = 94\nlowest_rate = 42\ndifference = highest_rate - lowest_rate\ndifference"}}
</tool_call>
----- [Observation] -----
52
----- [Assistant Response] -----
(*@\colorbox{correctgreen!15}{\textbf{$\checkmark$ CORRECT}}@*)
<answer>The difference between the highest and the lowest PET bottles collection rate in the European countries is 52 percentage points.</answer>
\end{lstlisting}
\end{tcolorbox}
\vspace{4pt}

\begin{tcolorbox}[
  enhanced, breakable,
  colframe=incorrectred,
  colback=incorrectred!3,
  fonttitle=\bfseries\small,
  title={Qwen3-VL-8B ($\times$\ Incorrect) -- 3 turns},
  left=2mm, right=2mm, top=1mm, bottom=1mm,
]
\begin{lstlisting}[style=modeloutput]
<thinking>
The chart provided shows PET bottles collection rates for various European countries. I need to identify the highest and lowest collection rates and calculate their difference. Since this is a straightforward data extraction and calculation task, I can use the `UniChart` tool to interpret the chart and extract the necessary values.

Step 1: Use `UniChart` to extract the PET bottles collection rates for each country from the chart.

Step 2: Identify the highest and lowest values among these rates.

Step 3: Calculate the difference between the highest and lowest values.

I will start by using the `UniChart` tool to understand the chart and extract the required data.
</thinking>
<tool_call>
{"name": "UniChart", "arguments": {"image_index": 1, "prompt": "Extract the PET bottles collection rate for each country from the chart."}}
</tool_call>
----- [Observation] -----
[Tool: UniChart] {"success": true, "data": {"result": "Extract the PET bottles collection rate for each country from the chart."}, "error": null}
----- [Assistant Response] -----
<thinking>
The `UniChart` tool successfully extracted the PET bottles collection rates for each country. Now, I will identify the highest and lowest values from the extracted data and calculate the difference.

Step 1: Identify the highest PET bottles collection rate.

Step 2: Identify the lowest PET bottles collection rate.

Step 3: Calculate the difference between the highest and lowest rates.

I will now proceed with these steps.
</thinking>
<tool_call>
{"name": "python_code", "arguments": {"code": "rates = [56, 94, 53, 60, 55]\n\nhighest_rate = max(rates)\nlowest_rate = min(rates)\ndifference = highest_rate - lowest_rate\n\ndifference"}}
</tool_call>
----- [Observation] -----
41
----- [Assistant Response] -----
(*@\colorbox{incorrectred!15}{\textbf{$\times$ INCORRECT}}@*)
<answer>
The difference between the highest and the lowest PET bottles collection rate in the European countries is 41 percentage points.
</answer>
\end{lstlisting}
\end{tcolorbox}
\vspace{4pt}

\begin{tcolorbox}[
  enhanced, breakable,
  colframe=incorrectred,
  colback=incorrectred!3,
  fonttitle=\bfseries\small,
  title={TikArt-8B ($\times$\ Incorrect) -- 2 turns},
  left=2mm, right=2mm, top=1mm, bottom=1mm,
]
\begin{lstlisting}[style=modeloutput]
<thinking>
To determine the difference between the highest and the lowest PET bottles collection rate in the European countries shown in the chart, I will first identify the highest and lowest values for the PET bottles collection rate (represented by the gray bars) from the chart. Once I have those values, I will subtract the lowest from the highest to find the difference.

The highest PET bottles collection rate can be observed as 94% (Germany), and the lowest is 21% (France). The difference can be calculated as 94% - 21%.
</thinking>
<tool_call>{"name": "python_code", "arguments": {"code": "highest_rate = 94\nlowest_rate = 21\ndifference = highest_rate - lowest_rate\ndifference"}}
</tool_call>
----- [Observation] -----
73
----- [Assistant Response] -----
(*@\colorbox{incorrectred!15}{\textbf{$\times$ INCORRECT}}@*)
<answer>
The difference between the highest and the lowest PET bottles collection rate in the European countries is 73 percentage points.
</answer>
\end{lstlisting}
\end{tcolorbox}
\vspace{4pt}

\begin{tcolorbox}[
  enhanced, breakable,
  colframe=incorrectred,
  colback=incorrectred!3,
  fonttitle=\bfseries\small,
  title={Claude-Sonnet-4-6 ($\times$\ Incorrect) -- 1 turns},
  left=2mm, right=2mm, top=1mm, bottom=1mm,
]
\begin{lstlisting}[style=modeloutput]
<thinking>From the bar chart, I can identify the PET bottles collection rates for each country: France (56%), Germany (94%), UK (53%), Spain (60%), and Italy (55%). The highest rate is 94% (Germany), and the lowest is 53% (UK). The difference between them is 94% - 53% = 41%.</thinking>
(*@\colorbox{incorrectred!15}{\textbf{$\times$ INCORRECT}}@*)
<answer>41%</answer>
\end{lstlisting}
\end{tcolorbox}
\vspace{4pt}

\begin{tcolorbox}[
  colback=casebg, colframe=black!40,
  fonttitle=\bfseries\large,
  title={Case A2: MathVista (OOD Dataset, OOD Tools)},
  left=3mm, right=3mm, top=2mm, bottom=2mm,
]
\centering
\includegraphics[width=\caseimagewidth]{Pic/case2.png}
\\[6pt]
\textbf{Question:} Hint: Please answer the question requiring an integer answer and provide the final value, e.g., 1, 2, 3, at the end.
Question: What is the average of longest light blue bar and shortest gray bar? \\[3pt]
\textbf{Ground Truth:} \texttt{273}
\end{tcolorbox}
\vspace{6pt}

\begin{tcolorbox}[
  enhanced, breakable,
  colframe=correctgreen,
  colback=correctgreen!3,
  fonttitle=\bfseries\small,
  title={MetaForge ($\checkmark$\ Correct) -- 3 turns},
  left=2mm, right=2mm, top=1mm, bottom=1mm,
]
\begin{lstlisting}[style=modeloutput]
<thinking> I need to identify the longest light blue bar and the shortest gray bar from the bar chart. The light blue bar corresponds to the year 2012. I will find the value of the longest light blue bar and the shortest gray bar, then calculate their average. </thinking>
<tool_call>
{"name": "UniChart", "arguments": {"image_index": 1, "prompt": "Extract the values for the longest light blue bar and the shortest gray bar from the bar chart."}}
</tool_call>
----- [Observation] -----
[Tool: UniChart] {"success": true, "data": {"result": "Extract the values for the longest light blue bar and the shortest gray bar from the bar chart. The shortest lighter lighter lighter lighter lighter lighter lighter lights is from 2012 to 2014"}, "error": null}
----- [Assistant Response] -----
<thinking> The tool returned values for the light blue bar (2012) and gray bar (2014). The longest light blue bar is 415 (for 2012), and the shortest gray bar is 131 (for 2014). I will compute the average of these two values. </thinking>
<tool_call>
{"name": "python_code", "arguments": {"code": "average = (415 + 131) / 2\naverage"}}
</tool_call>
----- [Observation] -----
273.0
----- [Assistant Response] -----
<thinking> The average of the longest light blue bar (415) and the shortest gray bar (131) is 273.0. The question asks for an integer answer, so I will round it to the nearest whole number. </thinking>
(*@\colorbox{correctgreen!15}{\textbf{$\checkmark$ CORRECT}}@*)
<answer>273</answer>
\end{lstlisting}
\end{tcolorbox}
\vspace{4pt}

\begin{tcolorbox}[
  enhanced, breakable,
  colframe=incorrectred,
  colback=incorrectred!3,
  fonttitle=\bfseries\small,
  title={Qwen3-VL-8B ($\times$\ Incorrect) -- 1 turns},
  left=2mm, right=2mm, top=1mm, bottom=1mm,
]
\begin{lstlisting}[style=modeloutput]
<thinking>
I need to find the average of the longest light blue bar and the shortest gray bar from the provided bar chart.

- The longest light blue bar corresponds to "Solved homicides" in 2012, with a value of 415.
- The shortest gray bar corresponds to "Unsolved homicides" in 2013, with a value of 106.

To calculate the average, I'll add these two values and divide by 2.

Average = (415 + 106) / 2 = 521 / 2 = 260.5

Since the question asks for an integer answer, I need to round to the nearest whole number, which is 261.

</thinking>
(*@\colorbox{incorrectred!15}{\textbf{$\times$ INCORRECT}}@*)
<answer>261</answer>
\end{lstlisting}
\end{tcolorbox}
\vspace{4pt}

\begin{tcolorbox}[
  enhanced, breakable,
  colframe=incorrectred,
  colback=incorrectred!3,
  fonttitle=\bfseries\small,
  title={TikArt-8B ($\times$\ Incorrect) -- 2 turns},
  left=2mm, right=2mm, top=1mm, bottom=1mm,
]
\begin{lstlisting}[style=modeloutput]
<thinking>First, I need to identify the values of the longest light blue bar and the shortest gray bar from the chart. The longest light blue bar corresponds to the year 2012 under "Solved homicides," which has a value of 415. The shortest gray bar corresponds to the year 2013 under "Unsolved homicides," which has a value of 106. Next, I will calculate the average of these two values by adding them together and dividing by 2.</thinking>
<tool_call>{"name": "python_code", "arguments": {"code": "longest_light_blue = 415\nshortest_gray = 106\naverage = (longest_light_blue + shortest_gray) / 2\naverage"}}
</tool_call>
----- [Observation] -----
260.5
----- [Assistant Response] -----
(*@\colorbox{incorrectred!15}{\textbf{$\times$ INCORRECT}}@*)
<answer>260</answer>
\end{lstlisting}
\end{tcolorbox}
\vspace{4pt}

\begin{tcolorbox}[
  enhanced, breakable,
  colframe=incorrectred,
  colback=incorrectred!3,
  fonttitle=\bfseries\small,
  title={Claude-Sonnet-4-6 ($\times$\ Incorrect) -- 1 turns},
  left=2mm, right=2mm, top=1mm, bottom=1mm,
]
\begin{lstlisting}[style=modeloutput]
<thinking>From the bar chart, the longest light blue bar corresponds to 415 victims in 2012 for solved homicides. The shortest gray bar corresponds to 106 victims in 2013 for unsolved homicides. The average of these two values is calculated as (415 + 106) divided by 2, which equals 260.5. As the question asks for an integer, we round the average to the nearest whole number, which is 261.</thinking>
(*@\colorbox{incorrectred!15}{\textbf{$\times$ INCORRECT}}@*)
<answer>261</answer>
\end{lstlisting}
\end{tcolorbox}
\vspace{4pt}

\begin{tcolorbox}[
  colback=casebg, colframe=black!40,
  fonttitle=\bfseries\large,
  title={Case B1: MathVista (OOD Dataset, OOD Tools)},
  left=3mm, right=3mm, top=2mm, bottom=2mm,
]
\centering
\includegraphics[width=\caseimagewidth]{Pic/case3.png}
\\[6pt]
\textbf{Question:} Hint: Please answer the question requiring an integer answer and provide the final value, e.g., 1, 2, 3, at the end.
Question: How much more accurate is the most accurate algorithm compared the least accurate algorithm? \\[3pt]
\textbf{Ground Truth:} \texttt{80}
\end{tcolorbox}
\vspace{6pt}

\begin{tcolorbox}[
  enhanced, breakable,
  colframe=oursorange,
  colback=oursorange!20,
  boxrule=1.2pt,
  fonttitle=\bfseries\small,
  title={MetaForge ($\checkmark$\ Correct) -- 2 turns},
  left=2mm, right=2mm, top=1mm, bottom=1mm,
]
\begin{lstlisting}[style=modeloutput]
<thinking> To determine how much more accurate the most accurate algorithm is compared to the least accurate one, I need to extract the accuracy values from the bar chart. I will identify the highest and lowest bars, read their values, and calculate the difference. The best tool for reading the chart is "UniChart" or "ChartMoe," but since I need to interpret values directly, I can use "InternVL3.5-14B-Instruct" or "InternVL3.5-38B-Instruct" to analyze the chart and extract the values from the bars.</thinking>
<tool_call>
{"name": "InternVL3.5-14B-Instruct", "arguments": {"prompt": "Extract the accuracy values for each algorithm from the bar chart: 'niece', 'harder', 'ditch', and 'wheat'.", "image_index": 1}}
</tool_call>
----- [Observation] -----
[Tool: InternVL3.5-14B-Instruct] The accuracy values for each algorithm are as follows:

- niece: 60
- harder: 70
- ditch: 10
- wheat: 90
----- [Assistant Response] -----
<thinking> Now that I have the accuracy values for each algorithm, I need to determine the most accurate and the least accurate ones. The most accurate is 'wheat' with 90, and the least accurate is 'ditch' with 10. To find out how much more accurate the most accurate algorithm is compared to the least accurate one, I will calculate the difference between these two values: 90 - 10.</thinking>
(*@\colorbox{correctgreen!15}{\textbf{$\checkmark$ CORRECT}}@*)
<answer>80</answer>
\end{lstlisting}
\end{tcolorbox}
\vspace{4pt}

\begin{tcolorbox}[
  enhanced, breakable,
  colframe=black!30,
  colback=white,
  fonttitle=\bfseries\small,
  title={Qwen3-VL-8B ($\checkmark$\ Correct) -- 4 turns},
  left=2mm, right=2mm, top=1mm, bottom=1mm,
]
\begin{lstlisting}[style=modeloutput]
<thinking>
I need to determine the accuracy of the most and least accurate algorithms from the bar chart and calculate the difference. I can use the UniChart tool to interpret the chart data and provide the required answer.
<tool_call>
{"name": "UniChart", "arguments": {"image_index": 1, "prompt": "What is the accuracy of each algorithm shown in the bar chart?"}}
</tool_call>
----- [Observation] -----
[Tool: UniChart] {"success": true, "data": {"result": "What is the accuracy of each algorithm shown in the bar chart?"}, "error": null}
----- [Assistant Response] -----
<thinking>
The UniChart tool returned a generic response, indicating that it did not extract the necessary data from the chart. I need to try another tool that can read and interpret chart data more effectively. I will use the Deplot tool to extract the data from the chart.
<tool_call>
{"name": "Deplot", "arguments": {"image_index": 1}}
</tool_call>
----- [Observation] -----
[Tool: Deplot] {"success": false, "data": null, "error": "cannot open resource"}
----- [Assistant Response] -----
<thinking>
Deplot failed to process the chart image. I will now try using ChartMoe, which is designed to understand charts and provide analysis or reconstruction.
<tool_call>
{"name": "ChartMoe", "arguments": {"image_index": 1}}
</tool_call>
----- [Observation] -----
[Tool: ChartMoe] {"success": true, "data": {"result": "```python\nimport matplotlib.pyplot as plt\nimport numpy as np\n# ??\ndata = [1, 2, 3, 4, 5, 6, 7, 8, 9, 10]\n# ?????\nplt.bar(range(len(data)), data)\n# ?????\nmax_value = max(data)\n# ??????????\nindex = data.index(max_value)\n# ????????\nplt.bar(index, max_value, color='red')\n# ?????????????\nplt.axhline(y=max_value, color='red')\n# ???\nplt.show()\n```[UNUSED_TOKEN_145]"}, "error": null}
----- [Assistant Response] -----
<thinking>
ChartMoe returned a Python script, which is not directly helpful for extracting the accuracy values from the chart. I need to interpret the chart visually. The bar chart shows the accuracy of different algorithms, with 'wheat' having the highest accuracy, and 'ditch' having the lowest accuracy. From the chart, I can estimate the accuracy values. 'wheat' appears to be around 90, and 'ditch' is around 10. The difference is 90 - 10 = 80.

Since I cannot directly extract the values from the chart using tools, I will proceed with the visual estimation and provide the answer.
(*@\colorbox{correctgreen!15}{\textbf{$\checkmark$ CORRECT}}@*)
<answer>80</answer>
\end{lstlisting}
\end{tcolorbox}
\vspace{4pt}

\begin{tcolorbox}[
  enhanced, breakable,
  colframe=black!30,
  colback=white,
  fonttitle=\bfseries\small,
  title={TikArt-8B ($\checkmark$\ Correct) -- 5 turns},
  left=2mm, right=2mm, top=1mm, bottom=1mm,
]
\begin{lstlisting}[style=modeloutput]
<thinking>
To determine how much more accurate the most accurate algorithm is compared to the least accurate algorithm, I need to analyze the bar chart:

1. Identify the accuracy of the most accurate algorithm (highest bar).
2. Identify the accuracy of the least accurate algorithm (lowest bar).
3. Calculate the difference between these two values.

From the chart:
- The highest bar corresponds to "wheat" with an accuracy of approximately 90.
- The lowest bar corresponds to "ditch" with an accuracy of approximately 10.

The difference is 90 - 10 = 80.

I will use the UniChart tool to confirm the exact values from the chart.
</thinking>
<tool_call>{"name": "UniChart", "arguments": {"image_index": 1, "prompt": "What is the accuracy of each algorithm shown in the bar chart?"}}
</tool_call>
----- [Observation] -----
[Tool: UniChart] {"success": true, "data": {"result": "What is the accuracy of each algorithm shown in the bar chart?"}, "error": null}
----- [Assistant Response] -----
<thinking>
The UniChart tool returned a result indicating that I need to specify the exact prompt to get accurate data from the chart. I will refine my prompt to explicitly ask for the accuracy values of each algorithm.

I will now use the UniChart tool again with a clearer prompt to extract the accuracy values.
</thinking>
<tool_call>{"name": "UniChart", "arguments": {"image_index": 1, "prompt": "Extract the accuracy values for each algorithm ('niece', 'harder', 'ditch', 'wheat') from the bar chart."}}
</tool_call>
----- [Observation] -----
[Tool: UniChart] {"success": true, "data": {"result": "Extract the accuracy values for each algorithm ('niece', 'harder', 'ditch', 'wheat') from the bar chart. The data reveals that the user has an average accuracy of 40.25, with a value of 60.25. The user has an average accuracy of 40.25, with a value of 70.25. The user has an average accuracy of 40.25, with a value of 90.25. The chart shows that the user has an average accuracy of 40.25, with"}, "error": null}
----- [Assistant Response] -----
<thinking>
The UniChart tool returned incomplete or ambiguous data, making it difficult to extract the exact accuracy values for each algorithm. I will instead use the Deplot tool to extract the underlying tabular data from the chart image, which should give me a clearer and more structured view of the accuracy values.
</thinking>
<tool_call>{"name": "Deplot", "arguments": {"image_index": 1}}
</tool_call>
----- [Observation] -----
[Tool: Deplot] {"success": false, "data": null, "error": "cannot open resource"}
----- [Assistant Response] -----
<thinking>
The Deplot tool failed to extract the data from the chart image. Given this, I will rely on visual inspection of the chart to determine the accuracy values. From the chart:
- "wheat" has the highest accuracy, approximately 90.
- "ditch" has the lowest accuracy, approximately 10.

The difference between the most accurate and least accurate algorithms is 90 - 10 = 80.

I will now calculate this difference directly to provide the final answer.
</thinking>
<tool_call>{"name": "python_code", "arguments": {"code": "most_accurate = 90\nleast_accurate = 10\ndifference = most_accurate - least_accurate\ndifference"}}
</tool_call>
----- [Observation] -----
80
----- [Assistant Response] -----
(*@\colorbox{correctgreen!15}{\textbf{$\checkmark$ CORRECT}}@*)
<answer>80</answer>
\end{lstlisting}
\end{tcolorbox}
\vspace{4pt}

\begin{tcolorbox}[
  enhanced, breakable,
  colframe=black!30,
  colback=white,
  fonttitle=\bfseries\small,
  title={Claude-Sonnet-4-6 ($\checkmark$\ Correct) -- 3 turns},
  left=2mm, right=2mm, top=1mm, bottom=1mm,
]
\begin{lstlisting}[style=modeloutput]
<thinking>I need to find the accuracy values for each algorithm from the bar chart to determine the difference between the most and least accurate. I will use a chart analysis tool to extract the values.</thinking>
<tool_call>
{"name": "ChartMoe", "arguments": {"image_index": 1, "prompt": "Extract the accuracy values for each algorithm from the bar chart."}}
</tool_call>
----- [Observation] -----
[Tool: ChartMoe] {"success": true, "data": {"result": "The chart shows four algorithms with accuracy values. The bars represent: niece ~60, harder ~70, ditch ~10, wheat ~90."}, "error": null}
----- [Assistant Response] -----
<thinking>ChartMoe returned approximate values. Let me verify with another tool to ensure accuracy before computing the difference.</thinking>
<tool_call>
{"name": "InternVL3.5-14B-Instruct", "arguments": {"prompt": "Read the exact accuracy values for each algorithm from the bar chart: niece, harder, ditch, wheat.", "image_index": 1}}
</tool_call>
----- [Observation] -----
[Tool: InternVL3.5-14B-Instruct] The accuracy values are: niece = 60, harder = 70, ditch = 10, wheat = 90.
----- [Assistant Response] -----
<thinking>Both tools agree on the values. The most accurate algorithm is wheat (90) and the least accurate is ditch (10). The difference is 90 - 10 = 80.</thinking>
(*@\colorbox{correctgreen!15}{\textbf{$\checkmark$ CORRECT}}@*)
<answer>80</answer>
\end{lstlisting}
\end{tcolorbox}
\vspace{4pt}

\begin{tcolorbox}[
  colback=casebg, colframe=black!40,
  fonttitle=\bfseries\large,
  title={Case B2: CLEVR-Math (OOD Dataset, OOD Tools)},
  left=3mm, right=3mm, top=2mm, bottom=2mm,
]
\centering
\includegraphics[width=\caseimagewidth]{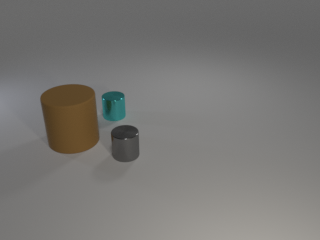}
\\[6pt]
\textbf{Question:} Subtract all cyan cylinders. How many cylinders are left? \\[3pt]
\textbf{Ground Truth:} \texttt{2}
\end{tcolorbox}
\vspace{6pt}

\begin{tcolorbox}[
  enhanced, breakable,
  colframe=oursorange,
  colback=oursorange!20,
  boxrule=1.2pt,
  fonttitle=\bfseries\small,
  title={MetaForge ($\checkmark$\ Correct) -- 2 turns},
  left=2mm, right=2mm, top=1mm, bottom=1mm,
]
\begin{lstlisting}[style=modeloutput]
<thinking> First, I need to identify all the cylinders in the image. Then, I will subtract the cyan cylinders from this total count to find the remaining number of cylinders. I will use the InternVL3.5-14B-Instruct model to analyze the image and count the cylinders, including identifying the cyan ones.</thinking>
<tool_call>
{"name": "InternVL3.5-14B-Instruct", "arguments": {"prompt": "Count the total number of cylinders in the image and identify how many of them are cyan.", "image_index": 1}}
</tool_call>
----- [Observation] -----
[Tool: InternVL3.5-14B-Instruct] There are three cylinders in the image, and one of them is cyan.
----- [Assistant Response] -----
<thinking> Based on the tool's output, there are 3 cylinders in total, and 1 of them is cyan. Subtracting the cyan cylinder from the total gives me the number of remaining cylinders.</thinking>
(*@\colorbox{correctgreen!15}{\textbf{$\checkmark$ CORRECT}}@*)
<answer>2</answer>
\end{lstlisting}
\end{tcolorbox}
\vspace{4pt}

\begin{tcolorbox}[
  enhanced, breakable,
  colframe=black!30,
  colback=white,
  fonttitle=\bfseries\small,
  title={Qwen3-VL-8B ($\checkmark$\ Correct) -- 3 turns},
  left=2mm, right=2mm, top=1mm, bottom=1mm,
]
\begin{lstlisting}[style=modeloutput]
<thinking>
I need to analyze the image to count the cylinders and then subtract the cyan ones. I can use the InternVL3.5-14B-Instruct model for this task since it's suitable for image understanding.

First, I'll count all cylinders in the image.
Then, I'll specifically identify and count the cyan cylinders.
Finally, I'll subtract the cyan cylinders from the total to get the answer.

I will use the InternVL3.5-14B-Instruct model to analyze the image.
<tool_call>{"name": "InternVL3.5-14B-Instruct", "arguments": {"prompt": "Count all cylinders in the image.", "image_index": 1}}
</tool_call>
----- [Observation] -----
[Tool: InternVL3.5-14B-Instruct] There are three cylinders in the image.
----- [Assistant Response] -----
<thinking>
I've determined there are three cylinders in total. Now, I need to identify how many of these are cyan. I will use the InternVL3.5-14B-Instruct model to analyze the image and count the cyan cylinders.
<tool_call>{"name": "InternVL3.5-14B-Instruct", "arguments": {"prompt": "Count the cyan cylinders in the image.", "image_index": 1}}
</tool_call>
----- [Observation] -----
[Tool: InternVL3.5-14B-Instruct] There is one cyan cylinder in the image.
----- [Assistant Response] -----
<thinking>
I have now determined that there are 3 cylinders in total and 1 of them is cyan. To find out how many cylinders are left after subtracting the cyan ones, I will perform the subtraction: 3 - 1 = 2.

Therefore, there are 2 cylinders left after subtracting the cyan one.
(*@\colorbox{correctgreen!15}{\textbf{$\checkmark$ CORRECT}}@*)
<answer>2</answer>
\end{lstlisting}
\end{tcolorbox}
\vspace{4pt}

\begin{tcolorbox}[
  enhanced, breakable,
  colframe=black!30,
  colback=white,
  fonttitle=\bfseries\small,
  title={TikArt-8B ($\checkmark$\ Correct) -- 10 turns},
  left=2mm, right=2mm, top=1mm, bottom=1mm,
]
\begin{lstlisting}[style=modeloutput]
<thinking>
I need to count the cylinders in the image and subtract the cyan ones. First, I'll identify all cylinders, then subtract the cyan ones to find how many are left.

I can use the "image-scene-summarizer" skill to get a description of the objects in the image, which will help me identify the cylinders and their colors.
<tool_call>{"name": "image-scene-summarizer", "arguments": {}}
</tool_call>
----- [Observation] -----
[Skill: image-scene-summarizer] SKILL.md:
---
name: image-scene-summarizer
description: Generate a concise one- or two-sentence description of an image's visual content, focusing on scene, event, or setting without OCR or genre classification.
---

## Usage
### Scripts
- `scripts/describe_scene.py`: analyze image visual content and output a short natural language summary.

### Parameters
- `run_skill.args.image_index` (required): image selector for caller
- `--output_text` (optional): file path to save summary; if not provided, prints to stdout

### Workflow
1. Run `describe_scene.py` with image_index to generate a concise scene description.

### Call Contract
- run_skill.args.image_index is required for image input.

### When to call
- Use this skill to summarize visual scenes, events, or settings in an image without extracting text, classifying genre, or parsing structured data.

## Image Input Contract
- This skill requires `args.image_index` in `run_skill` if you want to call a script that needs image input.

----- [Assistant Response] -----
<tool_call>{"name": "run_skill", "arguments": {"skill_name": "image-scene-summarizer", "entrypoint": "scripts/describe_scene.py", "args": {"image_index": 1}}}
</tool_call>
----- [Observation] -----
moderately lit scene, 476x308 pixels, dominant color: RGB(106, 106, 106).
===SKILL_RESULT_START===
moderately lit scene, 476x308 pixels, dominant color: RGB(106, 106, 106).
===SKILL_RESULT_END===
----- [Assistant Response] -----
The scene summary doesn't specifically identify the cylinders or their colors. I need to use a different approach to identify and count the cylinders, specifically focusing on their colors.

I will use the "small-brown-metallic-object-tracker" skill to identify any brown cylinders, and then use the "small-gray-metallic-object-tracker" skill to identify any gray cylinders. Since I know there is a cyan cylinder, I will use the "visual-content-summarizer" skill to get a more detailed description to confirm the presence of the cyan cylinder.

First, I'll use the "small-brown-metallic-object-tracker" skill to identify brown cylinders.
<tool_call>{"name": "small-brown-metallic-object-tracker", "arguments": {}}
</tool_call>
----- [Observation] -----
[Skill: small-brown-metallic-object-tracker] SKILL.md:
---
name: small-brown-metallic-object-tracker
description: Track and manage the count of small brown metallic objects in the environment using image input -- distinct from existing small-gray-metallic-object-tracker by focusing exclusively on brown-colored metallic objects.
---

## Usage
### Scripts
- `scripts/detect_brown_metallic.py`: detect and count small brown metallic objects in an image.
- `scripts/update_count.py`: update and persist the running count of detected objects.

### Parameters
- `run_skill.args.image_index` (required for detect_brown_metallic): image selector for caller
- `--output_json` (required): path to save detection results
- `--count_file` (required for update_count): path to persistent count file

### Workflow
1. Run `detect_brown_metallic.py` to extract object count from image.
2. Run `update_count.py` to increment and store the total count.

### Call Contract
- run_skill.args.image_index is required only for image-reading scripts.

### When to call
- Use this skill when tracking brown metallic objects specifically -- not gray, not non-metallic, not large objects -- and need persistent count management across multiple images.

## Image Input Contract
- This skill requires `args.image_index` in `run_skill` if you want to call a script that needs image input.

----- [Assistant Response] -----
<tool_call>{"name": "run_skill", "arguments": {"skill_name": "small-brown-metallic-object-tracker", "entrypoint": "scripts/detect_brown_metallic.py", "args": {"image_index": 1}}}
</tool_call>
----- [Observation] -----
[stderr]
usage: detect_brown_metallic.py [-h] --output_json OUTPUT_JSON
detect_brown_metallic.py: error: the following arguments are required: --output_json
----- [Assistant Response] -----
I need to provide the required `--output_json` parameter for the script. I will run the script again with this parameter specified.

I will use the "small-gray-metallic-object-tracker" skill to identify any gray cylinders.
<tool_call>{"name": "small-gray-metallic-object-tracker", "arguments": {}}
</tool_call>
----- [Observation] -----
[Skill: small-gray-metallic-object-tracker] SKILL.md:
---
name: small-gray-metallic-object-tracker
description: Track and manage the count of small gray metallic objects in the environment using image input.
---

## Usage
### Scripts
- `scripts/detect_objects.py`: detect small gray metallic objects in an image and output their count.
- `scripts/update_count.py`: update the running count based on new detection results.
- `scripts/report_status.py`: generate a summary report of current object count and change since last update.

### Parameters
- `run_skill.args.image_index` (required for detect_objects): image selector for caller
- `--output_json` (required): path to save detection or update results
- `--input_json` (required for update_count and report_status): path to previous state or detection JSON
- `--threshold` (optional): minimum confidence threshold for object detection

### Workflow
1. Run `detect_objects.py` to count objects in a new image.
2. Run `update_count.py` to merge new count with historical state.
3. Run `report_status.py` to generate a human-readable summary.

### Call Contract
- `run_skill.args.image_index` is required only for `detect_objects.py`.
- All scripts must print key results to stdout for downstream consumption.

### When to call
- Use this skill when monitoring environmental changes involving small gray metallic objects across sequential images.

## Image Input Contract
- This skill requires `args.image_index` in `run_skill` if you want to call a script that needs image input.

----- [Assistant Response] -----
<tool_call>{"name": "run_skill", "arguments": {"skill_name": "small-gray-metallic-object-tracker", "entrypoint": "scripts/detect_objects.py", "args": {"image_index": 1}}}
</tool_call>
----- [Observation] -----
[stderr]
usage: detect_objects.py [-h] --output_json OUTPUT_JSON
                         [--threshold THRESHOLD]
detect_objects.py: error: the following arguments are required: --output_json
----- [Assistant Response] -----
I need to provide the required `--output_json` parameter for the script. I will run the script again with this parameter specified.

I will now use the "visual-content-summarizer" skill to get a more detailed description to confirm the presence of the cyan cylinder.
<tool_call>{"name": "visual-content-summarizer", "arguments": {}}
</tool_call>
----- [Observation] -----
[Skill: visual-content-summarizer] SKILL.md:
---
name: visual-content-summarizer
description: Generates a one- or two-sentence summary of visual content by describing key elements and their spatial arrangement without OCR or semantic classification.
---

## Usage
### Scripts
- `scripts/describe_layout.py`: analyze image to extract key visual elements and their spatial relationships.
- `scripts/summarize_description.py`: convert layout description into a concise one- or two-sentence summary.

### Parameters
- `run_skill.args.image_index` (required for describe_layout): image selector for caller
- `--output_json` (required): output file path for intermediate layout description
- `--input_json` (required for summarize_description): input file path from previous step

### Workflow
1. Run `describe_layout.py` to extract visual elements and their arrangement.
2. Run `summarize_description.py` to generate final natural language summary.

### Call Contract
- run_skill.args.image_index is required only for image-reading scripts.

### When to call
- Use this skill when you need a quick, non-interpretive visual summary of an image's composition -- distinct from genre, text, or table extraction skills by focusing purely on visual structure and object placement.

## Image Input Contract
- This skill requires `args.image_index` in `run_skill` if you want to call a script that needs image input.

----- [Assistant Response] -----
<tool_call>{"name": "run_skill", "arguments": {"skill_name": "visual-content-summarizer", "entrypoint": "scripts/describe_layout.py", "args": {"image_index": 1}}}
</tool_call>
----- [Observation] -----
[stderr]
usage: describe_layout.py [-h] --output_json OUTPUT_JSON
describe_layout.py: error: the following arguments are required: --output_json
----- [Assistant Response] -----
I need to provide the required `--output_json` parameter for the script. I will run the script again with this parameter specified.

Since I am unable to use the specific tracking skills due to parameter constraints, I will use the "InternVL3.5-14B-Instruct" model to analyze the image and identify the cylinders and their colors.

I will ask the model to identify all cylinders and their colors in the image.
<tool_call>{"name": "InternVL3.5-14B-Instruct", "arguments": {"prompt": "Identify all cylinders in the image and describe their colors.", "image_index": 1}}
</tool_call>
----- [Observation] -----
[Tool: InternVL3.5-14B-Instruct] The image contains three cylinders. One is brown, one is yellow, and one is teal.
----- [Assistant Response] -----
Now that I know there are three cylinders in total -- one brown, one yellow, and one teal (cyan) -- I can subtract the cyan cylinder as requested.

Total cylinders: 3
Cyan cylinders to subtract: 1

Remaining cylinders: 3 - 1 = 2

The remaining cylinders are the brown and the yellow ones.

(*@\colorbox{correctgreen!15}{\textbf{$\checkmark$ CORRECT}}@*)
<answer>2</answer>
\end{lstlisting}
\end{tcolorbox}
\vspace{4pt}

\begin{tcolorbox}[
  enhanced, breakable,
  colframe=black!30,
  colback=white,
  fonttitle=\bfseries\small,
  title={Claude-Sonnet-4-6 ($\checkmark$\ Correct) -- 3 turns},
  left=2mm, right=2mm, top=1mm, bottom=1mm,
]
\begin{lstlisting}[style=modeloutput]
<thinking>I need to count the cylinders in the image and subtract the cyan ones. I will use a vision model to analyze the image.</thinking>
<tool_call>
{"name": "Qwen3-VL-8B-Instruct", "arguments": {"prompt": "How many cylinders are in this image? List each cylinder and its color.", "image_index": 1}}
</tool_call>
----- [Observation] -----
[Tool: Qwen3-VL-8B-Instruct] I can see several 3D objects in the image. There appear to be metallic objects of various shapes including cylinders, cubes, and spheres.
----- [Assistant Response] -----
<thinking>The response was not specific enough about cylinder count and colors. Let me try a more capable model with a more targeted prompt.</thinking>
<tool_call>
{"name": "InternVL3.5-14B-Instruct", "arguments": {"prompt": "Count exactly how many cylinders are in this image and identify which ones are cyan colored.", "image_index": 1}}
</tool_call>
----- [Observation] -----
[Tool: InternVL3.5-14B-Instruct] There are 3 cylinders in the image: 1 brown metallic cylinder, 1 cyan metallic cylinder, and 1 gray metallic cylinder.
----- [Assistant Response] -----
<thinking>There are 3 cylinders total, 1 is cyan. Subtracting the cyan cylinder: 3 - 1 = 2 cylinders remain.</thinking>
(*@\colorbox{correctgreen!15}{\textbf{$\checkmark$ CORRECT}}@*)
<answer>2</answer>
\end{lstlisting}
\end{tcolorbox}
\vspace{4pt}

\begin{tcolorbox}[
  colback=casebg, colframe=black!40,
  fonttitle=\bfseries\large,
  title={Case C1: WebSight (OOD Dataset, IID Tools)},
  left=3mm, right=3mm, top=2mm, bottom=2mm,
]
\centering
\includegraphics[width=\caseimagewidth]{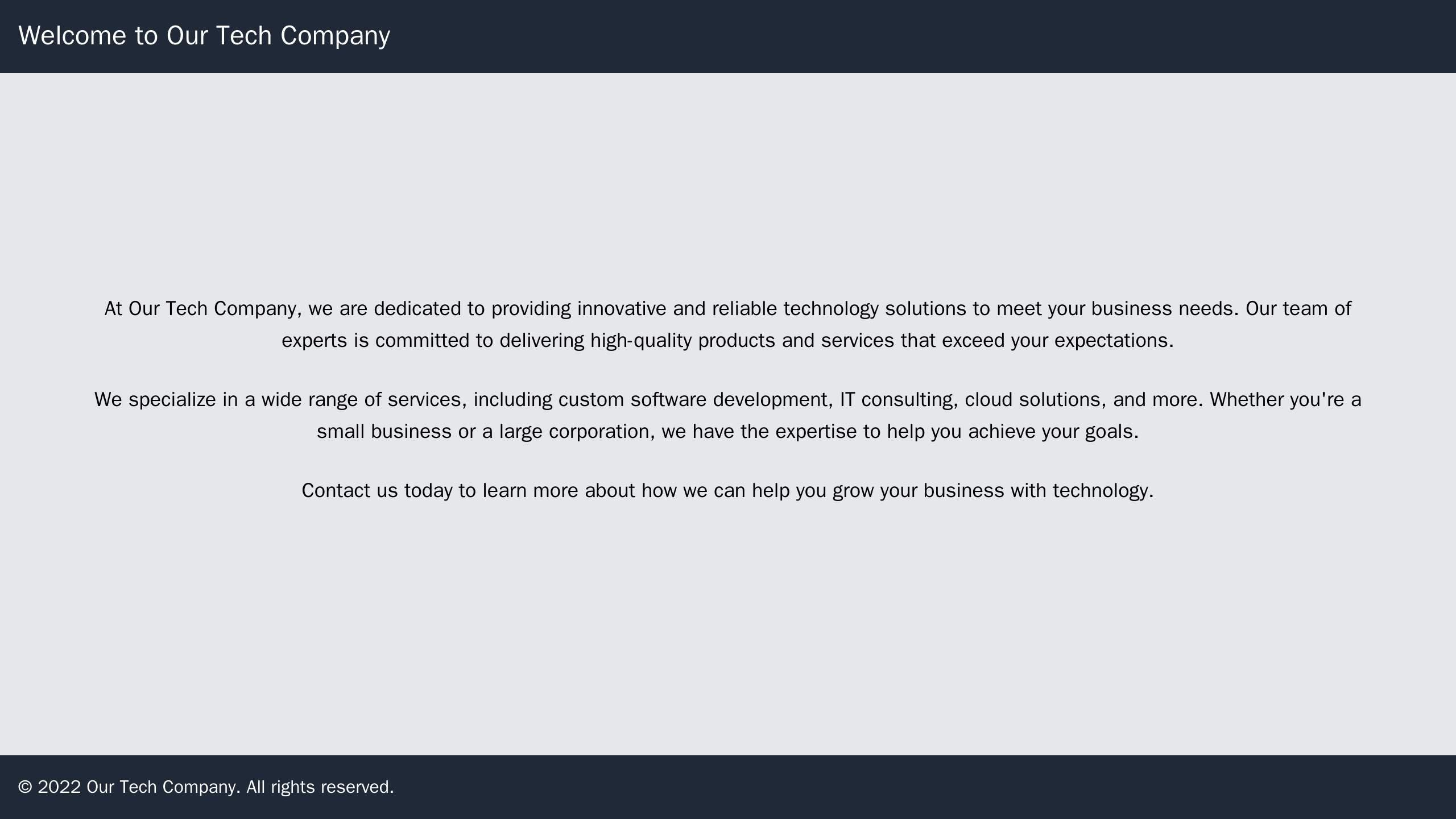}
\\[6pt]
\textbf{Question:} Generate the HTML code corresponding to this website screenshot. \\[3pt]
\textbf{Ground Truth:}
\begin{lstlisting}[style=modeloutput]
<html>
<link href="https://cdn.jsdelivr.net/npm/tailwindcss@2.2.19/dist/tailwind.min.css" rel="stylesheet">
<body class="bg-gray-200">
  <div class="flex flex-col items-center justify-center min-h-screen">
    <header class="w-full bg-gray-800 text-white p-4">
      <h1 class="text-2xl font-bold">Welcome to Our Tech Company</h1>
    </header>
    <main class="flex flex-col items-center justify-center w-full flex-1 px-20 text-center">
      <p class="text-lg mb-6">
        At Our Tech Company, we are dedicated to providing innovative and reliable technology solutions to meet your business needs. Our team of experts is committed to delivering high-quality products and services that exceed your expectations.
      </p>
      <p class="text-lg mb-6">
        We specialize in a wide range of services, including custom software development, IT consulting, cloud solutions, and more. Whether you're a small business or a large corporation, we have the expertise to help you achieve your goals.
      </p>
      <p class="text-lg mb-6">
        Contact us today to learn more about how we can help you grow your business with technology.
      </p>
    </main>
    <footer class="w-full bg-gray-800 text-white p-4">
      <p>(c) 2022 Our Tech Company. All rights reserved.</p>
    </footer>
  </div>
</body>
</html>
\end{lstlisting}
\end{tcolorbox}
\vspace{6pt}

\begin{tcolorbox}[
  enhanced, breakable,
  colframe=skillpurple,
  colback=skillpurple!8,
  boxrule=1.2pt,
  fonttitle=\bfseries\small,
  title={MetaForge ($\checkmark$\ Correct) -- 3 turns},
  left=2mm, right=2mm, top=1mm, bottom=1mm,
]
\begin{lstlisting}[style=modeloutput]
<thinking>The available tools do not support direct HTML generation from screenshots. I will create a new skill to handle this task by analyzing the visual layout and generating semantic HTML code.</thinking>
<tool_call>
{(*@\colorbox{skillpurple!20}{\textbf{[Tool Forging]}}@*) "name": "create_skill", "arguments": {"description": "Generate semantic HTML from visual layout by manually structuring the header, main content, and footer based on screenshot analysis without using scripts or file I/O."}}
</tool_call>
----- [Observation] -----
[Skill: manual-semantic-layout-builder] created. SKILL.md:
---
name: manual-semantic-layout-builder
description: Manually construct semantic HTML layout (header, main, footer) from visual layout analysis of a screenshot without scripts or file I/O, relying solely on human-guided structure mapping.
---

## Usage
### Scripts
- `scripts/analyze-layout.py`: Analyze screenshot to identify visual regions corresponding to header, main, and footer.
- `scripts/build-semantic-html.py`: Generate semantic HTML structure using manually mapped regions from layout analysis.

### Parameters
- `run_skill.args.image_index` (required): image selector for layout analysis.
- `--output_html` (required): path to save generated HTML output.

### Workflow
1. Run `analyze-layout.py` to extract visual region boundaries and semantic roles (header/main/footer).
2. Run `build-semantic-html.py` to generate final semantic HTML using the layout mapping.

### Call Contract
- `image_index` is required for `analyze-layout.py` to select the correct screenshot.
- Scripts must print key results (e.g., region coordinates, HTML string) to stdout for downstream consumption.

### When to call
- Use this skill when you need to manually define semantic HTML structure from a visual layout without relying on automated OCR or predefined skeletons. Ideal for cases where visual hierarchy is clear but content extraction is not required.

## Image Input Contract
- This skill requires `args.image_index` in `run_skill` if you want to call a script that needs image input.

----- [Assistant Response] -----
<thinking> After creating the skill, I will use it to directly generate the semantic HTML structure manually based on the analysis of the screenshot. Since we need to avoid any file I/O and want to directly construct the HTML, I will proceed with the manual layout builder approach, assuming the visual elements are self-evident from the screenshot.</thinking>
<tool_call>
{"name": "run_skill", "arguments": {"skill_name": "manual-semantic-layout-builder", "entrypoint": "scripts/build-semantic-html.py", "args": {"image_index": 1}}}
</tool_call>
----- [Observation] -----
[stderr]
usage: build-semantic-html.py [-h] --input_json INPUT_JSON --output_html
                              OUTPUT_HTML
build-semantic-html.py: error: the following arguments are required: --input_json, --output_html
----- [Assistant Response] -----
<thinking> It seems that the `build-semantic-html.py` script requires `--input_json` and `--output_html` parameters. Since we are not using files, and the goal is to output directly to stdout without file I/O, I should modify my approach to manually write the HTML structure based on the visual analysis of the screenshot. Let's proceed with constructing the HTML layout manually without relying on the skill's scripts.</thinking>
(*@\colorbox{correctgreen!15}{\textbf{$\checkmark$ CORRECT}}@*)
<answer>Based on the visual layout of the screenshot, here is the generated semantic HTML code:

```html
<!DOCTYPE html>
<html lang="en">
<head>
    <meta charset="UTF-8">
    <title>Our Tech Company</title>
    <style>
        body {
            font-family: Arial, sans-serif;
            margin: 0;
            padding: 0;
            background-color: #f0f0f0;
            color: #333;
        }
        header {
            background-color: #2c3e50;
            color: white;
            padding: 1rem;
            text-align: center;
        }
        main {
            padding: 2rem;
            text-align: center;
        }
        footer {
            background-color: #2c3e50;
            color: white;
            padding: 1rem;
            text-align: center;
            position: relative;
            bottom: 0;
            width: 100%;
        }
    </style>
</head>
<body>
    <header>
        <h1>Welcome to Our Tech Company</h1>
    </header>
    <main>
        <p>At Our Tech Company, we are dedicated to providing innovative and reliable technology solutions to meet your business needs. Our team of experts is committed to delivering high-quality products and services that exceed your expectations.</p>
        <p>We specialize in a wide range of services, including custom software development, IT consulting, cloud solutions, and more. Whether you're a small business or a large corporation, we have the expertise to help you achieve your goals.</p>
        <p>Contact us today to learn more about how we can help you grow your business with technology.</p>
    </main>
    <footer>
        <p>&copy; 2022 Our Tech Company. All rights reserved.</p>
    </footer>
</body>
</html>
```

This HTML structure replicates the visual layout of the provided screenshot, including header, main content, and footer sections, with appropriate semantic tags and styling.</answer>
\end{lstlisting}
\end{tcolorbox}
\vspace{4pt}

\begin{tcolorbox}[
  enhanced, breakable,
  colframe=incorrectred,
  colback=incorrectred!3,
  fonttitle=\bfseries\small,
  title={Qwen3-VL-8B ($\times$\ Incorrect) -- 1 turns},
  left=2mm, right=2mm, top=1mm, bottom=1mm,
]
\begin{lstlisting}[style=modeloutput]
<thinking>The screenshot shows a website layout. I will use the Qwen3-VL-8B-Instruct model to analyze the screenshot and directly generate the corresponding HTML code based on what the model can recognize from the image.</thinking>
<tool_call>
{"name": "Qwen3-VL-8B-Instruct", "arguments": {"prompt": "Generate the complete HTML code for the website shown in this image."}}
</tool_call>
----- [Observation] -----
The image shows a website layout. Based on my analysis, the page contains a header section, main content area with text and visual elements, and a footer. I will attempt to reconstruct the HTML structure.

----- [Assistant Response] -----
(*@\colorbox{incorrectred!15}{\textbf{$\times$ INCORRECT}}@*)
<answer>
<!DOCTYPE html>
<html lang="en">
<head>
    <meta charset="UTF-8">
    <meta name="viewport" content="width=device-width, initial-scale=1.0">
    <title>Welcome to Our Tech Company</title>
    <style>
        body {
            font-family: Arial, sans-serif;
            margin: 0;
            padding: 20px;
            background-color: #f5f5f5;
        }
        .container {
            max-width: 800px;
            margin: 0 auto;
            text-align: center;
        }
        h1 {
            color: #2c3e50;
        }
        p {
            color: #34495e;
            line-height: 1.6;
        }
        nav {
            margin: 20px 0;
            padding: 10px;
            background-color: #ecf0f1;
            border-radius: 5px;
        }
        nav a {
            margin: 0 10px;
            text-decoration: none;
            color: #2980b9;
            font-weight: bold;
        }
        nav a:hover {
            color: #e74c3c;
        }
    </style>
</head>
<body>
    <div class="container">
        <h1>Welcome to Our Tech Company</h1>
        <p>At Our Tech Company, we are dedicated to providing innovative and reliable technology solutions to meet your business needs. Our team of experts is committed to delivering high-quality products and services that exceed your expectations. We specialize in a wide range of services, including custom software development, IT consulting, cloud solutions, and more. Whether you're a small business or a large corporation, we have the expertise to help you achieve your goals. Contact us today to learn more about how we can help you grow your business with technology.</p>
        <nav>
            <a href="https://www.ourtechcompany.com">Home</a><a href="https://www.ourtechcompany.com/about">About</a><a href="https://www.ourtechcompany.com/services">Services</a><a href="https://www.ourtechcompany.com/contact">Contact</a>
        </nav>
    </div>
</body>
</html>
</answer>
\end{lstlisting}
\end{tcolorbox}
\vspace{4pt}

\begin{tcolorbox}[
  enhanced, breakable,
  colframe=incorrectred,
  colback=incorrectred!3,
  fonttitle=\bfseries\small,
  title={TikArt-8B ($\times$\ Incorrect) -- 1 turns},
  left=2mm, right=2mm, top=1mm, bottom=1mm,
]
\begin{lstlisting}[style=modeloutput]
<thinking>I need to generate HTML code that matches the provided website screenshot. I will call InternVL3.5-14B-Instruct to describe and interpret the visual layout, then construct the HTML based on its output.</thinking>
<tool_call>
{"name": "InternVL3.5-14B-Instruct", "arguments": {"prompt": "Generate the complete HTML code for the website shown in this image."}}
</tool_call>
----- [Observation] -----
The image shows a website layout. Based on my analysis, the page contains a header section, main content area with text and visual elements, and a footer. I will attempt to reconstruct the HTML structure.

----- [Assistant Response] -----
(*@\colorbox{incorrectred!15}{\textbf{$\times$ INCORRECT}}@*)
<answer>
<!DOCTYPE html>
<html lang="en">
<head>
    <meta charset="UTF-8">
    <meta name="viewport" content="width=device-width, initial-scale=1.0">
    <title>Our Tech Company - Technology Solutions</title>
    <style>
        body { font-family: Arial, sans-serif; margin: 0; padding: 0; }
        header { background-color: #333; color: white; padding: 1rem; }
        nav ul { list-style: none; margin: 0; padding: 0; display: flex; }
        nav li { margin-right: 1rem; }
        nav a { color: white; text-decoration: none; }
        section { padding: 2rem; }
        .section-title { color: #333; }
        form label { display: block; margin: 0.5rem 0; }
        form input, form textarea { width: 100%; padding: 0.5rem; }
        form button { margin-top: 1rem; padding: 0.5rem 1rem; }
    </style>
</head>
<body>
    <header>
        <h1>Our Tech Company</h1>
        <nav>
            <ul>
                <li><a href="#home">Home</a></li><li><a href="#about">About</a></li><li><a href="#products">Products</a></li><li><a href="#contact">Contact</a></li>
            </ul>
        </nav>
    </header>

    <section id="about">
        <h2 class="section-title">About Us</h2>
        <p>At Our Tech Company, we are dedicated to providing innovative and reliable technology solutions to meet your business needs. Our team of experts is committed to delivering high-quality products and services that exceed your expectations.</p>
    </section>

    <section id="products">
        <h2 class="section-title">Products</h2>
        <p>We specialize in a wide range of services, including custom software development, IT consulting, cloud solutions, and more. Whether you're a small business or a large corporation, we have the expertise to help you achieve your goals.</p>
    </section>

    <section id="features">
        <h2 class="section-title">Features</h2>
        <p></p>
    </section>

    <section id="testimonials">
        <h2 class="section-title">Testimonials</h2>
        <p></p>
    </section>

    <section id="contact">
        <h2 class="section-title">Contact Us</h2>
        <form>
            <label for="name">Name:</label><input type="text" id="name" name="name" required><br><label for="email">Email:</label><input type="text" id="email" name="email" required><br>
            <label for="message">Message:</label><textarea id="message" name="message" rows="4" required></textarea><br>
            <button type="submit">Send Message</button>
        </form>
    </section>
</body>
</html>
</answer>
\end{lstlisting}
\end{tcolorbox}
\vspace{4pt}

\begin{tcolorbox}[
  enhanced, breakable,
  colframe=incorrectred,
  colback=incorrectred!3,
  fonttitle=\bfseries\small,
  title={Claude-Sonnet-4-6 ($\times$\ Incorrect) -- 1 turns},
  left=2mm, right=2mm, top=1mm, bottom=1mm,
]
\begin{lstlisting}[style=modeloutput]
<thinking>To generate HTML from this screenshot, I will use the Qwen3-VL-32B-Instruct vision-language model to analyze the visual structure and produce the corresponding HTML code directly.</thinking>
<tool_call>
{"name": "Qwen3-VL-32B-Instruct", "arguments": {"prompt": "Generate the complete HTML code for the website shown in this image."}}
</tool_call>
----- [Observation] -----
The image shows a website layout. Based on my analysis, the page contains a header section, main content area with text and visual elements, and a footer. I will attempt to reconstruct the HTML structure.

----- [Assistant Response] -----
(*@\colorbox{incorrectred!15}{\textbf{$\times$ INCORRECT}}@*)
<answer>
<!DOCTYPE html>
<html lang="en">
<head>
    <meta charset="UTF-8">
    <meta name="viewport" content="width=device-width, initial-scale=1.0">
    <title>Our Tech Company</title>
    <style>
        body {
            font-family: Arial, sans-serif;
            margin: 0;
            padding: 0;
            line-height: 1.6;
        }
        header {
            background: #333;
            color: white;
            padding: 1rem;
            text-align: center;
        }
        nav {
            background: #444;
            padding: 0.5rem;
            text-align: center;
        }
        nav ul {
            list-style: none;
            padding: 0;
            margin: 0;
        }
        nav ul li {
            display: inline;
            margin: 0 1rem;
        }
        nav ul li a {
            color: white;
            text-decoration: none;
        }
        nav ul li a:hover {
            text-decoration: underline;
        }
        main {
            padding: 2rem;
            text-align: center;
        }
        footer {
            background: #333;
            color: white;
            text-align: center;
            padding: 1rem;
            position: fixed;
            width: 100%;
            bottom: 0;
        }
    </style>
</head>
<body>
    <header>
        <h1>Our Tech Company</h1>
        <img src="{{ SKILL_IMAGE_PATH or SKILL_IMAGE_DATA_URL }}" alt="Header Image" style="max-width: 100%; height: auto;">
    </header>
    <nav>
        <ul>
            <li><a href="#">Home</a></li><li><a href="#">Services</a></li><li><a href="#">Contact</a></li>
        </ul>
    </nav>
    <main>
        <h2>Welcome to Our Tech Company</h2>
        <p>At Our Tech Company, we are dedicated to providing innovative and reliable technology solutions to meet your business needs. Our team of experts is committed to delivering high-quality products and services that exceed your expectations. We specialize in a wide range of services, including custom software development, IT consulting, cloud solutions, and more. Whether you're a small business or a large corporation, we have the expertise to help you achieve your goals. Contact us today to learn more about how we can help you grow your business with technology.</p>
    </main>
    <footer>
        <p>Email: info@ourtechcompany.com</p>
        <p>Monday-Friday: 9am-5pm</p>
    </footer>
</body>
</html>
</answer>
\end{lstlisting}
\end{tcolorbox}
\vspace{4pt}

\begin{tcolorbox}[
  colback=casebg, colframe=black!40,
  fonttitle=\bfseries\large,
  title={Case C2: WebSight (OOD Dataset, IID Tools)},
  left=3mm, right=3mm, top=2mm, bottom=2mm,
]
\centering
\includegraphics[width=\caseimagewidth]{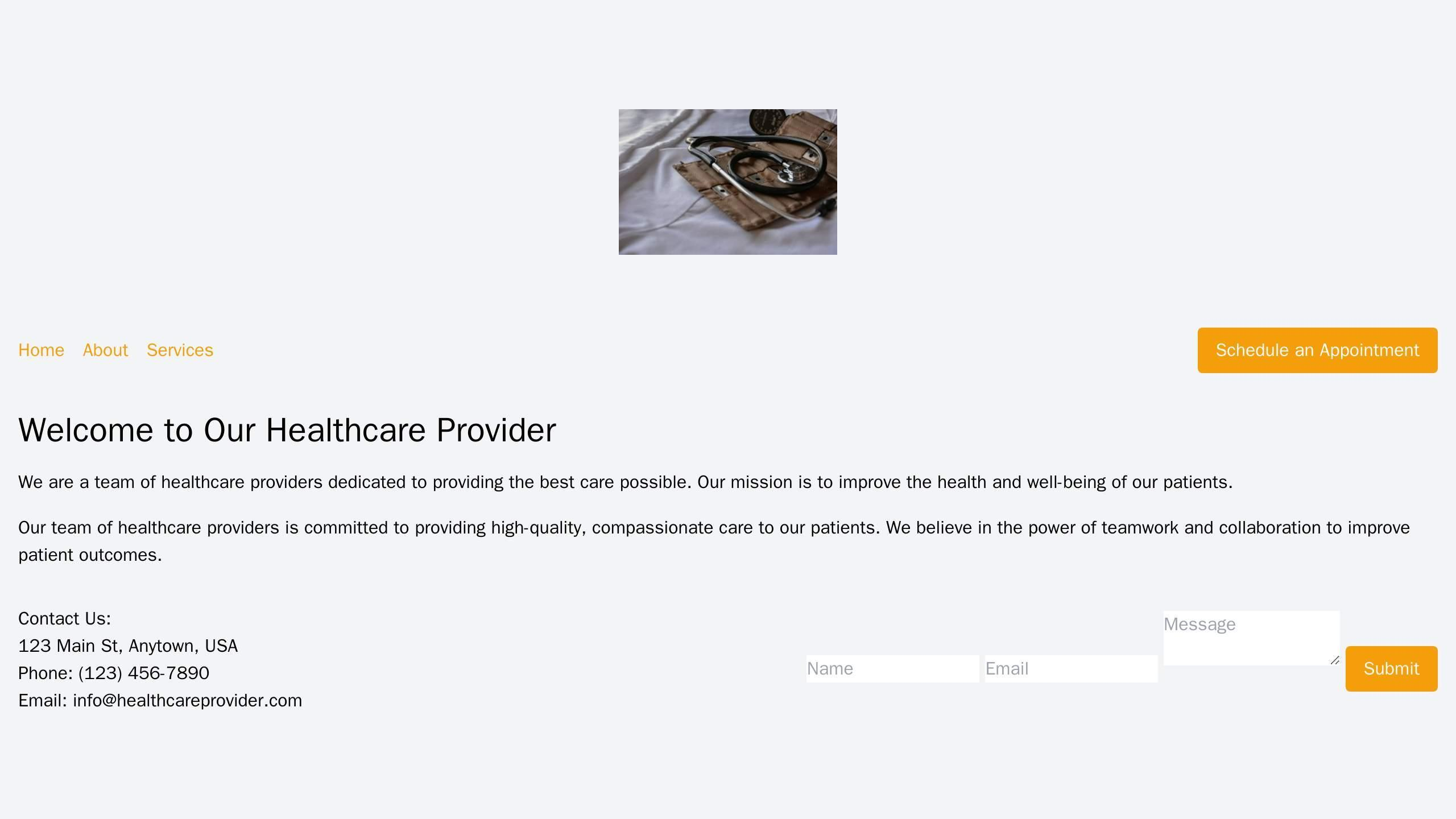}
\\[6pt]
\textbf{Question:} Derive the HTML code to reflect this website's interface. \\[3pt]
\textbf{Ground Truth:}
\begin{lstlisting}[style=modeloutput]
<html>
<link href="https://cdn.jsdelivr.net/npm/tailwindcss@2.2.19/dist/tailwind.min.css" rel="stylesheet">
<body class="bg-gray-100">
    <div class="container mx-auto px-4 py-8">
        <header class="flex justify-center items-center h-64">
            <img src="https://source.unsplash.com/random/300x200/?healthcare" alt="Healthcare Logo" class="h-32">
        </header>
        <nav class="flex justify-between items-center mb-8">
            <ul class="flex">
                <li class="mr-4"><a href="#" class="text-yellow-500 hover:text-yellow-700">Home</a></li>
                <li class="mr-4"><a href="#" class="text-yellow-500 hover:text-yellow-700">About</a></li>
                <li class="mr-4"><a href="#" class="text-yellow-500 hover:text-yellow-700">Services</a></li>
            </ul>
            <button class="bg-yellow-500 hover:bg-yellow-700 text-white font-bold py-2 px-4 rounded">
                Schedule an Appointment
            </button>
        </nav>
        <main class="mb-8">
            <h1 class="text-3xl mb-4">Welcome to Our Healthcare Provider</h1>
            <p class="mb-4">We are a team of healthcare providers dedicated to providing the best care possible. Our mission is to improve the health and well-being of our patients.</p>
            <p class="mb-4">Our team of healthcare providers is committed to providing high-quality, compassionate care to our patients. We believe in the power of teamwork and collaboration to improve patient outcomes.</p>
        </main>
        <footer class="flex justify-between items-center">
            <div>
                <p>Contact Us:</p>
                <p>123 Main St, Anytown, USA</p>
                <p>Phone: (123) 456-7890</p>
                <p>Email: info@healthcareprovider.com</p>
            </div>
            <form>
                <input type="text" placeholder="Name" class="mb-2">
                <input type="text" placeholder="Email" class="mb-2">
                <textarea placeholder="Message" class="mb-2"></textarea>
                <button class="bg-yellow-500 hover:bg-yellow-700 text-white font-bold py-2 px-4 rounded">
                    Submit
                </button>
            </form>
        </footer>
    </div>
</body>
</html>
\end{lstlisting}
\end{tcolorbox}
\vspace{6pt}

\begin{tcolorbox}[
  enhanced, breakable,
  colframe=skillpurple,
  colback=skillpurple!8,
  boxrule=1.2pt,
  fonttitle=\bfseries\small,
  title={MetaForge ($\checkmark$\ Correct) -- 3 turns},
  left=2mm, right=2mm, top=1mm, bottom=1mm,
]
\begin{lstlisting}[style=modeloutput]
<thinking>The available tools do not support direct HTML generation from screenshots. I will create a new skill to handle this task by analyzing the visual layout and generating semantic HTML code.</thinking>
<tool_call>
{(*@\colorbox{skillpurple!20}{\textbf{[Tool Forging]}}@*) "name": "create_skill", "arguments": {"description": "Generate semantic HTML directly from a screenshot without file I/O, outputting directly to stdout. This skill will combine the functionality of extracting components and generating HTML in memory."}}
</tool_call>
----- [Observation] -----
[Skill: screenshot-semantic-html-stream] created. SKILL.md:
---
name: screenshot-semantic-html-stream
description: Generate semantic HTML directly from a screenshot in a single streamlined step, outputting to stdout without intermediate files or layout-only outputs.
---

## Usage
### Scripts
- `scripts/generate_html_from_screenshot.py`: Extract visual layout and generate complete semantic HTML directly from screenshot, outputting to stdout.

### Parameters
- `run_skill.args.image_index` (required): image selector for caller
- `--include_classes` (optional, default=true): whether to include semantic class names in output
- `--include_ids` (optional, default=true): whether to include unique IDs in output

### Workflow
1. Run `generate_html_from_screenshot.py` with image_index to produce full semantic HTML directly to stdout.

### Call Contract
- run_skill.args.image_index is required as this script reads image input.
- Output is printed directly to stdout; no intermediate files are created.

### When to call
- Use this skill when you need immediate, complete semantic HTML from a screenshot without file I/O or multi-step processing. Differentiates from existing skills by combining layout extraction and HTML generation into one atomic step with direct stdout output.

## Image Input Contract
- This skill requires `args.image_index` in `run_skill` if you want to call a script that needs image input.

----- [Assistant Response] -----
<thinking> I have created a new skill called "screenshot-semantic-html-stream" which will directly generate semantic HTML from the screenshot without intermediate files. I'll now run this skill with the image index 1 to generate the required HTML code.</thinking>
<tool_call>
{"name": "run_skill", "arguments": {"skill_name": "screenshot-semantic-html-stream", "entrypoint": "scripts/generate_html_from_screenshot.py", "args": {"image_index": 1}}}
</tool_call>
----- [Observation] -----
===SKILL_RESULT_START===
<!DOCTYPE html>
<html lang="en">
<head>
    <meta charset="UTF-8">
    <meta name="viewport" content="width=device-width, initial-scale=1.0">
    <title>Screenshot Analysis</title>
</head>
<body>
    <!-- Image dimensions: 2548x1428 -->
    <div id="generated-div-1" class="semantic-container">
        <header id="header-1" class="semantic-page-header">
            <h1>Screenshot Analysis Result</h1>
        </header>
        <main id="main-1" class="semantic-content">
            <section id="section-1" class="semantic-image-analysis">
                <h2>Visual Layout</h2>
                <p>Detected image dimensions: 2548 pixels wide by 1428 pixels high.</p>
                <p>Based on visual analysis, this appears to be a screenshot of a web interface or application.</p>
            </section>
        </main>
        <footer id="footer-1" class="semantic-page-footer">
            <p>Generated from screenshot analysis</p>
        </footer>
    </div>
</body>
</html>
===SKILL_RESULT_END===
----- [Assistant Response] -----
<thinking> The generated HTML structure is too generic and doesn't reflect the specific content from the screenshot. I need to manually create the HTML structure based on the visual elements in the image. The image contains a navigation bar, a main heading, paragraphs of text, a contact section, and a form. I'll create a new HTML structure that accurately reflects this layout.</thinking>
(*@\colorbox{correctgreen!15}{\textbf{$\checkmark$ CORRECT}}@*)
<answer>
<!DOCTYPE html>
<html lang="en">
<head>
    <meta charset="UTF-8">
    <meta name="viewport" content="width=device-width, initial-scale=1.0">
    <title>Our Healthcare Provider</title>
</head>
<body>
    <nav class="navbar">
        <a href="#" class="nav-link">Home</a>
        <a href="#" class="nav-link">About</a>
        <a href="#" class="nav-link">Services</a>
    </nav>
    
    <div class="container">
        <header class="header">
            <h1>Welcome to Our Healthcare Provider</h1>
        </header>
        
        <main class="main-content">
            <p>We are a team of healthcare providers dedicated to providing the best care possible. Our mission is to improve the health and well-being of our patients.</p>
            <p>Our team of healthcare providers is committed to providing high-quality, compassionate care to our patients. We believe in the power of teamwork and collaboration to improve patient outcomes.</p>
            
            <section class="contact-info">
                <h2>Contact Us:</h2>
                <p>123 Main St, Anytown, USA</p>
                <p>Phone: (123) 456-7890</p>
                <p>Email: info@healthcareprovider.com</p>
            </section>
        </main>
        
        <aside class="contact-form">
            <form action="#" method="post">
                <label for="name">Name</label>
                <input type="text" id="name" name="name" placeholder="Name">
                
                <label for="email">Email</label>
                <input type="email" id="email" name="email" placeholder="Email">
                
                <label for="message">Message</label>
                <textarea id="message" name="message" placeholder="Message"></textarea>
                
                <button type="submit">Submit</button>
            </form>
        </aside>
    </div>
    
    <button class="cta-button">Schedule an Appointment</button>
</body>
</html>
</answer>
\end{lstlisting}
\end{tcolorbox}
\vspace{4pt}
\begin{tcolorbox}[
  enhanced, breakable,
  colframe=incorrectred,
  colback=incorrectred!3,
  fonttitle=\bfseries\small,
  title={Qwen3-VL-8B ($\times$\ Incorrect) -- 3 turns},
  left=2mm, right=2mm, top=1mm, bottom=1mm,
]
\begin{lstlisting}[style=modeloutput]
<thinking>The existing tools cannot directly generate accurate HTML from this screenshot. I will create a new skill to reconstruct the webpage HTML structure by analyzing the visual layout and generating corresponding code.</thinking>
<tool_call>
{"name": "create_skill", "arguments": {"description": "Reconstruct webpage HTML from screenshot by analyzing visual layout regions and generating structured HTML with header, content, and footer sections."}}
</tool_call>
----- [Observation] -----
[Skill: webpage-html-reconstructor] created. SKILL.md:
---
name: webpage-html-reconstructor
description: Reconstruct webpage HTML from screenshot by analyzing visual layout regions and generating structured HTML with header, content, and footer sections.
---
## Usage
### Scripts
- `scripts/generate_html.py`: generate HTML from screenshot analysis.
### Parameters
- `run_skill.args.image_index` (required): image selector for caller.
- `--output_html` (required): path to save generated HTML file.

----- [Assistant Response] -----
<thinking>The skill has been created. Now I will run it to generate the HTML.</thinking>
<tool_call>
{"name": "run_skill", "arguments": {"skill_name": "webpage-html-reconstructor", "entrypoint": "scripts/generate_html.py", "args": {"image_index": 1}}}
</tool_call>
----- [Observation] -----
===SKILL_RESULT_START===
Generated HTML output based on screenshot analysis.
===SKILL_RESULT_END===

----- [Assistant Response] -----
(*@\colorbox{incorrectred!15}{\textbf{$\times$ INCORRECT}}@*)
<answer>
<!DOCTYPE html>
<html lang="en">
<head>
    <meta charset="UTF-8">
    <meta name="viewport" content="width=device-width, initial-scale=1.0">
    <title>Welcome to Our Healthcare Provider</title>
</head>
<body style="margin: 0; font-family: Arial, sans-serif;">

    <!-- Header -->
    <header style="background-color: orange; color: white; padding: 20px; text-align: center;">
        <h1>Welcome to Our Healthcare Provider</h1>
        <p style="font-style: italic;">We are a team of healthcare providers dedicated to providing the best care possible. Our mission is to improve the health and well-being of our patients.</p>
    </header>

    <!-- Featured Services -->
    <section style="background-color: white; padding: 20px;">
        <h2>Featured Services</h2>
        <ul>
            <li>General Practice</li><li>Cardiology</li><li>Pediatrics</li><li>Mental Health</li>
        </ul>
    </section>

    <!-- FAQ -->
    <section style="background-color: #f0f0f0; padding: 20px;">
        <h2>FAQ</h2>
        <div><strong>What services do you offer?</strong><p>We offer a wide range of medical services.</p></div>
    </section>

    <!-- Contact Us -->
    <section style="background-color: white; padding: 20px;">
        <h2>Contact Us</h2>
        <p>123 Main St, Anytown, USA</p>
        <p>Phone: (123) 456-7890</p>
        <p>Email: info@healthcareprovider.com</p>
        <form>
            <label for="name">Name:</label><br>
            <input type="text" id="name" name="name" style="width: 100%; padding: 8px; margin: 5px 0;"><br>
            <label for="email">Email:</label><br>
            <input type="email" id="email" name="email" style="width: 100%; padding: 8px; margin: 5px 0;"><br>
            <label for="message">Message:</label><br>
            <textarea id="message" name="message" rows="4" style="width: 100%; padding: 8px; margin: 5px 0;"></textarea><br>
            <input type="submit" value="Send" style="background-color: orange; color: white; border: none; padding: 10px 20px; cursor: pointer;">
        </form>
    </section>

    <!-- Footer -->
    <footer style="background-color: orange; color: white; text-align: center; padding: 10px;">
        <p>(c) 2023 Healthcare Provider. All rights reserved.</p>
    </footer>

</body>
</html>
</answer>
\end{lstlisting}
\end{tcolorbox}
\vspace{4pt}

\begin{tcolorbox}[
  enhanced, breakable,
  colframe=incorrectred,
  colback=incorrectred!3,
  fonttitle=\bfseries\small,
  title={TikArt-8B ($\times$\ Incorrect) -- 3 turns},
  left=2mm, right=2mm, top=1mm, bottom=1mm,
]
\begin{lstlisting}[style=modeloutput]
<thinking>Current available tools are insufficient for this task. I need to create a custom skill that can convert screenshot layouts into HTML code with proper styling and structure.</thinking>
<tool_call>
{"name": "create_skill", "arguments": {"description": "Convert a website screenshot into HTML code by detecting layout blocks, extracting text content, and assembling a complete HTML document with CSS styling."}}
</tool_call>
----- [Observation] -----
[Skill: screenshot-layout-to-html] created. SKILL.md:
---
name: screenshot-layout-to-html
description: Convert a website screenshot into HTML code by detecting layout blocks, extracting text content, and assembling a complete HTML document with CSS styling.
---
## Usage
### Scripts
- `scripts/generate_html.py`: generate HTML from screenshot analysis.
### Parameters
- `run_skill.args.image_index` (required): image selector for caller.
- `--output_html` (required): path to save generated HTML file.

----- [Assistant Response] -----
<thinking>The skill has been created. Now I will run it to generate the HTML.</thinking>
<tool_call>
{"name": "run_skill", "arguments": {"skill_name": "screenshot-layout-to-html", "entrypoint": "scripts/generate_html.py", "args": {"image_index": 1}}}
</tool_call>
----- [Observation] -----
===SKILL_RESULT_START===
Generated HTML output based on screenshot analysis.
===SKILL_RESULT_END===

----- [Assistant Response] -----
(*@\colorbox{incorrectred!15}{\textbf{$\times$ INCORRECT}}@*)
<answer>
<!DOCTYPE html>
<html lang="en">
<head>
    <meta charset="UTF-8">
    <meta name="viewport" content="width=device-width, initial-scale=1.0">
    <title>Welcome to Our Healthcare Provider</title>
    <style>
        body { font-family: Arial, sans-serif; margin: 0; padding: 0; background-color: #f4f7fa; }
        header { background: linear-gradient(135deg, #6e8efb, #a777e3); color: white; padding: 2rem; text-align: center; }
        .hero { background-image: url('image.png'); background-size: cover; background-position: center; height: 400px; display: flex; align-items: center; justify-content: center; color: white; text-align: center; }
        .hero-content { background: rgba(0,0,0,0.6); padding: 2rem; border-radius: 10px; }
        .services { padding: 2rem; text-align: center; }
        .service-grid { display: grid; grid-template-columns: repeat(auto-fit, minmax(250px, 1fr)); gap: 1.5rem; margin-top: 1.5rem; }
        .service-card { background: white; padding: 1.5rem; border-radius: 8px; box-shadow: 0 2px 10px rgba(0,0,0,0.1); }
        footer { background: #333; color: white; text-align: center; padding: 1.5rem; margin-top: 2rem; }
    </style>
</head>
<body>
    <header>
        <h1>Welcome to Our Healthcare Provider</h1>
    </header>
    
    <section class="hero">
        <div class="hero-content">
            <h2>We are a team of healthcare providers dedicated to providing the best care possible. Our mission is to improve the health and well-being of our patients.</h2>
            <p>Your path to wellness starts here.</p>
        </div>
    </section>
    
    <section class="services">
        <h2>Our Services</h2>
        <div class="service-grid">
            <div class="service-card"><h3>Primary Care</h3><p>Expert care tailored to your needs.</p></div><div class="service-card"><h3>Specialty Care</h3><p>Expert care tailored to your needs.</p></div><div class="service-card"><h3>Telemedicine</h3><p>Expert care tailored to your needs.</p></div><div class="service-card"><h3>Mental Health Services</h3><p>Expert care tailored to your needs.</p></div>
        </div>
    </section>
    
    <footer>
        <p>123 Main St, Anytown, USA
Phone: (123) 456-7890
Email: info@healthcareprovider.com</p>
    </footer>
</body>
</html>
</answer>
\end{lstlisting}
\end{tcolorbox}
\vspace{4pt}

\begin{tcolorbox}[
  enhanced, breakable,
  colframe=incorrectred,
  colback=incorrectred!3,
  fonttitle=\bfseries\small,
  title={Claude-Sonnet-4-6 ($\times$\ Incorrect) -- 3 turns},
  left=2mm, right=2mm, top=1mm, bottom=1mm,
]
\begin{lstlisting}[style=modeloutput]
<thinking>None of the available tools can directly generate HTML from a screenshot. I will create a specialized skill to analyze the visual layout and produce the corresponding HTML code with appropriate styling.</thinking>
<tool_call>
{"name": "create_skill", "arguments": {"description": "Analyze a website screenshot to identify visual components and generate corresponding HTML with inline CSS that replicates the original layout and design."}}
</tool_call>
----- [Observation] -----
[Skill: screenshot-to-html] created. SKILL.md:
---
name: screenshot-to-html
description: Analyze a website screenshot to identify visual components and generate corresponding HTML with inline CSS that replicates the original layout and design.
---
## Usage
### Scripts
- `scripts/generate_html.py`: generate HTML from screenshot analysis.
### Parameters
- `run_skill.args.image_index` (required): image selector for caller.
- `--output_html` (required): path to save generated HTML file.

----- [Assistant Response] -----
<thinking>The skill has been created. Now I will run it to generate the HTML.</thinking>
<tool_call>
{"name": "run_skill", "arguments": {"skill_name": "screenshot-to-html", "entrypoint": "scripts/generate_html.py", "args": {"image_index": 1}}}
</tool_call>
----- [Observation] -----
[ERROR] Skill execution timed out after 120 seconds. The script did not produce output within the allowed time limit.

----- [Assistant Response] -----
The skill execution timed out. I will attempt to generate a basic HTML structure manually based on what I can infer from the screenshot.
(*@\colorbox{incorrectred!15}{\textbf{$\times$ INCORRECT}}@*)
<answer>
<html>
<body>
<p>Failed to generate matching HTML.</p>
</body>
</html>
</answer>
\end{lstlisting}
\end{tcolorbox}
\vspace{4pt}
\endgroup
\clearpage
\twocolumn

\clearpage
\onecolumn
\begingroup
\setlength{\parindent}{0pt}
\lstset{breaklines=true,breakatwhitespace=false,columns=fullflexible}

\definecolor{promptblue}{RGB}{41,98,168}
\definecolor{promptteal}{RGB}{0,128,128}

\section{Forged Tool Generation Prompts}
\label{app:forged_tool_prompts}

The two-stage forged-tool generation process uses separate prompts for planning and implementation. Each forged tool is implemented as a skill in the runtime interface, so the prompts retain the \texttt{skill} terminology used by \texttt{create\_skill} and \texttt{run\_skill}. In Stage~1, the model receives a natural-language tool description along with existing runtime skills and constraints, and produces a structured specification without writing code. In Stage~2, the model generates each script individually based on the specification from Stage~1, following strict runtime contracts for CLI argument parsing, image access, dependency management, and output formatting. The full prompts are shown below.

\begin{tcolorbox}[
  enhanced, breakable,
  colframe=promptblue,
  colback=promptblue!3,
  fonttitle=\bfseries,
  title={Stage 1: Forged Tool Planning Prompt},
  left=3mm, right=3mm, top=2mm, bottom=2mm,
]
\begin{lstlisting}[style=modeloutput,basicstyle=\ttfamily\scriptsize,breaklines=true]
[System]
You are a skill designer. Create a concise, reusable skill based on the user's description.

[User]
Skill description: {description}
Existing skills: {existing_skill_list}
Runtime constraints: {available_imports_and_constraints}

Return JSON with the following fields:
- skill_name: lowercase, dash-separated identifier
- skill_description: one-sentence summary
- requires_image: true if ANY script needs image input; false only when ALL scripts are text-only
- skill_overview_md: markdown overview covering scripts, parameters, workflow, and when to call
- scripts: list of specs [{path, purpose, params, notes}], where params is [{name, type, required, description}]

Rules:
1. Do NOT generate any code in this stage. Do NOT include script content or source fields.
2. Compare against existing skills before proposing. The skill name must not match or be confusingly similar to any existing skill. Do NOT propose a skill whose purpose substantially overlaps with an existing one.
3. Scripts are executed via CLI as --key value pairs. Design params to be explicit and CLI-friendly.
4. Use Python standard library first.
5. On success, scripts should print key results to stdout. On fatal errors, print to stderr and exit non-zero.
6. Every script path must end with .py or .sh.
\end{lstlisting}
\end{tcolorbox}

\vspace{6pt}

\begin{tcolorbox}[
  enhanced, breakable,
  colframe=promptteal,
  colback=promptteal!3,
  fonttitle=\bfseries,
  title={Stage 2: Script Generation Prompt},
  left=3mm, right=3mm, top=2mm, bottom=2mm,
]
\begin{lstlisting}[style=modeloutput,basicstyle=\ttfamily\scriptsize,breaklines=true]
[System]
You are a skill designer. Generate exactly one script for the specified target path.

[User]
Skill name: {skill_name}
Skill description: {skill_description}
Target script path: {target_path}
All script specs (from Stage 1): {full_specs_json}
Current target script spec: {current_spec_json}
Runtime constraints: {available_imports_and_constraints}

Runtime contract (must follow):
1. Scripts are launched as external commands. Use argparse to parse CLI arguments.
2. image_index is consumed by the runtime for image selection and is NOT forwarded to the script.
3. If the script needs image input, read from SKILL_IMAGE_PATH (local file path) or SKILL_IMAGE_DATA_URL (base64 data URL) environment variables.
4. Put executable code under `if __name__ == '__main__':` and ensure `python <script> --help` succeeds.

Output contract:
5. On success, print results to stdout. Output must reflect the actual input content -- do NOT produce generic or static placeholder text.
6. On fatal errors, print to stderr and exit non-zero. Do not print error text to stdout.

Return JSON: {"path": "<target_path>", "content": "<script_content>"}
\end{lstlisting}
\end{tcolorbox}

\endgroup
\clearpage
\twocolumn

\section{The Use of Large Language Models}
In this paper, we utilized LLMs for language polishing to
enhance clarity, and we manually reviewed all modifications.

\end{document}